\documentclass[longauth]{aa}

\usepackage[varg]{txfonts}
\usepackage[]{graphicx}
\usepackage{subfigure}
\usepackage{transparent}
\usepackage{natbib}
\usepackage{bm}
\usepackage[hang,multiple]{footmisc}
\usepackage[normalem]{ulem}
\bibpunct{(}{)}{;}{a}{}{,}

\providecommand{\abs}[1]{\left\lvert#1\right\rvert} 
\newcommand{\dd}{\mathrm{d}} 
\newcommand{\e}[1]{\times 10^{#1}} 
\newcommand{\ml}[1]{\mathrm{#1}} 
\newcommand{\dg}{^{\circ}}
\newcommand{\specialcell}[2][l]{%
  \begin{tabular}[#1]{@{}l@{}}#2\end{tabular}}
\graphicspath{ {./Figures/} }
\newcommand\elec{\mathrm{e^-}}
\newcommand{\bit}{\begin{itemize}}
\newcommand{\bits}{\begin{itemize}\addtolength{\itemsep}{-0.5\baselineskip}} 
\newcommand{\ei}{\end{itemize}}
\newcommand{\me}{\mathrm{M}_{\oplus}} 
\newcommand{\ms}{\mathrm{M}_{\odot}} 

\begin{document}
\title{A detector interferometric calibration experiment for high precision astrometry.}

\subtitle{}

\author{A. Crouzier\inst{1} \and F. Malbet \inst{1} \and F. Henault\inst{1} \and A. Léger\inst{3} \and C. Cara\inst{2} \and J. M. LeDuigou\inst{4} \and O. Preis\inst{1} \and P. Kern\inst{1} \and A. Delboulbe\inst{1} \and G. Martin\inst{1} \and P. Feautrier\inst{1} \and E. Stadler\inst{1} \and S. Lafrasse\inst{1} \and S. Rochat\inst{1} \and C. Ketchazo\inst{2} \and M. Donati\inst{2} \and E. Doumayrou\inst{2} \and P. O. Lagage\inst{2} \and M. Shao \inst{5} \and R. Goullioud\inst{5} \and B. Nemati\inst{5} \and C. Zhai\inst{5} \and E. Behar\inst{1} \and S. Potin\inst{1} \and M. Saint-Pe\inst{1} \and J. Dupont\inst{1}
}

\offprints{R. Plemmons, \email{plemmons@...}}

\institute{Institut de Planétologie et d'Astrophysique de Grenoble, 414, Rue de la Piscine
Domaine Universitaire, 38400 St-Martin d’Hères, France
\and Commisariat à l'Energie Atomique et aux Energies Alternatives, Saclay, centre d'études nucléaires de Saclay, Paris, France
\and Institut d'Astrophysique Spatiale, Centre universitaire d'Orsay, Paris, France
\and Centre National d'Etudes Statiales, 2 place Maurice Quentin, Paris, France
\and Jet Propulsion Laboratory, 4800 Oak Grove Drive, Pasadena, CA, U.S.A. 91109
}

\date{Received xx January 20xx / Accepted xx January 20xx}

\abstract
{Exoplanet science has made staggering progress in the last two decades, due to the relentless exploration of new detection methods and refinement of existing ones. Yet astrometry offers a unique and untapped potential of discovery of habitable-zone low-mass planets around all the solar-like stars of the solar neighborhood. To fulfill this goal, astrometry must be paired with high precision calibration of the detector.}
{We present a way to calibrate a detector for high accuracy astrometry. An experimental testbed combining an astrometric simulator and an interferometric calibration system is used to validate both the hardware needed for the calibration and the signal processing methods. The objective is an accuracy of $5\e{-6}$ pixel on the location of a Nyquist sampled polychromatic point spread function.} 
{The interferometric calibration system produced modulated Young fringes on the detector. The Young fringes were parametrized as products of time and space dependent functions, based on various pixel parameters. The minimization of function parameters was done iteratively, until convergence was obtained, revealing the pixel information needed for the calibration of astrometric measurements.} 
{
The calibration system yielded the pixel positions to an accuracy estimated at $4\e{-4}$ pixel. After including the pixel position information, an astrometric accuracy of $6\e{-5}$ pixel was obtained, for a PSF motion over more than five pixels. In the static mode (small jitter motion of less than $1\e{-3}$ pixel), a photon noise limited precision of $3\e{-5}$ pixel was reached.
} 
{}


\maketitle 


\section{Introduction}\label{sec:Introduction}


The year 2014 was marked by a symbolic yet significant milestone, the number of confirmed exoplanets exceeded 1000. The pace of discoveries is accelerating: at the time of writing, the exoplanet.eu database shows more than 3400 confirmed planets \citep{2011AA...532A..79S}, the last recent increase being mostly due to the success of the Kepler mission \citep{2014ApJ...784...45R}. Another interesting trend has been the discovery of increasingly smaller planets, down to the terrestrial ones. In some specific cases, we can detect these terrestrial planets in the habitable-zones of their stars, for example with transits \citep{2015arXiv150101101T} or for M stars \citep[][hereafter B13]{2013AA...549A.109B}.

However, in the present state of exoplanet detection techniques, most likely none of the rocky planets of the Solar System would be discovered, even around a star as close as $\alpha$ Centauri, our closest Sun-like neighbor, located at the distance of 1.34 pc \citep{exoplanetDetMethods}. Rocky planets would only be found if the observer (located in a random direction near the Solar System) was lucky enough to observe their transits. Yet the rocky planets are a very strong constraint on the scenarios of formation of planetary systems \citep{2012AREPS..40..251M}. 

While for the question of planet formation, it is possible to rely on the power of statistics and the increasing number of detections, the question of life remains unanswered. Finding potentially habitable Earths twins in the Solar neighborhood would be a major step forward for exoplanet detection and these planets would be prime targets for attempting to find life outside of the Solar System \citep{2014SPIE.9148E..20M}. The next step is to search for bio-markers in their atmospheres by spectroscopy \citep{2010ARAA..48..631S}.


Astrometry, by measuring the gravitational perturbation of planets on their central host stars, can determine the mass of planets and their orbits. From space, differential astrometry at a sub micro arcsecond ($\mu$as) accuracy around nearby solar-type stars can detect exoplanets down to one Earth mass in habitable-zone \cite[][hereafter M14]{2014SPIE.9143E..2LM}. 

The angular amplitude of the gravitational perturbation (the astrometric signal) is given by:
\begin{equation}\label{eq:astrometric_signal}
A = 3\,\mu \ml{as} \times \frac{M_{\ml{Planet}}}{\me} \times \left(\frac{M_{\ml{Star}}}{\ms}\right)^{-1} \times \frac{a}{1\ml{AU}} \times \left(\frac{D}{1\ml{pc}}\right)^{-1}.
\end{equation}

Where $D$ is the distance between the Sun and the observed star, $M_{\ml{Planet}}$ is the exoplanet mass, $a$ is the exoplanet semi major axis and $M_{\ml{Star}}$ is the mass of the observed host star. The constant $3$~$\mu$as corresponds to the signal of an exo-Earth observed from a distance of one parsec. In order to look for exo-Earths around Sun-like stars up to 10 pc (about one hundred targets), the signal to be detected is 0.3 $\mu$as. A crucial advantage expected for this method is that the astrometric jitter from stellar activity is small. Solar observations coupled with numerical models showed that the jitter should be smaller than the signal of an exo-Earth except for very active stars \citep{Makarov2010}, at least five times more active than the Sun \citep{Lagrange2011}.

Given the current biases and limitations of the two major exoplanet detection techniques in use today (radial velocities and transits), current knowledge of exoplanets around nearby stars is still incomplete. 
Out of the 455 main sequence stars of the Hipparcos catalog located at a distance of less than 20 pc from the Sun, only 43 (9.5\%) have known exoplanets. This statistic was obtained from a crossmatch between Hipparcos and the exoplanet.eu database, updated on 09 Mar 2016 \citep{2015PhDT.........7C}. The true occurrence rate of planets is significantly higher than 9.5\% \citep{Wolfgang2011,2013ApJ...764..105S}, therefore many more planets remain to be discovered. Past surveys only probed a small part of the orbital parameter space and we suspect that most of those stars have planets.

%

\begin{figure*}[t]
\sidecaption
\includegraphics[width = 12cm]{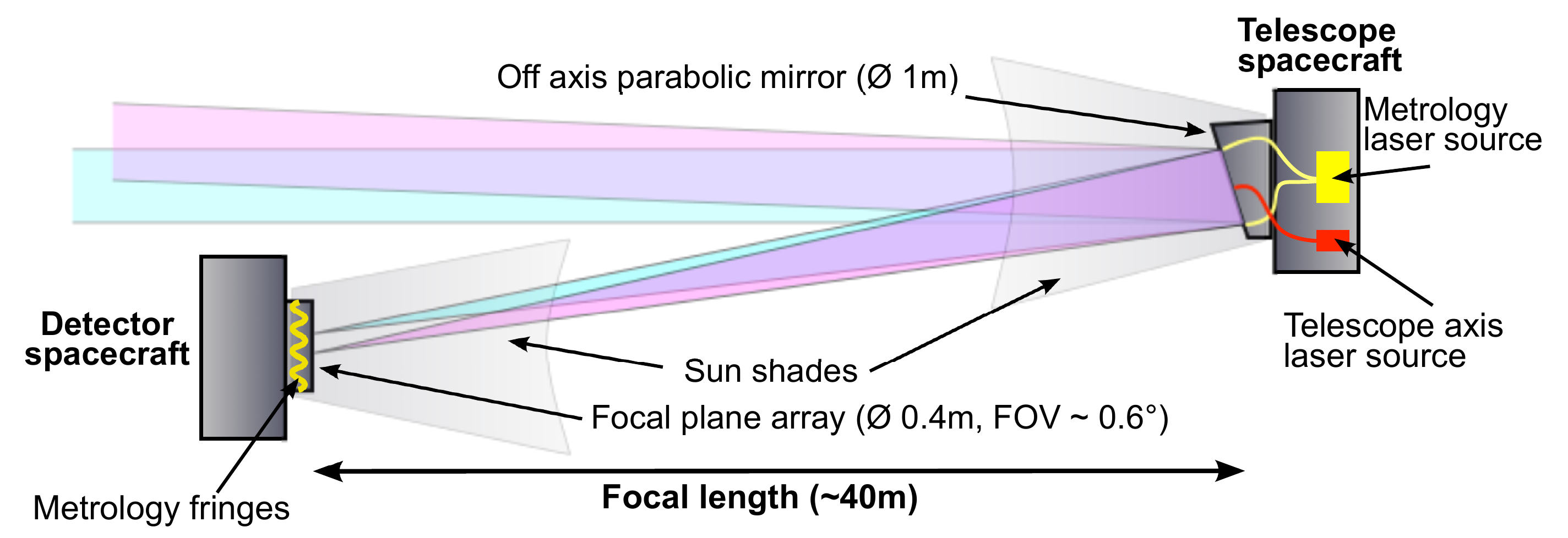}
\caption{Schematic of the NEAT formation flying spacecraft. NEAT is composed of a mirror spacecraft and a detector spacecraft. The optical configuration is an off-axis parabola so there is no vignetting of the FoV. Sun shades prevent direct or reflected Sun light from reaching the CCD. A metrology system with laser beams launched from fibers located on the mirror projects dynamic Young fringes on the detector. The fringes allow a very precise calibration of the CCD. All three aspects (metrology, no vignetting, and Sun shades) are critical to reach micro arcsecond accuracy.}
\label{neat_concept_diagram_v3}
\end{figure*}
This paper is about DICE, an interferometric calibration experiment of a (visible light) detector, which primary goal is to demonstrate the feasibility of sub-$\mu$as astrometry. The experiment is carried with a testbed that was assembled and operated at IPAG and funded by CNES in the context of the NEAT mission proposal to ESA in 2010 \citep{2012ExA....34..385M,2015PhDT.........7C}. The experiment only tackles the detector calibration issue (not the optical aspect). We have already presented the progress of the experiment \citep{crouzier12,crouzier13,crouzier14}. Here we present the scientific context (Sect. \ref{sec:Scientific context}), the experiment goal and principle (Sect. \ref{sec:Calibration experiment}), the data processing methods and their validation using numerical simulations (respectively Sect. \ref{sec:Data processing} and \ref{sec:Numerical simulations for DICE}) and the latest results obtained with testbed data (Sect. \ref{sec:Analysis of experimental data}).

\section{Scientific context}\label{sec:Scientific context}

\subsection{The case for sub micro-arcsecond astrometry in the current context}

The case for sub-$\mu$as astrometry resides in our current difficulties to find exo-Earths around Sun-like stars in the solar neighborhood (distance $<20$ pc) and to measure their masses. Next generation RVs instruments (e.g. EXPRESSO, CODEX) aim at a precision of 0.1 m/s \citep{2014AN....335....8P}, which is the level required for an exo-Earth. However, below 1 m/s, the noise due to stellar activity and stellar spots is much larger than the instrument noise for most targets, as is illustrated by the case of $\alpha$ Cen Bb \citep{2013ApJ...770..133H,2016MNRAS.456L...6R}.

PLATO \citep{2013arXiv1310.0696R} and TESS \citep{2010AAS...21545006R} will discover many transiting planets closer to our Sun than those that Kepler has already found. JWST will obtain transiting and eclipse spectra, down to a few Earth masses \citep{2012ExA....34..311T,2009PASP..121..952D}. But for very close stars, transits are impaired by the geometric transit probability: there are only about 400 Sun-like stars (main sequence F, G and K spectral types) closer than 20 pc. The frequency of Earth analogs per Sun-like star, defined as $\zeta_{0.1}$, the terrestrial planet occurrence rate with radius within 20\% of Earth radius and period within 20\% of Earth period, per star, derived from the Kepler data, is still a hotly debated number, with a wide uncertainty from 0.01 to 2 per star \citep{Petigura2013,2014ApJ...795...64F,2015ApJ...809....8B}. Unless believing the extremely optimistic range of estimates, with a transit probability of 0.5\% for an Earth analog, there may be very few or no nearby transiting exo-Earth (around a Sun-like star) to detect. This will not prevent space born missions to successfully survey bright stars over all the sky, but to overcome the transit probability they have to mostly look at targets further out than 20 pc.

At last, direct imaging: this method has the capability of both finding our closest planetary neighbors and performing spectroscopy to characterize their atmospheres and surface properties. But the angular separation and contrast requirements restrict us to close stars, at about 10 pc \citep{2006ApJS..167...81G}. This distance limit still leaves only one hundred main sequence stars as potential targets\footnote{Based on the Hipparcos catalog \citep{1997AA...323L..49P}, which includes spectral types from A to a few early M stars for these distances}. The most serious issue with a detection by direct imaging alone is that it gives a weak constraint on the planet mass. The radius has to be constrained from the measured flux, assuming or guessing a planetary albedo. Then mass limits could only be estimated from the radius, using more models and/or mass-radius scatter diagrams of exoplanets for which both mass and radius have been measured. The end products are highly model dependent approximative mass limits. Having both mass and radius gives the planet mean density and allows distinction between rocky planets, water ocean-planets, and planets with hydrogen rich atmospheres. A good constraint on mass is therefore critical to have before making exobiological statements. Given the limitations of the other methods discussed above, astrometry is an interesting alternative to develop.

\subsection{Short-term perspectives for space astrometry}

Despite the great potential of exoplanet astrometric detection, all current astrometric instruments are still far from sub-$\mu$as accuracy. This is the consequence of the extreme requirement and associated technical challenge. The current cornerstone astrometric mission, Gaia, is performing a global astrometric survey with expected end of mission accuracies of 10~$\mu$as in the best cases, for visible magnitudes between 6 and 13 \citep{2010IAUS..261..296L}. Gaia has a bright limit caused by saturation at $V_{\ml{mag}}$~=~6 \citep{2014SPIE.9143E..0YM}, which corresponds to the Sun at 10 pc. For faint stars ($V_{\ml{mag}}$ > 13) the accuracy limited by the photon noise. The exoplanet yield from the Gaia mission hold great promise, with a predicted number of detections of 21,000$\pm$6,000 around Sun-like stars \citep{2014ApJ...797...14P}, plus $\sim$2600 around M dwarfs \citep{2014MNRAS.437..497S}, as well as up to hundreds around binaries \citep{2015MNRAS.447..287S}. The total is an order of magnitude more than the current number of confirmed detections. However, because of this 10 $\mu$as threshold, Gaia will only discover gas giants. There is an ongoing study to see whether a special electronic readout mode could be used to measure stars brighter than $V_{\ml{mag}}$~=~6, but the accuracy would still be around the 10 $\mu$as mark, as opposed to degraded accuracy \citep{2016arXiv160508347S}.


\subsection{Technical developments for high precision astrometry}
To find exo-Earths up to 10 pc, the requirement is an end of mission accuracy below 1 $\mu$as, for which no instrument is currently planned. Progress was made in this direction through multiple technology demonstrations during the preparation of the SIM-Lite mission \citep{2008PASP..120...38U}, which was a space interferometer capable of reaching sub-$\mu$as accuracy. SIM-Lite was however canceled in 2010 by NASA, leaving no perspective for sub-$\mu$as astrometry in the near future. As an alternative \citet[hereafter M12]{2012ExA....34..385M} proposed the Nearby Earth Astrometric Telescope (hereafter NEAT) to ESA in 2010. Figure \ref{neat_concept_diagram_v3} is a schematic of the spacecraft, which is designed specifically for sub-$\mu$as differential astrometry over a small field of view (hereafter FoV) of 0.6$\dg$. Early 2015, a revised concept for differential astrometry called Theia has been proposed (M4 candidates), using a three mirror Korsch configuration on a single spacecraft instead of formation flying (M14).



The direct imager concepts presented above require special kinds of calibrations to reach sub-$\mu$as accuracy. This is needed for their advertised scientific objectives, in particular nearby exo-Earth detection. For NEAT, a calibration of the detector is required, to control errors caused by pixel non-homogeneous responses (see Sect. \ref{subsec:The calibration requirement}). For Theia, calibration of the detector and also the optical field distortion is necessary \citep{2015pathwaysMalbet,Malbet2016}.

\section{Detector interferometric calibration experiment (DICE)}\label{sec:Calibration experiment}

DICE specifically uses interference fringes of a coherent source to calibrate a visible detector. The focus of the experiment is on astrometric performance (with enhancement by the interferometric calibration). Demonstration of a precision sufficient for nearby exo-Earth astrometric detection is the main and original motivation, but many other applications are possible. The same calibration technique can also be used to characterize the gain in photometric accuracy (e.g. for transit detection) or for high resolution spectrometry (e.g. for RVs). In general, this technique is a powerful tool to finely characterize detectors.

\subsection{Introduction: precursor experiments, state of the art calibration techniques, alternatives methods}

We have taken advantage of past experience at the JPL where a similar testbed called MCT (Micropixel Centroid Testbed) has been developed. The MCT testbed also included both artificial stars (for astrometry) and an interferometric metrology system for calibration. With static stars (common motion of only $2.5\e{-4}$ pixel), an astrometric precision of $3\e{-5}$ pixel was obtained \citep{Nemati11}.

DICE is a very similar and successor experiment to MCT, the main difference being in the metrology system, which is entirely made of integrated components for DICE. IPAG has a lot of experience with integrated optics and this new configuration avoids polarization stability issues encountered by MCT. In this context, the need for the DICE experiment was driven by mastering of the technology in Europe (for ESA proposals) and to improve upon the result obtained with MCT in order to reach the NEAT astrometric specification.

A first experiment of stand-alone calibration (without astrometric simulator) using Young interference fringes was done by \cite{1995ApOpt..34.6672S} and has yielded pixel positions with an accuracy of 0.01 pixel. Prior to our experiment and its JPL homologue (MCT), this was the best known accuracy (for pixel positions) using this technique.

Several other techniques for detector calibration exist. An intuitive way to obtain very detailed information on the pixel response profile (hereafter PRF) is to perform a spot scan. \cite{1998OptEn..37..948K} measured a high resolution PRF of a single pixel (plus some adjacent pixels for crosstalk), using a scanning electron microscope to project a small light beam (Ø 0.5 $\mu$m). However this technique is impractical to scan a large number of pixels because it is very slow.

To mitigate the speed issue, a team located at CEA Saclay is currently developing a different calibration technique using spot arrays \citep{2014SPIE.9154E..1YK} generated by self imaging gratings \citep{Guerineau:01}. The goals for the calibration are more general than precision astrometry, it includes for example photometry of diffuse objects.  The technique is well suited to obtain an intrapixel calibration (knowledge of PRFs), which is needed to improve the photometry \citep{2012Ingalls}. However the pixel positions cannot yet be measured at high precision with this technique, which is still in development. The technique is also sensitive to optics alignments.


Although it does not include any special optical device to perform calibrations beyond flat field, the calibration process used for the Kepler spacecraft \citep{2010SPIE.7740E..1XQ} is worth mentioning. Because of the high SNR required to detect small transiting exoplanets, a calibration that goes beyond the basic dark and flat fields is used. Numerous smaller systematics or parasitic effects are removed, such as cosmic rays, smear, electronic undershoot/overshoot and non linearity. Developing, validating and continuously maintain and upgrading such a sophisticated model requires a highly integrated architecture, from raw detector values to scientific observables, with end to end simulations to validate the entire pipeline. 

In the DICE experiment we tend toward this kind of ideal situation: this is precisely why the experiment combines a calibration system coupled with an astrometric simulator, plus a numerical model allowing to inject simulated data. The ultimate validation of the interferometric calibration technique is to verify that it improves the astrometric accuracy.



\subsection{The calibration requirement}\label{subsec:The calibration requirement}

Actual CCD and CMOS detectors are not perfectly homogeneous: pixel quantum efficiency (hereafter QE) and gain are non uniform among all pixels and the pixel layout is not perfectly regular. QE and gain non uniformity is calibrated by the usual flat field technique, but for sub-$\mu$as astrometry the pixel (spatial) offsets, that is the distances between the true pixel positions and a perfectly regular layout, needs to be measured. The ‘‘spatial offset'' is not to be confused with pixel ‘‘electronic offsets''. When using the term ‘‘pixel offset'' we mean spatial offset, not electronic offset. The challenge is not so much in the need to measure these pixel offsets, it is set by the accuracy at which both types of calibration (flat-field and offsets) must be done. 

The exact calibration requirements are derived from the angular accuracy needed for exo-Earth detection, the geometric optical parameters of the telescope considered for the astrometric mission and the Nyquist sampling condition. Equation \ref{eq:astrometric_signal} gives the smallest astrometric signal to be detected: 0.3 $\mu$as, i.e an exo-Earth at 10 pc. The corresponding end of mission accuracy needed is $ \sigma_{\ml{f}}=$ 0.05 $\mu$as when considering a signal to noise ratio of 6. The number of single epoch measurements ($N_{\ml{meas}}$) can be set conservatively as twice the number of parameters needed for the astrometric fit ($N_{\ml{param}}$). Assuming an average of $p=3$ planets per star, the astrometric fit requires $5 + 7p=26$ parameters (M12). The corresponding single epoch accuracy ($\sigma_{\ml{epoch}}$) follows from the relation:
\begin{equation}
 \sigma_{\ml{f}} = \frac{\sigma_{\ml{epoch}}}{\sqrt{N_{\ml{meas}}-N_{\ml{param}}}} =  \frac{\sigma_{\ml{epoch}}}{\sqrt{N_{\ml{param}}}}.
\end{equation}
This angular value $\sigma_{\ml{epoch}}$ is then converted in pixel units, using the size of the pixel on the sky and the Nyquist relation: $2e = \lambda f/D$, where $e$ is the pixel size, $f$ the telescope focal, $D$ the primary mirror diameter and $\lambda$ the blue cut-off wavelength. For the NEAT concept, the residual calibration error when calculating the centroid location from the PSF must be smaller than $5\e{-6}$ pixel (M12). For Theia the requirement is nearly identical: $10^{-5}$ pixel \citep{2015pathwaysMalbet,Malbet2016}. 

\subsection{Experiment concept and simplified principle}\label{subsec:Experiment concept and simplified principle}

\begin{figure}[t]
\begin{center}
\includegraphics[width = 0.54\textwidth]{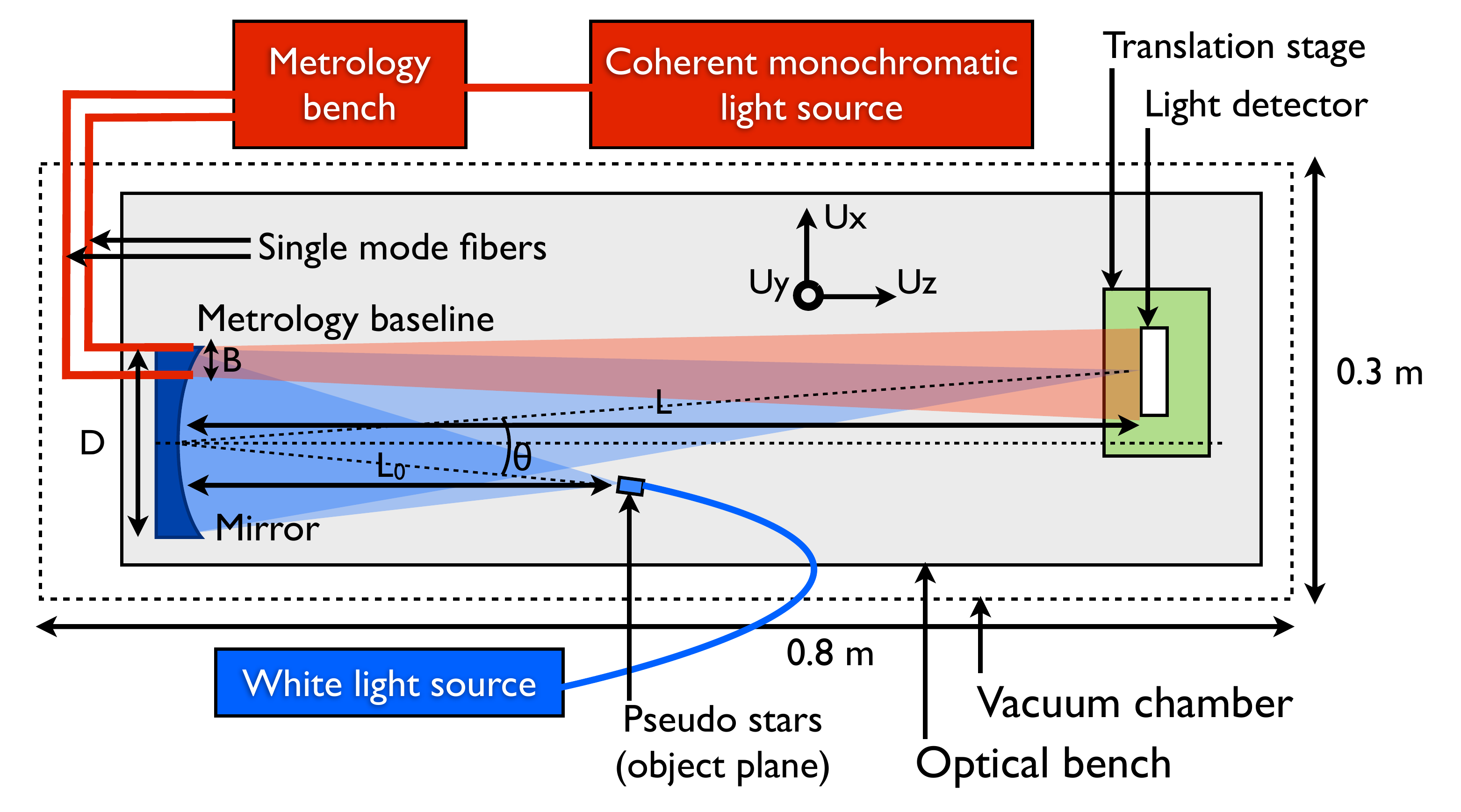}
\caption{\label{base_concept} DICE testbed concept (top view, not to scale). The elements associated with the metrology, the pseudo stars and the mechanical supports are respectively in red, blue and grey. The (X,Y,Z) axis indicated on the figure will be used consistently throughout the paper to indicate directions.}
\end{center}
\end{figure}

Figure~\ref{base_concept} is a conceptual diagram of the experiment testbed. The metrology fibers create either vertical or horizontal Young fringes on the detector (exactly two fibers are turned on simultaneously). The phase is modulated between the fibers to create moving fringes. The pixel offsets are obtained by comparing the phase of the sinewaves observed on individual pixels with the phase of the global system of fringes. The pixel response non uniformity (hereafter PRNU) could also be derived from the metrology data, however in our experiment flat fields obtained in broad band light are used. The details about how to measure the pseudo stars positions, how to estimate the astrometric error, how to obtain the calibration (pixels QEs and offsets) and use it to improve the accuracy are developed in Sect. \ref{sec:Data processing} (data processing).

\begin{figure}[t]
\begin{center}
\includegraphics[width = 0.30\textwidth]{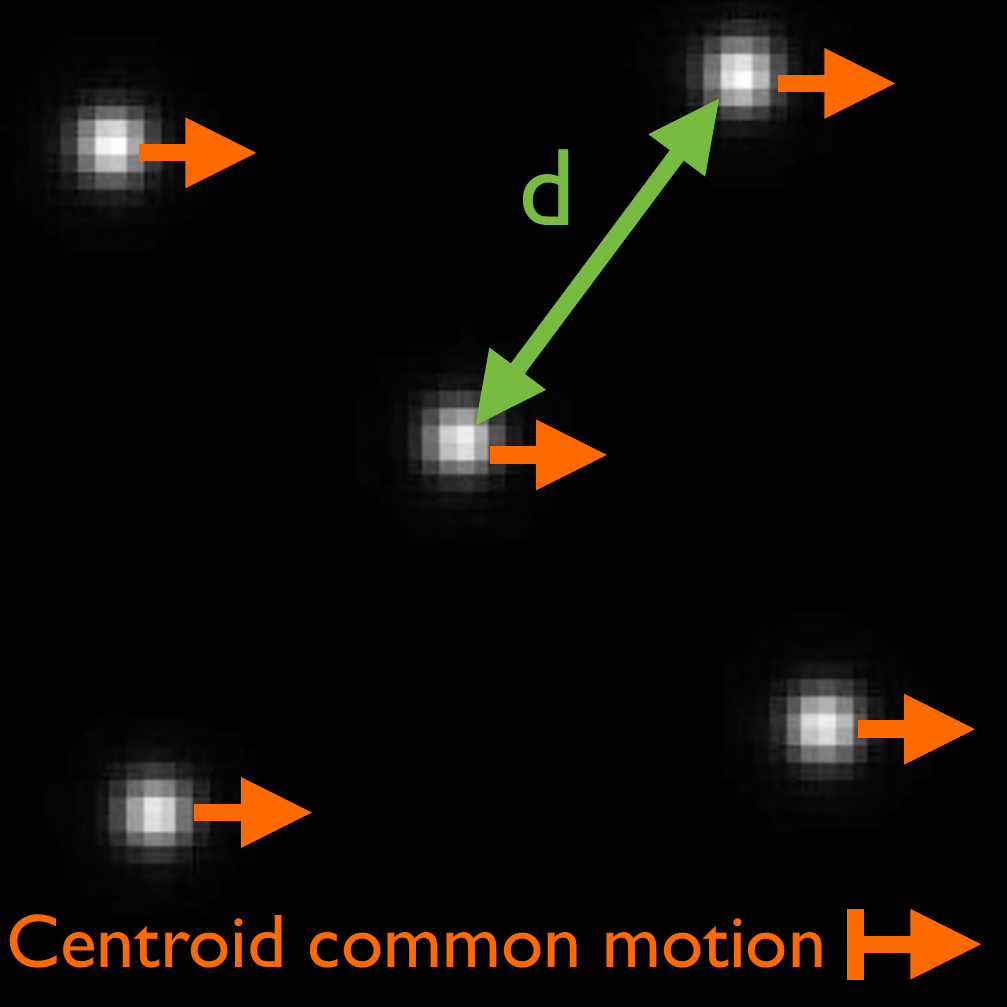}
\caption{\label{simplified_experiment_principle} Simplified experiment principle. The orange arrows represent the CCD motion, which appear as a common motion of all centroids. The final astrometric accuracy is measured as the standard deviation of the distance. The location of each centroid is measured by ‘‘autocorrelation after resampling by Fourier transform''}
\end{center}
\end{figure}

Figure~\ref{simplified_experiment_principle} explains the basic principle of the experiment. The final astrometric accuracy is measured as the standard deviation of the distances between the pseudo star centroids for different positions of the detector. Each pseudo star centroid is affected by different pixelation errors. The pixelation error is defined as the random astrometric shift caused by pixel sensitivities and offsets. In order to obtain uncorrelated pixelation errors, the motion has to be larger than the PSF diameter (\!\!~$\sim$5 pixels). Without calibration, astrometric errors larger than the requirement are expected. The role of the calibration is precisely to provide information about the pixels. By integrating this new information the errors can be brought down, hopefully down to the NEAT requirement of $5\e{-6}$ pixel.

\subsection{Testbed specifications and design}

The design and specification presented here are a condensed version. For more details refer to \cite{crouzier12} and \cite{2015PhDT.........7C}. The testbed configuration has seen substantial changes, the version presented here is up to date and shows the configuration consistent with the experimental data presented in Sect. \ref{sec:Analysis of experimental data}, which was obtained just before the testbed was shut down. Table \ref{tab:design parameters} summarizes the notations for the critical dimensions and parameters that will be used consistently throughout this document. The last column shows the values chosen for the design. 

\begin{table}[t]
\hspace{1cm}
\footnotesize
\centering
\caption{\label{tab:design parameters}Design parameters.}
\begin{center}
\begin{tabular}{lll}
  \hline
  Parameter & Notation & Value\\
  \hline
  \hline
distance mirror to CCD & $L$ & 600 mm\\\hline
\specialcell{distance mirror to\\pseudo star objects} & $L_0$ & 300 mm\\\hline
\specialcell{min/max wavelength\\of pseudo stars} & $\lambda_{\ml{min}} / \lambda_{\ml{max}}$ & 400 nm/800 nm\\\hline
entrance pupil diameter & $D$ & $5\pm1$ mm\\\hline
mirror focal length & $f$ & $200\pm5$ mm\\\hline
\specialcell{separation between\\pseudo stars objects} & $s$ & 240 $\mu$m\\\hline
pseudo star pinhole diameter & $d$ & 15 $\mu$m\\\hline
off axis angle (pseudo stars) & $\theta$ & 2$\dg$\\\hline
metrology source wavelength  & $\lambda_\ml{m}$ & 633 nm\\\hline
metrology baseline & $B$ & from 1 to 6 mm\\\hline
pixel size & $e$ & 24 $\mu$m\\\hline
size of the detector matrix & $N_{\ml{pix}}$ & 80x80 pixels\\\hline
effective quantum well size & $w$ & 200,000 $\elec$\\\hline
\end{tabular}
\end{center}
\vspace{-0.4cm}
\end{table}

\begin{figure*}[t]
\begin{center}
\includegraphics[width = 0.8\textwidth]{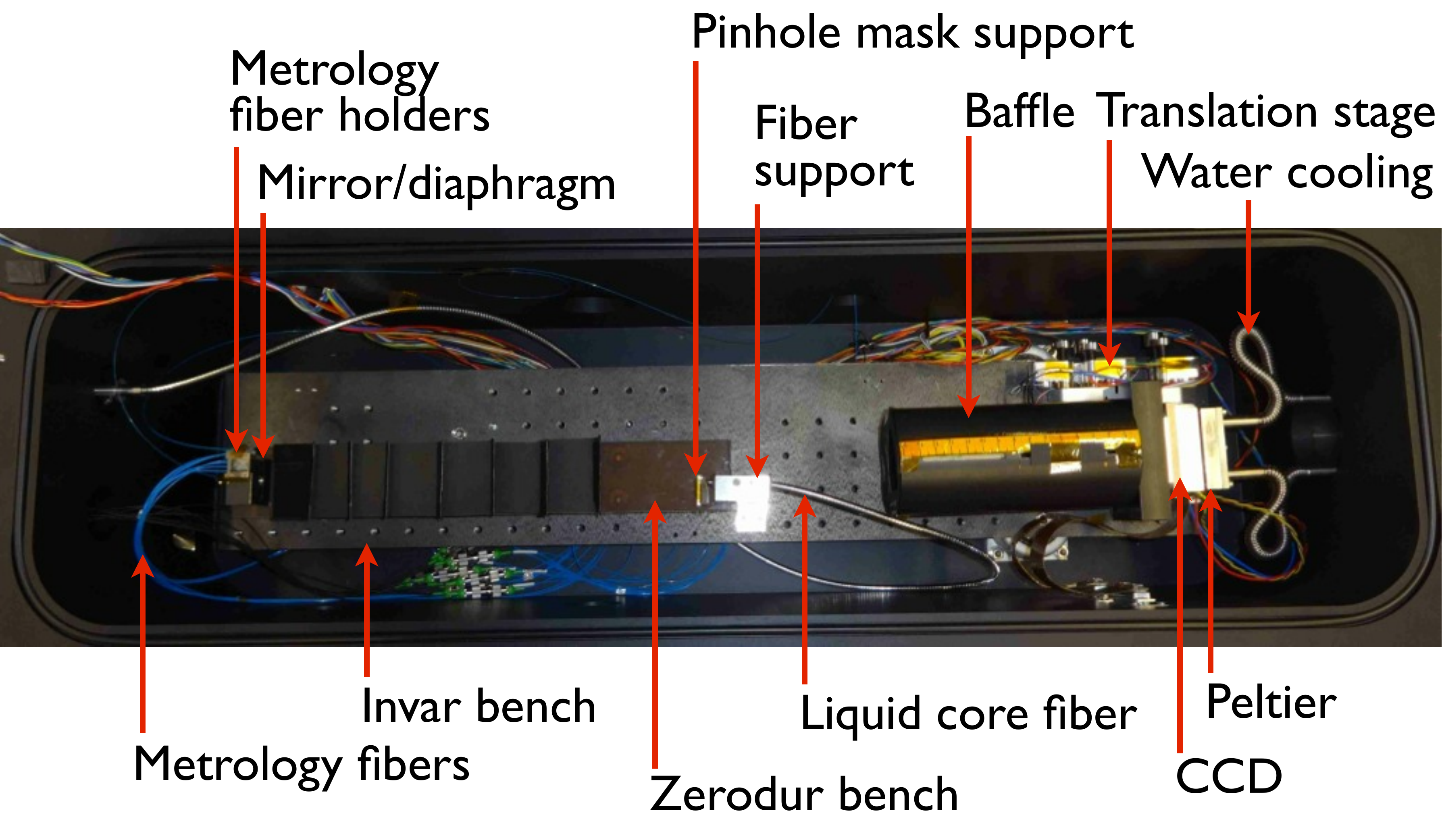}
\caption{\label{schematic_vacuum_chamber_v2}Optical bench inside the vacuum chamber.}
\vspace{-0.5cm}
\end{center}
\end{figure*}

Fig.~\ref{schematic_vacuum_chamber_v2} is a labeled picture of the interior of the vacuum chamber.

\subsubsection{List of high-level specifications}

The complete list of specifications with associated design constraints and compliance tests is not presented here, it can be found in the PhD thesis manuscript \citep{2015PhDT.........7C}. Only the most critical aspects are mentioned here, for each specification (delimited by quotes), there is a short explanation of how it was determined. For some specifications calculations could be made, but in other cases we had to rely on past experience at JPL. For some of the design counter-parts, a best effort approach was used, coupled with final compliance tests. One important goal of the experiment is precisely to determine what are the conditions needed to reach micro-pixel accuracy. The specifications are the following:
\bit
\item ‘‘There is only one optical surface and no vignetting of the FoV.'' Just like for the NEAT concept, beam walk errors are avoided in this experiment. The experiment goal is to characterize detector pixelation errors only, not optical distortion.
\item ‘‘The thermal stability of the focal plane is below $10^{-2}$ K.'' With an expansion coefficient on the order of $3\e{-6}$, thermal expansion of the chip is lower than $3\e{-8}$. The platescale is calibrated directly from the centroid measurements: the final error is caused by temperature inhomogeneities only. As a worst case a temperature gradient of $10^{-2}$ K between the top and the bottom of the chip is considered. The resulting error would have a trapezoidal pattern (not corrected by the metrology/platescale), of an amplitude of $3\e{-8} N_{\ml{pix}} /2 = 1.2\e{-6}$ pixel ($N_{\ml{pix}} =80$ in our setup).
\item ‘‘The PSF at the focal plane is Nyquist sampled.'' The centroid measuring technique uses resampling in the Fourier space. In order to be accurate with this method, the PSF has to be Nyquist sampled. As the pseudo stars are in broadband light (400 to ~900 nm) we set the Nyquist condition for the shortest wavelength at 400 nm. In practice we adapt the size of a diaphragm on the mirror to the pixel size to have: $2e = \lambda L /D$.
\item ‘‘The mechanical centroid jitter is below $10^{-2}$ pixel.'' High mechanical stability is useful to integrate a large number of photons at a given position, with static pixellation errors and thus keeping the overall pseudo stars processing simple. The target is a stability of 1\% of a pixel, which is significantly lower than the PSF size (a few pixels wide at Nyquist sampling).
\item ‘‘Pseudo stars (objects) are not spatially resolved.'' They are diffraction limited. To properly emulate punctual stars, the geometric image of the pinhole size and optical aberrations must be smaller than the diffraction limit.
\item ‘‘The integration time needed to reach a photon noise below $5\e{-6}$ pixel is a few minutes.''
To repeat the experiment quickly and test different parameters, a high photon collecting speed is needed. Both the metrology calibration and the pseudo star photon noise must reach micro pixel photon noise in a few minutes. Two factors can limit the integration speed, the capacity of the detector to absorb photons (quantum well multiplied by frame rate) or the photon flux itself. The choice of detector, its readout electronics and the light sources has to be made accordingly.
\item ‘‘The CCD is photon noise limited.'' In the final images the dark and read noises are lower than the photon noise (the latter is typically 30 counts or greater).
\ei

\subsubsection{Subsystem design: light detector and readout electronics}

The detector is a crucial part of our experiment. Our final choice was the ‘‘CCD39-02'' from e2v. The readout electronics and software have been developed and tested by CEA (Commissariat à l'Energie Atomique et aux Energies Alternatives). The CCD 39-02 is a back illuminated visible CCD, with a frame size of 80 by 80 pixels. The physical pixel size is 24 $\mu$m by 24 $\mu$m, making for a total sensitive area of 1.92 by 1.92 mm. 

Our initial choice was to use the ‘‘CCD39-01'' version of the detector, with four amplifiers, to enable high frame rate operation (up to 1000 Hz). In this version the imaging area is split into four square areas which are called quadrants, of 40 by 40 pixels. Each quadrant has a separate readout channel, that is different wires and electronics components. However electronic ghosts generated by symmetry relative to the quadrant boundaries, at the $1\e{-3}$ relative intensity level, impaired the analysis and forced us to reconsider this choice. The most recent data from the experiment which is presented in this article was obtained after a significant hardware alteration. It was done in order to switch to the single amplifier version. The readout speed used is 108 Hz.

\subsubsection{Subsystem design: pseudo stars}

The object sources are a pinhole mask made in Zerodur. The mask is coated on one face, with 12 micro holes in the coating. The optical configuration corresponds to a magnification factor of two and an off axis angle of two degrees. This allows the installation of the pseudo stellar sources and the detector without any beam obstruction with some margins to accommodate the supports. Additionally, with an aperture as small as 5 mm, a spherical surface is sufficient to obtain optical aberrations that produce a spot diagram smaller than the diffraction pattern in the whole field of view.

For high stability of the pseudo stars, the pinhole mask holder and the mirror are grouped into a single Zerodur block (by molecular adhesion). The Zerodur bench and the CCD translation stage are held on a invar bench. Vibration damping of the pseudo stars is based on a best effort approach. Four stages of vibration damping are stacked in order to stabilize the core of the experiment: a standard optical table with pneumatic suspension, damping supports between the table and the vacuum chamber, silent blocks (another kind of damping support) between the vacuum chamber and the invar bench and there is a last stage of damping between the invar and the Zerodur bench.

\subsubsection{Subsystem design: flat field projector}


The system consists of a high power broadband white light LED (400 to 700 nm) connected to a multimode fiber which tip is taped on the mirror block and oriented toward the CCD. The light source is located outside the vacuum chamber, a custom made feed-through enables operation in vacuum. For experimental tests and reliability (by redundancy), there are two fibers with respectively 200 $\mu$m and 365 $\mu$m diameter cores.

An incoherent source is used because stray light is much harder to control in coherent light than in incoherent light. Noting $I_0$ the direct intensity from the light source and $I_0$ the stray light intensity after (multiple) reflection(s), the perturbation in coherent light is $\propto\sqrt{I_1/I_0}$, instead of directly proportional to $I_1/I_0$ \citep{2015PhDT.........7C}. The main reason for having a multimode fiber is the luminous flux needed to operate the detector in a photon noise limited regime and to accumulate photons at a decent rate, resulting in a total integration time similar to the other systems (a few minutes). With a single mode fiber (hereafter SMF), a superluminescent type of source would have been required. 

\subsubsection{Subsystem design: metrology}

\begin{figure}[t]
\begin{center}
\includegraphics[width = 0.5\textwidth]{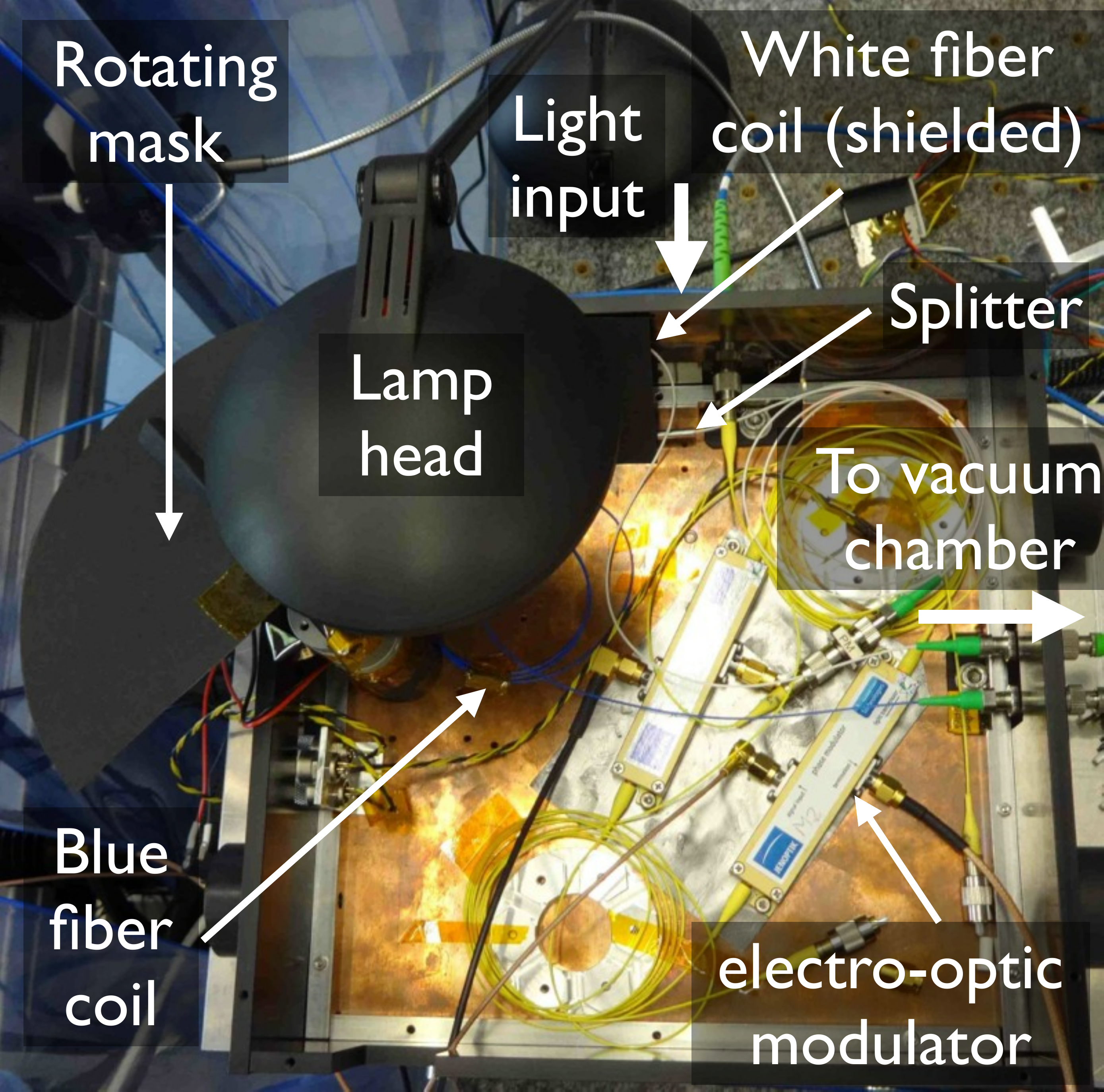}
\caption{\label{fig:metrology_box}Picture of the metrology box.}
\vspace{-0.3cm}
\end{center}
\end{figure}

\begin{figure*}[t]
\centering
\subfigure[Schematic of the metrology. The axis are indicated on the figure: Z is aligned with the optical axis, X and Y are aligned with the horizontal and vertical directions within the focal plane.]{\label{schematic_metrology_subsystem}
\includegraphics[width=0.74\textwidth]{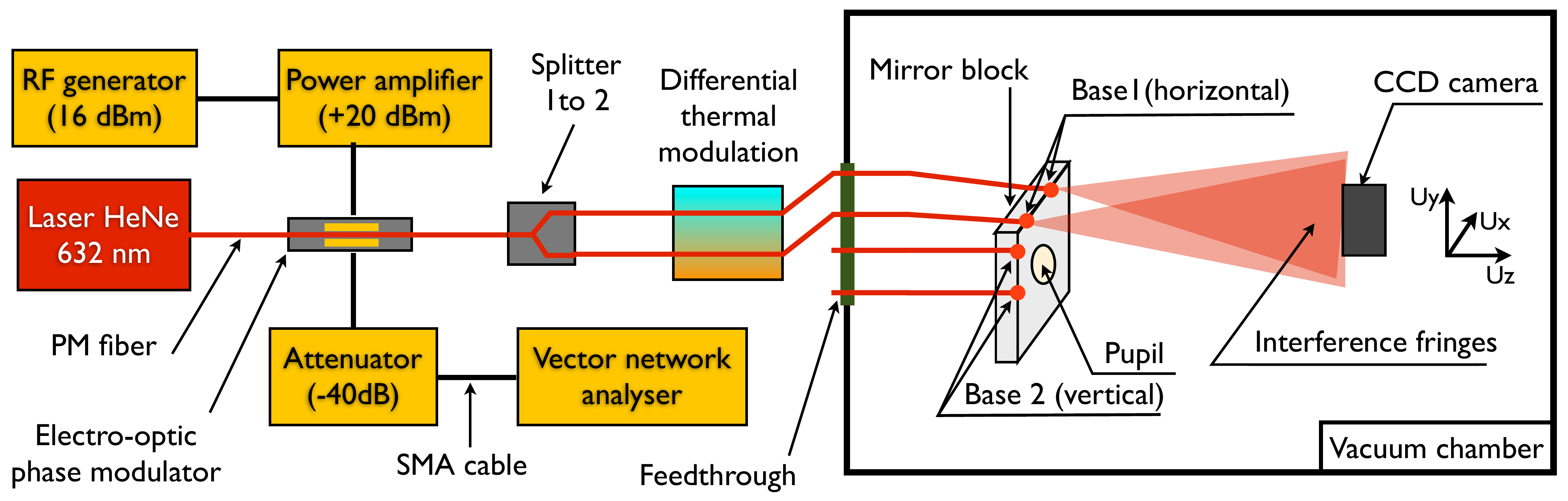}}
\subfigure[Positions of the metrology launching fibers that define the baselines.]{\label{metrology_baselines}
\includegraphics[width=0.24\textwidth]{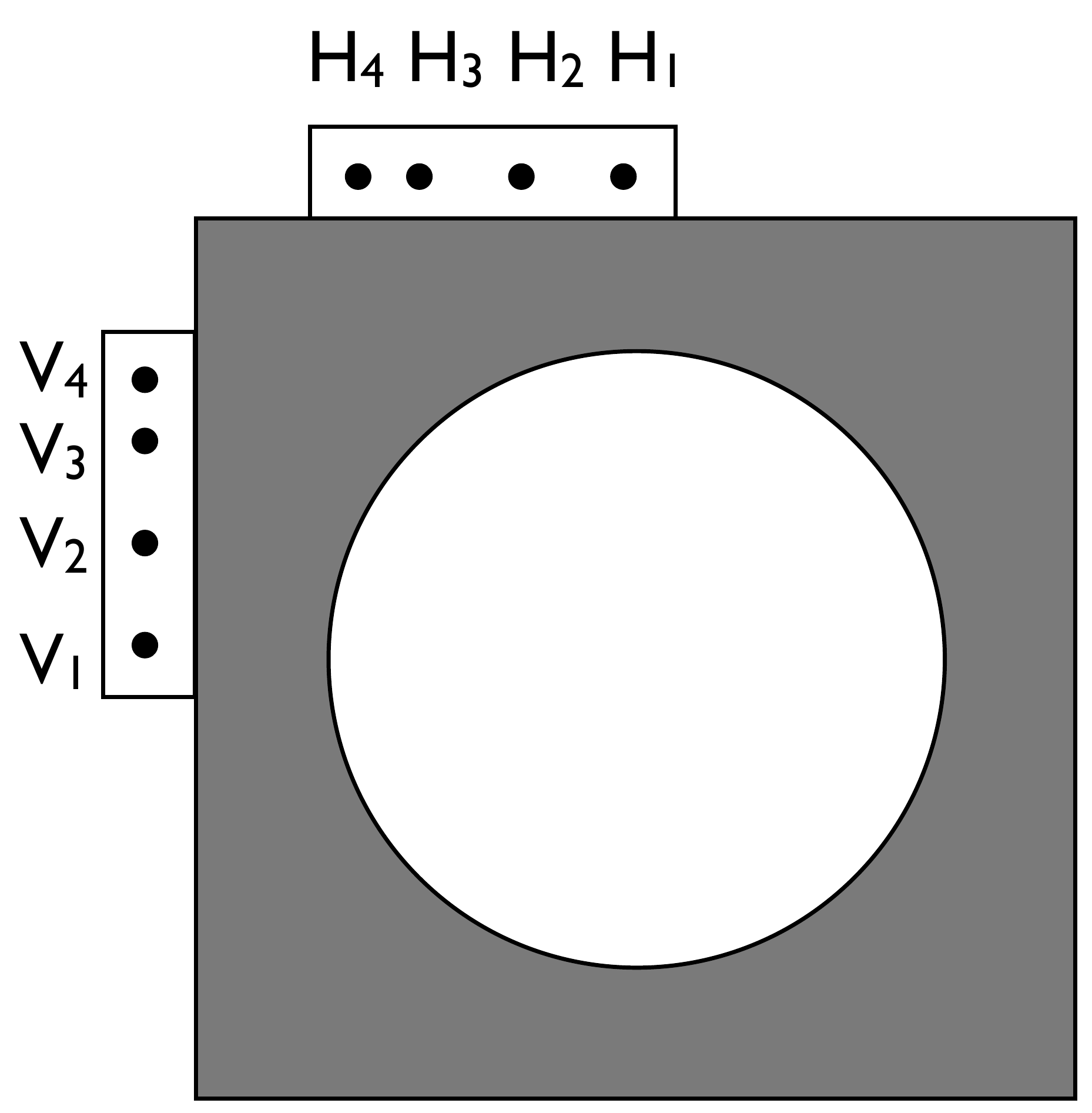}}
\vspace*{-2mm}
\caption{\label{metrology_subsystem}The metrology subsystem.}
\end{figure*}

\begin{figure*}[t!]
\begin{center}
\includegraphics[width = 160mm]{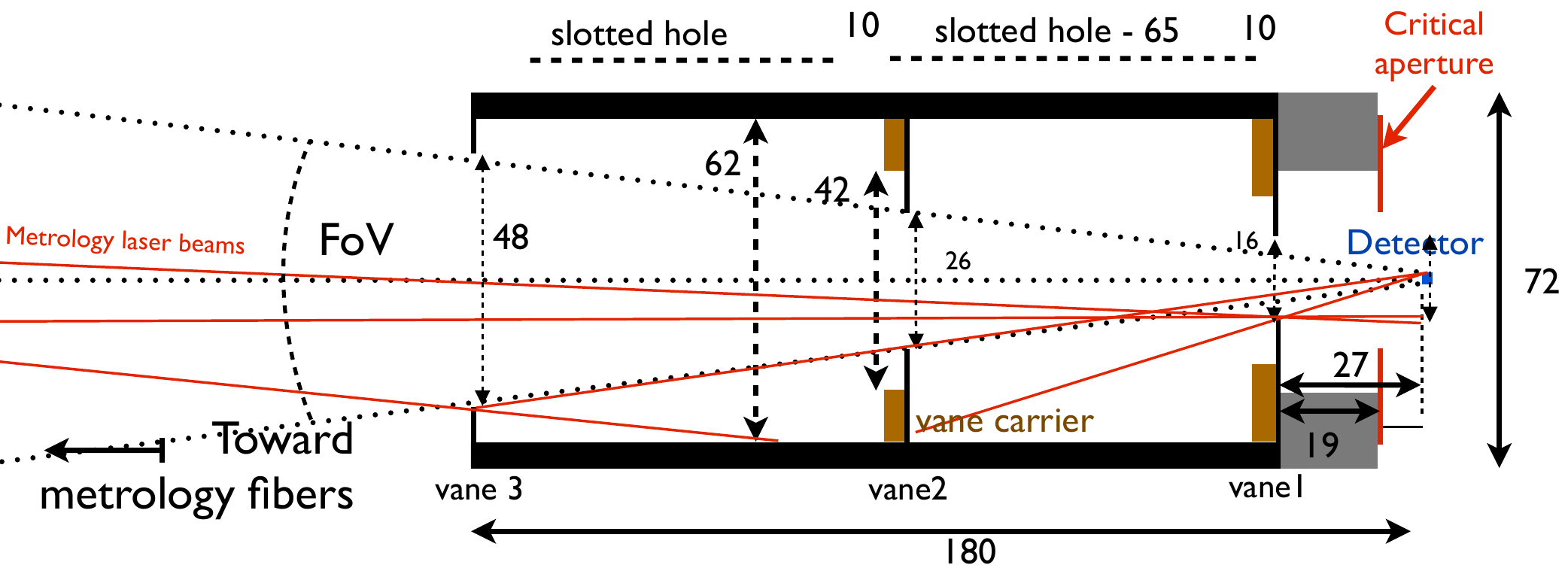}
\caption{\label{fig:baffle_simplified_schematic_v4}Schematic of the baffle. The dimensions are proportional (indicated in mm). The red rays coming from the left are from the metrology fibers. Only the most geometrically constraining rays are represented.}
\vspace{-0.3cm}
\end{center}
\end{figure*}

The metrology is composed of integrated components from the laser source to the fiber tips that forms the baselines (see Fig.~\ref{schematic_metrology_subsystem}). The source for the metrology is a HeNe laser. The light goes first through an electro-optic phase modulator, which operates at 3 GHz. This high frequency phase modulation creates a spectral broadening, thereby reducing the coherence length to 1 cm (which is the smallest coherence attainable with our hardware). The limit is set by the amplitude of the radio frequency signal injected into the modulator. This makes stray light reflections at large optical path differences (hereafter OPD) incoherent and do not create any adverse effect on the fringes (which have OPD~$<$~0.2mm). The light is then split into two fibers, which undergo a slow differential phase modulation through thermal effects (>0.1 Hz). A lamp with rotating shadow is used to create a cyclic thermal modulation on one fiber. This differential modulation makes the fringes slowly sweep across the detector (back and forth).

During the metrology calibration phase, two vertically and horizontally aligned fibers are selected successively to project respectively horizontal and vertical dynamic Young fringes. The baselines are chosen among the possible combinations offered by the layout of the fiber tips on the mirror (see Fig. \ref{metrology_baselines}). Linearly polarized laser and polarization maintaining fibers are used all the way from the laser to the fiber tips to ensure good fringe contrast. The fiber splitter and the electro-optic and thermal phase modulators are packed into a box (see Fig.~\ref{fig:metrology_box}).

\subsubsection{Subsystem design: light baffles}

The purpose of baffling is to mitigate stray light inside the vacuum chamber, in priority for the metrology which is the most sensitive to stray light as explained in the metrology sub-section. The situation with respects to stray light is unusual, because the main source of light to worry about is on axis and coherent. The attenuation factor looked for in this case is thus for on-axis light, after diffusions inside the baffle and vacuum chamber. We have been unable to derive useful ways to specify the baffling inside the vacuum chamber in early phases of the testbed conception for practical reasons (lack of local expertise), we have relied mostly on an experimental approach using the data from the experiment itself (trial and error process).

The baffle is modular: it is a cylindrical tube, in which we can insert several movable vane holders. We can thus easily change the vane configuration. A schematic of the 4th version of the baffle is shown in Fig. \ref{fig:baffle_simplified_schematic_v4}. The geometry of the baffle is such that there is no critical reflection of stray light (i.e. light arriving on the detector after only one diffusion), other than on the vane edges, this is illustrated by the red rays. The vane edges are made with standard commercial ‘‘double edge'' razor blades, which radius of curvature are less than 400 nm (verified with an scanning electron microscope). This prevents significant critical diffusions from the edges unto the detector. These diffusions were creating measurable effects with previous baffle versions. The vane apertures are octagonal with a 14$\dg$ FoV, resulting from a trade-off between two opposite needs: opening the FoV enough to avoid diffraction from the vane edges and preventing critical diffusions on the inner tube wall. The tube diameter is just large enough to allow this configuration to work, and is also close to its maximum physical dimension limited by mechanical obstruction. Circular apertures would be more efficient for this trade-off but would not be possible with razor blades.

The inside of the baffle and vacuum chamber are covered with high performance diffusive-absorbent materials, with a total hemispherical reflectance of 1\% in the visible\footnote{Acktar Metal Velvet\circledR}. Additionally, during the metrology data acquisition, we clean-up the baffle FoV of most physical elements, and in particular all elements angularly too close to the metrology fibers. The Zerodur bench (holding the mirror for pseudo stars) is one of these elements that must be removed, as it produces detectable stray light, even with complete optical protection. Thus we do not exactly use the configuration intended in the initial design, as presented in Fig.~\ref{schematic_vacuum_chamber_v2}, we have to do a manual operation to switch between metrology and pseudo star configurations. 

\section{DICE model and data processing}\label{sec:Data processing}


\begin{figure}[t]
\begin{center}
\includegraphics[width = 0.5\textwidth]{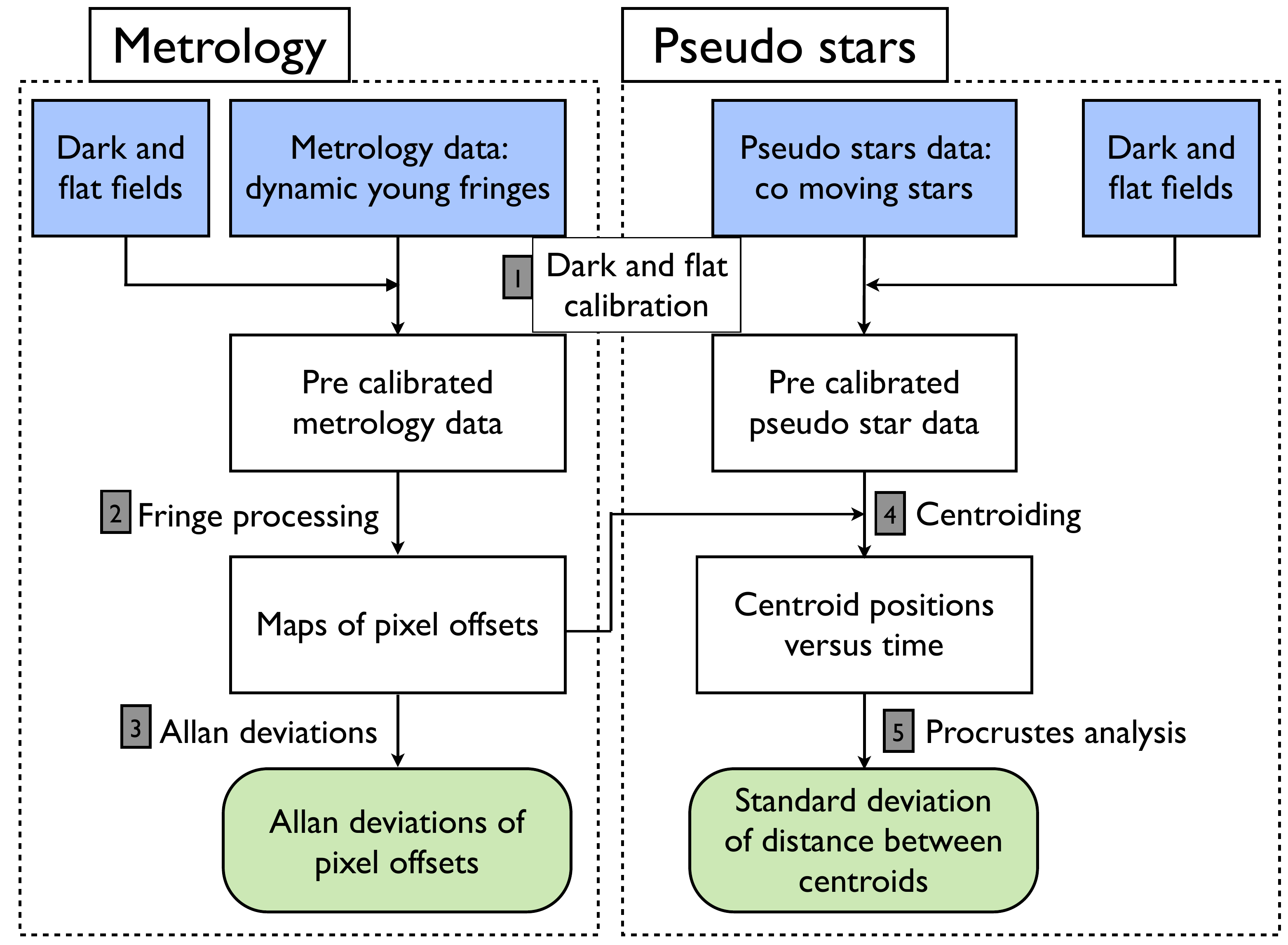}
\caption{\label{schematic_neat_demo_data_processing}Overview of data reduction process for the NEAT testbed. Steps 1 to 5 of the process are described in the next subsections.}
\end{center}
\end{figure}

This section presents the formal model and the methods used to analyze different types of data from the testbed (flat-fields, metrology and pseudo stars). Figure \ref{schematic_neat_demo_data_processing} is a diagram summarizing the different steps involved. It shows how the metrology and pseudo star processing are linked together.

\subsection{Dark subtraction and flat-field correction}\label{subsec:Data analysis: methods - Dark and flat fields}

The first step of the data reduction is the standard dark subtraction (of a temporally averaged dark frame) and division by the PRNU map, also called flat-field correction. For both pseudo star and metrology data, dark subtraction is systematic, while the flat-field correction is optional. The application to the data is straightforward, the reduced data cube is simply: $I' = (I - I{_\ml{dark}})/I_{\ml{PRNU}}$ (this operation is pixel-wise). The delicate part here is obtaining a high quality PRNU map in the first place.

A raw flat-field is a data cube, that is a stack of images, with a relatively constant illumination level (spatially and temporally). The corrected flat-field (PRNU map), is derived from this raw data, which can have some bias that are specific to the hardware. In our case, flat-fields are obtained in broadband light. The system consists of a multimode fiber facing the detector and a broadband LED source. The vacuum chamber does not allow for an integrating sphere, because the light from pseudo stars and the metrology would be blocked. The multimode fiber produces a fairly flat intensity pattern on the detector: the intensity profile is a Gaussian beam with a waist of about 10 cm which is much larger than the detector field (2 mm). The detector thus sees an intensity gradient with a possible slight curvature. The method used to derive the PRNU map consists of the following simple steps:
\begin{enumerate}
\item Dark subtraction (of each flat-field frame)
\item Temporal normalization: each frame is scaled to have an average flux equal to the average intensity level of the first frame.
\item Calculation of the temporal mean frame
\item Suppression of the image gradient in the mean frame
\item Normalization of the mean pixel value to 1
\end{enumerate}

Step 2 suppresses variations of the light source. This allows a precise estimation of experimental temporal noise, which must be at first order equal to photon noise. Our LED broadband source shows variations with an RSD of about $5\e{-4}$ during typical integration durations (2 minutes). The impact of this step on the final result is thus very weak (the final mean is slightly different, as frame weights are changed). Step 4 sets the gradient to 0. This step has no effect on the final differential astrometric accuracy, as the effect of a PNRU gradient is an rigorously homogeneous centroid offset (which cancels out in a difference). When comparing different flat fields, or analyzing the stability of a flat field versus time, effects like source intensity variation and illumination gradient variations often dominate the dynamics of any flat-field difference, while having no consequence on the differential astrometry. Hence the choice to automatically suppress them in the pipeline.

These dark and flat processing methods are straightforward and do not induce significant astrometric errors. The quality (accuracy and stability) of dark and flat-fields obtained experimentally is verified, in order to estimate the amplitude of systematic errors and in fine their impact on astrometric accuracy. More details will be given in section \ref{sec:Numerical simulations for DICE} (numerical simulations) and section \ref{xpdata:dark_and_flat} (experimental dark/flat quality tests).

The flat has higher SNR in broadband incoherent light, than if derived from the metrology data. Coherent stray light produces relative intensity variations $\propto \sqrt{I_1/I_0}$, instead of $\propto I_1/I_0$, because of interferences ($I_0$ is the intensity of the direct beam from the fiber, $I_1$ is the intensity after a parasite reflection, see appendix \ref{append:Coherent and incoherent stray light} for details). The interference pattern is complex and has spatial features unresolved by the pixels for large angular separations. This is the case for example for reflections on the edges of the stop apertures of the baffle. That is why we use the method presented above to derive PRNU maps instead of relying on metrology data.

\subsection{Metrology}\label{subsec:Data analysis: methods - Metrology}

\subsubsection{Global solution}

The interference pattern created at the detector with a monochromatic source of wavelength $\lambda_{\ml{met}}$ and for given metrology baseline $B$ of coordinates $(B_x,B_y)$ is:
\begin{equation}\label{interference_pattern}
I(x,y,t) \propto 2I_0\left[1+V\cos\left(\phi_0 + \Delta \phi(t) + \frac{2\pi(x B_x + y B_y)}{\lambda_{\ml{met}} L}\right)\right].
\end{equation}
Where $I_0$ is the average intensity at the focal plane for one fiber, $L$ is the distance between the fibers and the detector, $\phi_0$ is a static phase difference and $\Delta \phi (t)$ is the modulation applied between the lines. Although the exact shape of the fringes is hyperbolic, at the first order the fringes can be considered straight and aligned with the direction perpendicular to the metrology baseline. Assuming that the point sources are of equal intensity and that the intensity created at the focal plane is uniform gives a fringe visibility of $V=1$. In reality, the visibility is affected by the intensity and polarization mismatch between the point sources. Each fiber launches a Gaussian beam and the beams are not co-centered.

Because all pixels see different visibilities and different average intensities, the solution for the cube of metrology data is written under the following form:
\begin{equation}
\begin{split}
&I(i,j,t) = B(t)\iota(i,j) + A(t)\alpha(i,j)\\
&\sin\left[iK_x(t)+jK_y(t) + \phi(t) + \delta_x(i,j)K_x(t) + \delta_y(i,j)K_y(t)\right],
\end{split}
\end{equation}
where $i,j$ are integer pixel position indexes and $\delta_x$ and $\delta_y$ are pixel offsets, that is the difference between the pixel true locations and an ideal regularly spaced grid. Time and spatial variations are decoupled in the equation. $t$ has been implicitly transformed into a discrete index representing a frame number (it naturally carries the connotation of a dimension associated with time). The meaning of all remaining variables is explained in Table \ref{metrology_analysis_variable}. The table also mention which kind of noise are absorbed by the variables. 

\begin{table}[t]
\caption{\label{metrology_analysis_variable} Table of metrology variables for data analysis. The last columns gives the types of noises that affect the values of each variable.}
\begin{center}
\small
    \begin{tabular}{lll}
    \hline
Notation & Name & Absorbed noises \\ \hline \hline
$B(t)$&	Average intensity & \specialcell{Laser flux,\\offset fluctuations}\\ \hline
$A(t)$& Amplitude & \specialcell{Laser flux\\polar. fluctuations}\\ \hline
$K_x(t)$& Metrology wavevector ($x$ proj.) & \specialcell{Laser freq. fluctuation,\\thermal expansion} \\ \hline
$K_y(t)$& Metrology wavevector ($y$ proj.) & \specialcell{Laser freq. fluctuation,\\thermal expansion} \\ \hline
$\phi(t)$& Differential phase & \specialcell{phase jitter (thermal\\and mechanical)}\\ \hline
$\iota(i,j)$& Pixel relative intensity & PRNU\\ \hline
$\alpha(i,j)$& Pixel relative amplitude & \specialcell{PRNU,\\visibility vs. space}\\ \hline
$\delta_x(i,j)$ & Pixel offsets ($x$ proj.) & -  \\ \hline
$\delta_y(i,j)$ & Pixel offsets ($y$ proj.) & -  \\ \hline
    \end{tabular}
\end{center}
\end{table}

To avoid degenerated solutions, the following normalization constraints are added:
\begin{equation}	
\left\{
\begin{array}{l}
\sum_{i=1}^{n} \sum_{j=1}^{m}\iota (i,j) = nm\\
\sum_{i=1}^{n} \sum_{j=1}^{m}\alpha (i,j) = nm\\
\sum_{i=1}^{n} \sum_{j=1}^{m}\delta_x (i,j) = 0\\
\sum_{i=1}^{n} \sum_{j=1}^{m}\delta_y (i,j) = 0\\
\sum_{i=1}^{n} \sum_{j=1}^{m} i\delta_y (i,j) = 0\\
\sum_{i=1}^{n} \sum_{j=1}^{m} j\delta_y (i,j) = 0
\end{array}
\right.
.
\end{equation}

\subsubsection{Minimization strategy}

\begin{figure}[t]
\begin{center}
\includegraphics[width = 0.45\textwidth]{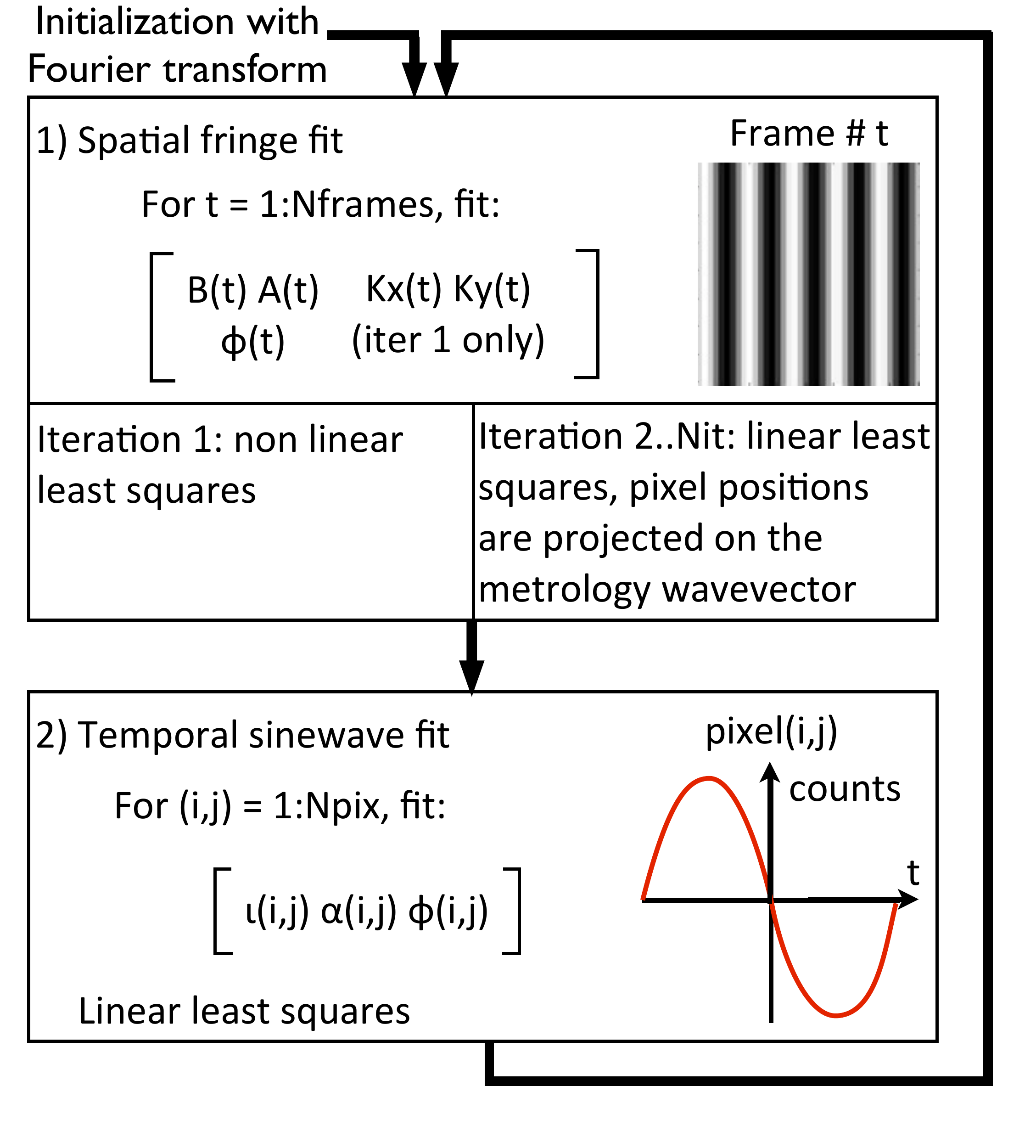}
\caption{\label{metrologyFringeFit} Iterative process used to fit the metrology fringes (step 2 in Fig.~\ref{schematic_neat_demo_data_processing}). The difference between the measured phase of a pixel ($\phi(i,j)$) and the phase expected (global fringe phase) is the phase offset caused by the pixel offset projected in the direction of the wavevector.}
\end{center}
\end{figure}

A set of metrology data cube has a typical size of about $N_{\ml{pix}}^2\times N_{\ml{frames}}$ (number of frames), a cube can contain as many as 200000 frames and for our CCD, the matrix size is $N_{\ml{pix}}=80$. The problem is non linear as the fringe spacing is a free parameter. The total number of fit parameters (80$\times$80+5$N_{\ml{frames}}$) is not practical for a direct least square minimization of the whole cube, that is why an iterative process is used. First a spatial fitting procedure is performed on each frame to constrain the time dependent variables and then a temporal fitting procedure in performed on each pixel to constrain the space (or pixel) dependent variables (see Fig.~\ref{metrologyFringeFit}). This order is highly preferable as the global phase is very noisy and can easily be fit by the spatial fit, but not by the temporal fit. After four iterations convergence down to $5\e{-6}$ pixel of the final astrometric solution is obtained.


\paragraph{- Non linear spatial 2D sine wave fit\\}
The spatial fit is a non linear least square minimization for the first iteration, using a Levenberg-Marquardt optimization procedure. The fit is initialized with parameters obtained from a Fourier transform of the first frame. From one frame to the next all parameters can be reused for initialization. The parameters being optimized at this step are: $A(t)$, $B(t)$, $\phi(t)$, $K_x(t)$ and $K_y(t)$. For the first iteration, all other parameters have perfect pixel case values: for all $(i,j)$, $\iota = 1$, $\alpha = 1$, $\delta_{x}=0$ and $\delta_{y}=0$. 

From iteration number 2 until $N_{\ml{it}}$, the fit can be linearized by fixing the wavevector to its temporal average: $K_x=\langle K_x(t)\rangle$ and $K_y=\langle K_y(t)\rangle$. The pixel locations are then projected onto the wavevector and the remaining parameters are obtained through a process analogous to the linear 1D fit described below.

\paragraph{- Linear temporal 1D sine wave fit\\}
The temporal sine wave fit is always a linear one. The method is very similar to the standard linear least square fit of a sine wave of known period (i.e. the optimization of average, amplitude and phase parameters only). There is however one important difference: instead of the period, only the phase of the 2D carrier sine wave for each frame is known. The phase modulation is in practice a piecewise exponential function with added thermal and mechanical noise. In fact it can be any function, as long as the wrapped phase is properly sampled between 0 and $2\pi$. The temporal signal versus the phase can be reconstructed as a pure sine wave after normalization to average fringe intensity $B=\langle B(t)\rangle$ and average fringe amplitude $A=\langle A(t)\rangle$. The phase of the resulting sine wave carries information on the pixel location projected along the modulation direction:
%
\begin{equation}\label{temporal_sinwave}
I(i,j,t) = B\,\iota(i,j) + A\,\alpha(i,j) \sin\left[\phi(t) + \phi(i,j)\right] .
\end{equation}

The parameters are not solved directly, but the least square fit is linearized by rewritting $I(i,j,t)$:
\begin{equation}\label{temporal_sinwave_linearized_fit}
I(i,j,t) = a_{i,j} \sin(\phi(t)) + b_{i,j} \cos(\phi(t)) + c_{i,j}.
\end{equation}
where $\iota(i,j)$, $\alpha(i,j)$ and $\phi(i,j)$ are derived from the coefficients $a_{i,j}$, $b_{i,j}$ and $c_{i,j}$ (see appendix \ref{append:Linear sine wave fit} for details).

The pixel phase $\phi(i,j)$ contains information about the pixel true location:
\begin{equation}\label{pixel_phase}
\phi(i,j) = iK_x(t)+jK_y(t) + \delta_x(i,j)K_x(t) + \delta_y(i,j)K_y(t).\;\;
\end{equation} 

\paragraph{- Deprojection of pixel offsets\\}
The true pixel offset vector $\pmb{\delta}=(\delta_x(i,j)$,$\delta_y(i,j))$ is derived from the pixel phase. The previously described metrology reduction process applies to a single set of data with a quasi-constant $\vec{K}(t) = (K_x(t),K_y(t)) \approx (K_x,K_y)$ metrology wavevector. The testbed is designed to ensure that $\vec{K}$ is highly stable in amplitude and direction. From the first iteration, we obtain $\vec{K}(t)$ so we can assess the stability experimentally. After assuming $\vec{K}$ constant, the $iK_x+jK_y$ term can be calculated and subtracted (it corresponds to the phase offset for a location of a perfect pixel on a regular grid). However the remaining difference is a scalar, whereas the true offset is a pair (a vector in a 2D space): $\vec{\delta_p}(i,j) \cdot \vec{K} = \delta_x(i,j)K_x + \delta_y(i,j)K_y$. In order words, with each baseline, only the projection of the true pixel offset onto $\vec{K}$ (noted $\pmb{\delta_p}(i,j)$), can be measured.

To solve the degeneracy and retrieve $\delta_x$ and $\delta_y$, the iterative analysis presented above is repeated on two sets of metrology fringes (with noncolinear wavectors): two maps of projected pixel offsets are obtained ($\pmb{\delta_{p,1}} \pmb{\delta_{p,2}}$). The wavelength vectors of each data set are not strictly perpendicular but fairly close in practice. From this two maps, true $x$ and $y$ offsets (i.e. coordinates in a standard orthonormal basis aligned with the pixel grid) are derived by finding for each pixel the point that generates the right projected offset coordinates. Expressing $\delta_x$ and $\delta_y$ as functions of known parameters is a simple problem of Euclidian geometry. A figure illustrating the principle of the deprojection and the derivation of the solution are presented in appendix \ref{append:Deprojection of pixel offsets}.

\subsubsection{Allan deviations}

\begin{figure}[t]
\centering
\includegraphics[width = 0.45\textwidth]{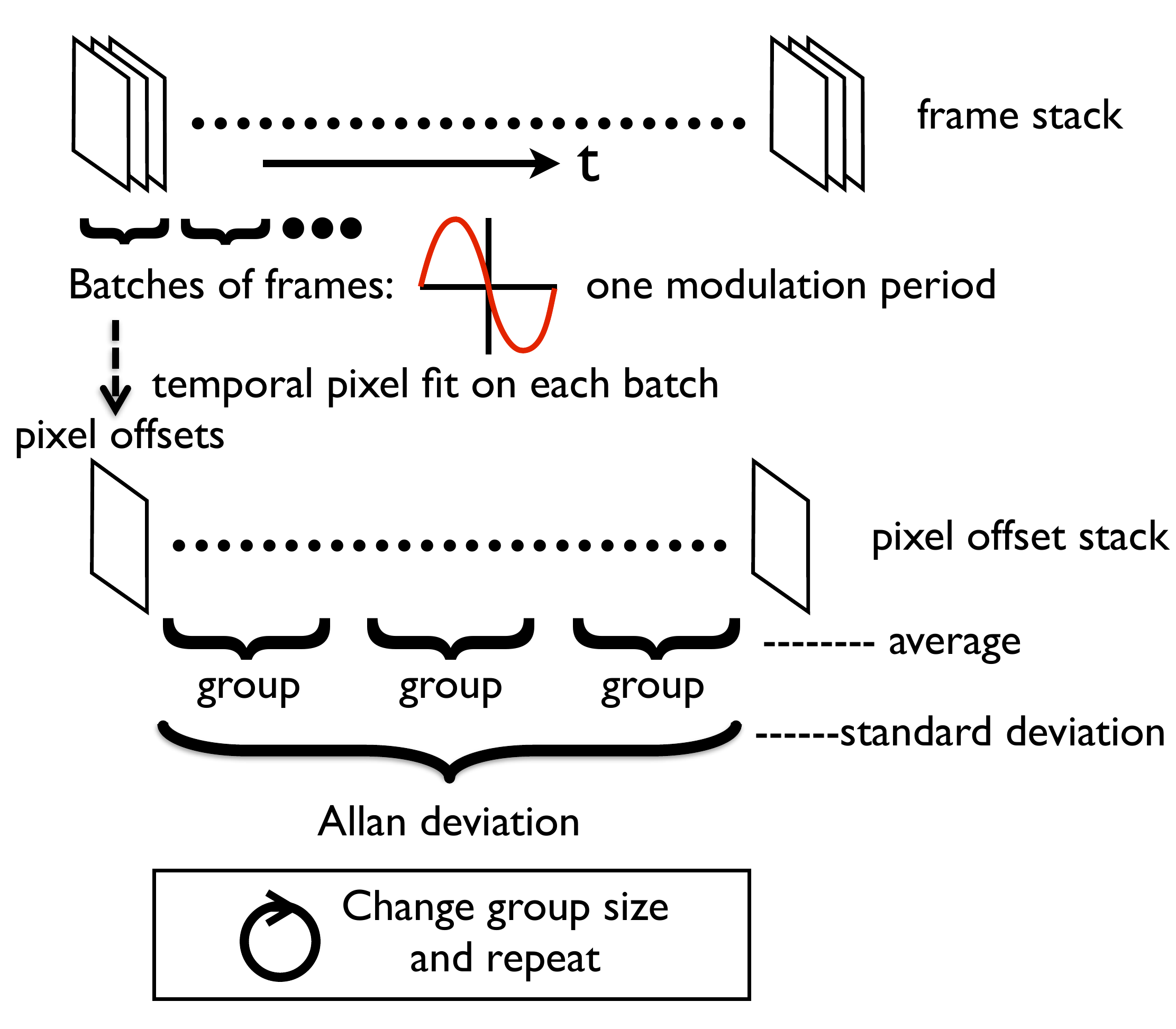}
\caption{How to calculate the Allan deviations.}
\label{allan_deviation_diagram}
\end{figure}

The second result given by the metrology, besides the pixel offsets, is the temporal Allan deviation of these offsets \citep{1966IEEEP..54..221A}. This method gives an estimate of the precision of the measured pixel offsets (temporal deviation) and detects the presence of correlated noise. The basic principle of the method is to simulate having done several identical successive experiments (instead of one) and looking how the accuracy changes with the experiment duration. By splitting the data into several subsets, only a single experiment is actually needed. The Allan deviation result is almost insensitive to the basis in which the coordinates are expressed, as long as the vectors of the basis are close to perpendicular. Figure \ref{allan_deviation_diagram} shows how the Allan deviations are calculated from the accumulated frames.

The analysis starts with dark subtracted frames and the final solution of the spatial fits from the previous iterative process. Temporal pixel fits are performed on small parts of the data called ‘‘batches'', instead of the whole data cube. The number of frames in each batch is calculated so that the temporal signal seen by each pixel covers at least one sine wave period in order to have a well constrained fit. One map of projected pixel offsets is obtained for each batch.

For the second step, the Allan deviations per say are applied on the cube of pixel offsets. The principle is to form groups of pixel offsets maps (in fact group of batches), to calculate the average within each group and then the standard deviation between the groups. The final standard deviation depends on how many batches each group has, that is the group size. 

Each group size corresponds to one point on the Allan deviation plot, so the second step is repeated for different group sizes to obtain a curve. The maximum group size is when the standard deviation is calculated on only two groups. 


To interpret the Allan deviation plot, the results have to be compared to the photon noise limit. Let us consider a sine wave sampled by a punctual pixel: $f(r) = A \sin(2\pi r/\lambda) = A\sin(k r)$. The value $\frac{\dd f}{\dd r}(0) = A k$, is thus the gradient of the sine wave seen by a pixel at position 0. This position is optimal for the measurement because the photon count is most sensitive to the pixel offset, and inversely the offset is least sensitive to photon noise. For an individual frame and an optimally located pixel, the error on the projected pixel offset (noted $\Delta r$) as a function of the photon noise (noted $\Delta f$) is thus: $\Delta r = \frac{\Delta f}{A k}$.

In reality a sine wave is fit on a batch of frames covering a period. Considering that the error on the pixel position is mostly constrained by the frame near the optimal points, we estimate that the photon noise decrease as $\propto\sqrt{N_{\ml{frames}}/2}$ to take into account that there are frames for which the gradient is near 0. This factor of two is true (empirically verified) in our numerical simulations, the exact coefficient depends on the type of temporal fit used, for example the application of statistical weights could improve slightly the performance. The final relation is: 
\begin{equation}
\Delta r = \frac{\Delta f\sqrt{N_{\ml{frames}}/2}}{A k}.
\end{equation}

\subsection{Pseudo stars}\label{subsec:Data analysis: methods - Pseudo stars}

Two different centroiding methods are used. The first one is a simple Gaussian fit with 7 parameters: background level, intensity, position $X$, position $Y$, width $X$, width $Y$, angle. The second one is needed to reach high accuracy on actual data, it is a Fourier-resampling technique. The principle is to measure the displacement between two images by resampling with a phase ramp in the Fourier domain. It uses the following property of the Fourier transform (noted $\ml{FT}$):
\begin{equation}\label{fourier_resampling}
\ml{FT}\left[\ml{PSF}(x-x_0,y-y_0)\right] = \exp{[-i 2\pi (x_0 x + y_0 y)]} \ml{FT}\left[\ml{PSF}(x,y)\right].
\end{equation}

To find the displacement between two identical PSFs, a translation vector $(x_0,y_0)$ is found by a minimum search. The vector for which the residual image between the first PSF and the resampled second PSF is minimal (in the least square sens) is the displacement:
\begin{multline}\label{fourier_resampling}
(x_0,y_0) = \min_{x_t,y_t}\\
\sum_{x,y} \Big[   \ml{PSF_1}(x,y)\;-\;\ml{FT}^{-1}\big[\exp{\big(-i 2\pi (x_t x + y_t y)\big)} \ml{FT} [\ml{PSF_2}(x,y) ]\big] \Big]^2
\end{multline}
In Eq. \ref{fourier_resampling}, the $\ml{PSF}_{1/2}$ notation represent the pixel values inside each one of the fitting windows ($x\in [x_\ml{min}..x_\ml{max}], y\in [y_\ml{min}..y_\ml{max}]$), which are centered around the PSFs maximum. At this stage the pixel values are already dark-subtracted and also (if chosen to) flat-field corrected. 

When done on Nyquist sampled data, the FT - phase ramp - $\ml{FT}^{-1}$ series of operations is equivalent to a perfect interpolation. Another advantage of this method is that the data itself is used to reconstruct the PSF (no model is needed), thus avoiding potential errors caused by a model/real PSF mismatch. But this method has a drawback: it only gives relative displacement. In order to know the distance from one centroid to another, an autocorrelation between two distinct centroids is used. However, because the optical configuration is with only one optical surface and no obscuration of the FoV, the PSF is expected to be nearly invariant. The errors of this process should be mainly caused by the pixels, which are calibrated by the metrology plus flat-field.

To take the pixel offsets into account, an intermediate step is added. Before calculating the offset with Eq. (\ref{fourier_resampling}), the PSFs are corrected by finding their theoretical shapes for null offsets. Eq. (\ref{fourier_resampling}) is then applied between the corrected PSFs. This is done by using the detector model, generated from the pixel offsets. The corrected PSF ($\ml{PSF}_{c}$) is found by minimization of the expression:


\begin{equation}\label{offset_correction}
\min\limits_{(\ml{PSF}_c(x),\ml{PSF}_c(y))}\;\sum_{x,y} \Big[ \ml{PSF}(x,y) - \ml{PRF}(x,y)\times\ml{PSF}_{c}(x,y) \Big]^2.
\end{equation}

$\ml{PRF}(x,y)$ is the pixel response function of pixel $(x,y)$. The second term in the sum, $\ml{PRF}(x,y)\times\ml{PSF}_{c}(x,y)$, is not a 
straightforward product. Indeed, $\ml{PSF}_c(x,y)$ is a scalar value, whereas $\ml{PRF}(x,y)$ is a matrix representing the pixel response function. The latter product notation is a simplified way to represent the convolution products between PRF and $\ml{PSF}_c$, evaluated for each pixel in the window (at their fixed local pixel coordinates). In order to compute this convolution product, $\ml{PSF}_c$ is first oversampled by FFT - zero padding - $\ml{FFT}^{-1}$ to match exactly with the PRF sampling and integrated over each pixel.

\subsection{Differential astrometry dispersion metrics}\label{Differential astrometric dispersion metrics}

We have shown in the previous section how we measure the locations of the pseudo stars. These locations are expressed as pairs of X and Y pixel coordinates. By abuse of language, we call these locations centroids. The literal meaning of ‘‘centroid'' is closer to a geometric mean, that is a barycenter. In our case the location is not obtained from a barycenter of pixel values (it would be an inaccurate method), but by a more sophisticated fit. In essence it is an improved barycenter. So the centroids are essentially the resulting astrometric measurements. 

Thus, after completion of the pseudo star processing we have a measure of the centroids (one per star, 5 in total), either versus time, or for several positions on the detector, depending on whether the translation stage supporting the detector was moved between the different acquisitions or not. As previously explained by Fig.~\ref{simplified_experiment_principle} (Sect. \ref{subsec:Experiment concept and simplified principle}), moving the detector produces a common translation of all 5 centroids (we never move the CCD during the acquisitions). 

The final step of the analysis is to produce dispersion metrics of the measured centroids. The metrics should represent accurately the stability of their relative locations: the dispersion of the absolute position of a given centroid (i.e. the position in the pixel grid referential) gives the setup stability, in other words how well the testbed as been stabilized against external perturbations, such as thermal expansion or mechanical vibrations. Although the absolute position dispersion is interesting to know, it is very different from the relative astrometric accuracy. In contrast, the relative astrometric accuracy (or dispersion) is analogous to the standard deviation (hereafter SD) of the distances between the centroids. But it is not exactly that (see Procrustes analysis shortly after). 

There are two useful kinds of relative dispersion (depending if the the translation stage supporting the detector was moved between the different acquisitions), so then the relative astrometric accuracy is calculated (respectfully either versus time or versus the detector position), it define two distinct modes of observation, which serve different purposes. 


We call the first mode where the detector does not move the \textit{single-position} analysis. In this case, we obtain a measure that is somewhat sensitive to some environmental factors such as mechanical stability, air turbulence etc... but if the absolute positions are stable enough, the pixelation errors are nearly constant and therefore do not affect the measures significantly. In practice the absolute positions are stable to better than 1\% of a pixel (and caused mostly by mechanical vibrations). 


The second mode is called \textit{multi-position} analysis: the centroids are placed at different positions on the detector with the translation stage. The amplitude of the motion between each position can be controlled and can range between 1\% of a pixel to several pixels. In this mode, the relative positions of the centroids are be strongly affected by pixel responses when the distance between two detector positions is several pixels. 

\begin{figure}[t]
\centering
\includegraphics[width = 0.45\textwidth]{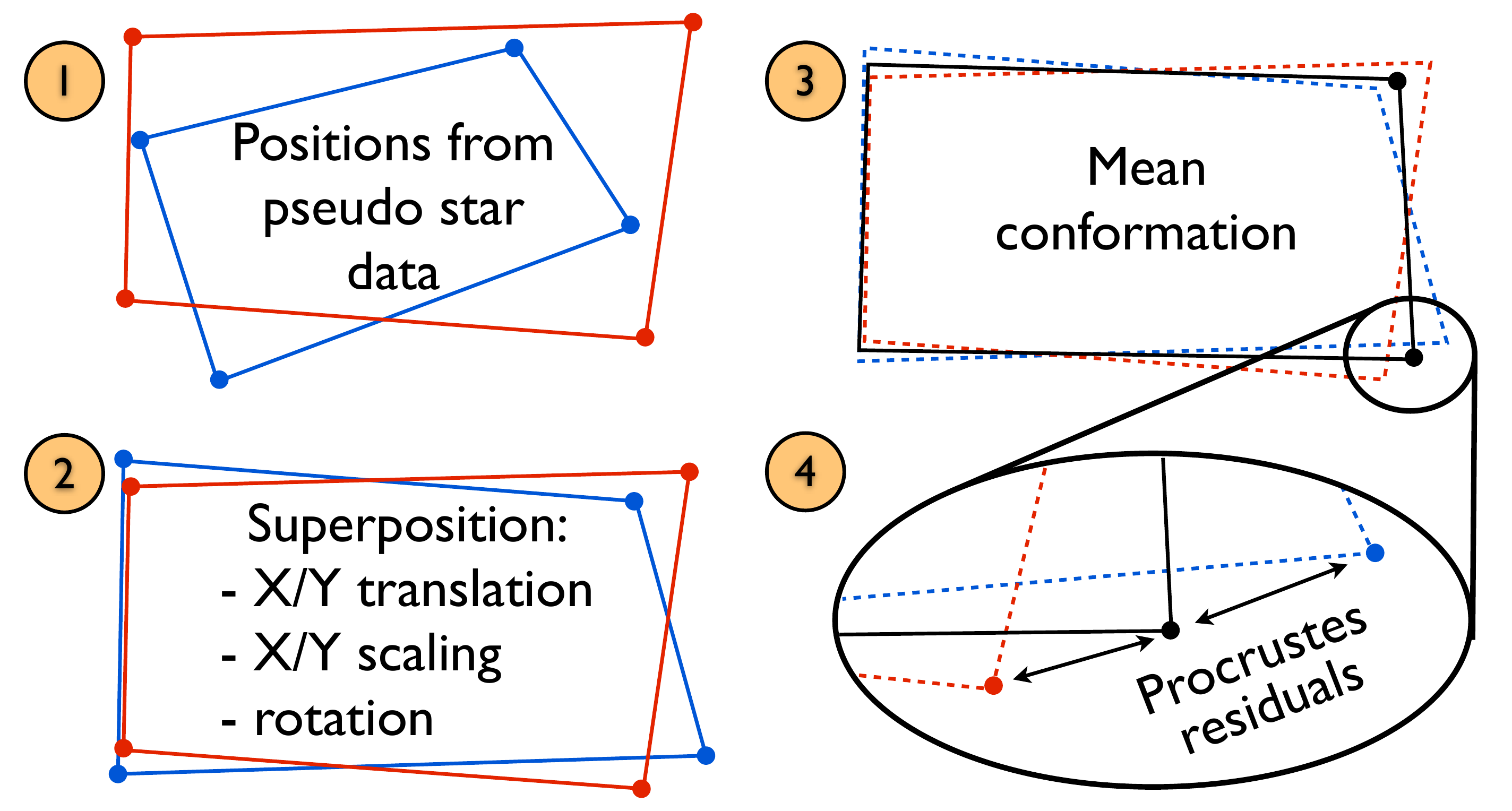}
\caption{Diagram of the Procrustes analysis. The relative centroid positions for two geometric conformations, each corresponding to one detector position, are represented respectively in red and blue. The black conformation (plain black line at step 3) is the average between the red and the blue ones.}
\label{fig:procrustes_analysis}
\end{figure}

However there is an additional critical effect that occurs in this case: when moved, the translation stage induces large tip-tilt errors. This produces vertical and horizontal scale changes have to be taken into account. To correct these a Procrustes superimposition procedure is used. The principle is to find the geometric transformation that results in the closest overlap of the measured centroid positions. The residuals between the overlaps indirectly yields the final accuracy. Five parameters are needed to define the transformation: translation $X$ and $Y$, scaling $X$ and $Y$, rotation. This is less than the number of data points; 2 axes $\times$ 5 centroids for each position. Figure \ref{fig:procrustes_analysis} illustrates how the Procrustes analysis is done and how the residuals are obtained. 

\section{Numerical simulations}\label{sec:Numerical simulations for DICE}

\begin{figure}[t]
\begin{center}
\includegraphics[width = 0.35\textwidth]{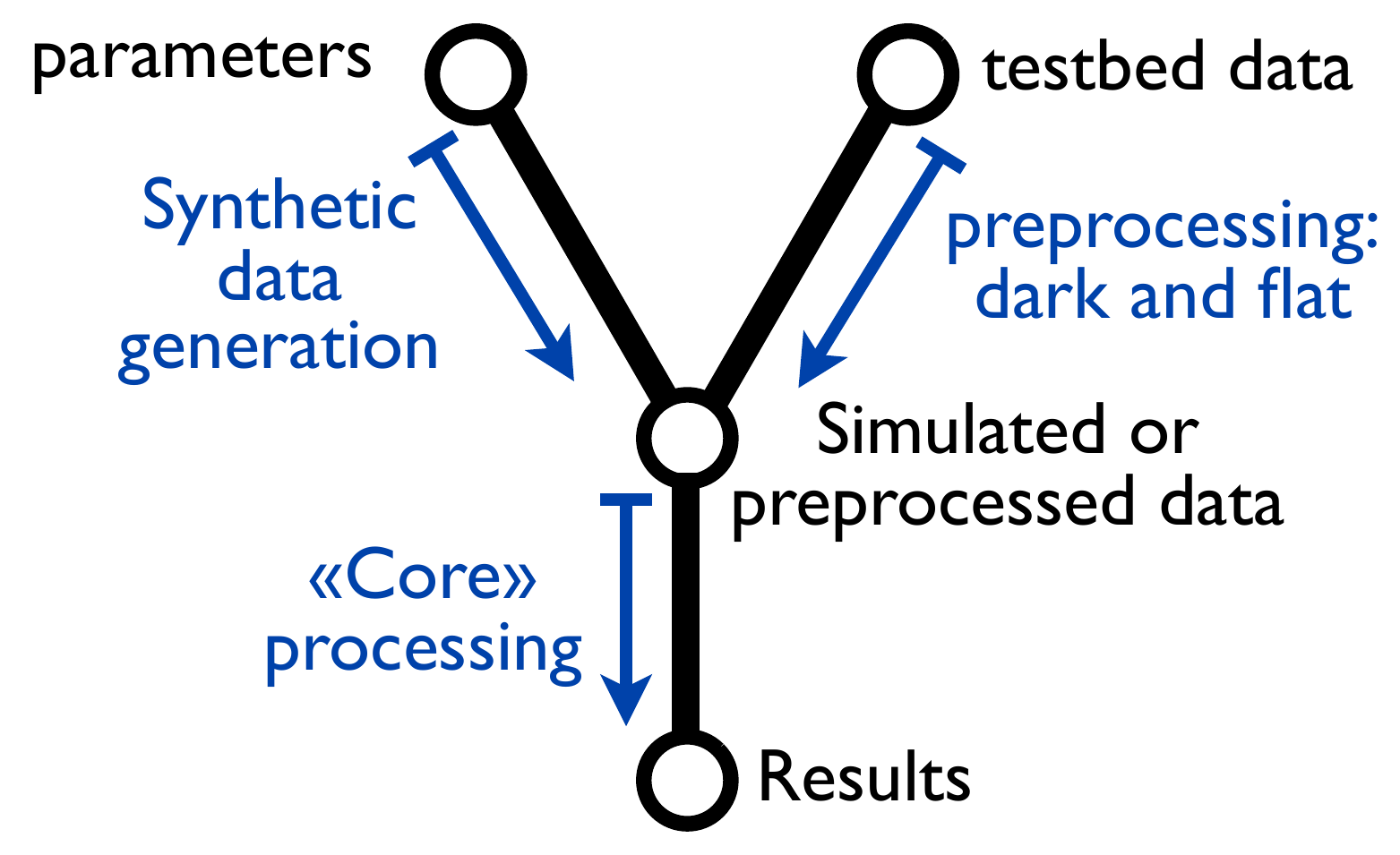}
\caption{\label{Yarchitecture} Actual data and synthetic data flow diagram. The pseudo star and metrology data analysis scripts both follow this architecture. An option setting parameter can be used to switch between synthetic and actual data.}
\end{center}
\end{figure}

In addition to the data reduction process itself which yields pixel positions and star positions (i.e. centroids) from raw images, our set of tools also includes a numerical model, which is used for debugging, checking the data processing for artifacts and characterizing error propagation from parameters uncertainties to final accuracy. It has been especially useful for errors that are hard to assess analytically. Synthetic data is generated and plugged at a specific point into the data analysis pipeline, that is right after the dark and flat calibrations, as illustrated by Fig. \ref{Yarchitecture}.

\subsection{Data pipeline}

The numerical model does not include a detailed solution of the detector behavior (e.g. including bias, dark current...), which could be used to simulate some subtle effects like uniform or non uniform (different for each pixel) changes of the bias, or of the gain, or others complex phenomena that could result in unexpected systematic effects in the measurements. In our numerical model, photo-electron counts are directly converted from a photon count estimation by an homogeneous scaling factor. The photon count estimation comes directly from integrating a theoretical illumination pattern (either for the star or the metrology configuration) over the pixels. In order words, we make the working assumption that the detector is well behaved, at least to the extent that it would not adversely affect the metrology and astrometric measurements.

However, readout noise and PNRU noise (caused by either by an absence of calibration or a residual error after calibration), can be simply simulated by direct application on the photo-electron counts.
Only homogeneous Gaussian PRNU noise and readout noise are applied onto synthetic data. The treatment is slightly different between synthetic and actual data: for the latter additional processing steps are required, precisely dark subtraction and flat-field calibration. In the case of synthetic data, the PRNU noise is not meant to be calibrated but is used to simulate a possible flat-field calibration residual error. 
One critical point: in this configuration, the amplitude of PRNU map errors, which includes possible errors from imperfect flat-field processing, can not be estimated with the numerical model. However these errors can be derived using the difference of actual PRNU maps taken in different conditions (e.g. source fiber position).

The core of the processing, that is the fringe fit and derivation of pixel offsets for metrology data and the PSF resampling for pseudo star data, is common to actual data and synthetic data. This enables:
\bit
\item a reliable debugging of the core of the processing. When synthetic data is injected, the exact solution is known and is used to determine true errors (computed solutions, i.e. outputs, minus the exact ones, i.e. inputs). Under ideal conditions (no noise), the true error should be close to zero, ideally within the numerical precision.
\item characterization of artifacts introduced by the processing. In practice true errors do not have to be at numerical precision, only below the level required for $5\e{-6}$ pixel final error on centroids. True errors can still be determined when any kind of perturbation (random noises or systematics) are added to the synthetic data. Thus the conditions under which the resulting accuracy is compatible with the experiment objective can be determined.
\item confirmation and extension of the analytical model. Applied in Monte Carlo simulations, the same error analysis process can be used to test each noise source separately, for different noise amplitudes, yielding empirical relations between the noise sources and the final accuracy. The consistency between the analytical and numerical models can be checked and more complex noise sources can be characterized.
\ei

\subsection{Data generation}

The element needed to generate data that incorporates pixelation errors is a model of the detector, which consists of PRFs concatenated together. We model PRFs by truncated parametric hyper-Gaussians (Eq. (\ref{eq:PRF})), which parameters vary from pixel to pixel.

\begin{equation}\label{eq:PRF}
\ml{PRF}(x,y) = C*\exp{\left[\frac{(x-x_0)^n}{2\sigma_x^n}+\frac{(y-y_0)^n}{2\sigma_y^n}\right]}.
\end{equation}

The pixel sensitivity corresponds to the sum of all the elements in the PRF, the pixel offset corresponds to its barycenter and the width is its standard deviation. The use of hyper-Gaussians allows for an easy modeling of the pixel global parameters and provides a smooth function with little high frequency content. The metrology measurements are not sensitive to high frequencies within pixels and are thus not relevant in the model.

%

Simulated metrology fringes are generated by multiplying the detector model above with an ideal and oversampled pattern of fringes. The pixel values are obtained by summing over each pixel area. The frames are generated one by one, by shifting the fringe phase. The result is a simulated metrology data cube of moving fringes carrying the information of pixel sensitivities ($C$), offsets ($x_0,y_0$) and widths ($\sigma_x,\sigma_y$).

For the generation of pseudo stars, PSFs are approximated with Gaussian functions whose width is equivalent to an Airy spot at the average wavelength for our experiment (about 600 nm). The reason for using a Gaussian instead of a more complex PSF shape, such as a polychromatic sum of Airy functions or a PSF derived from the data is simplicity and efficiency. Our experiment has a highly stable optical configuration with only one optical surface, the PSFs are quasi invariant and are not expected to contribute significant errors, whatever their exact shapes are. A simplified model is thus sufficient. 

The goal of the pseudo stars model is to estimate the relations between the uncertainties on various parameters (e.g. PRNU, pixel offsets...) and the centroiding accuracy. This process effectively tells us what error sources dominate and should be addressed in priority. For this we also use the detector model. To generate pixelated pseudo stars, the product between oversampled Gaussian centroids and the detector model is computed (pixel values are obtained by summing over each pixel area.).

\subsection{Metrology simulation results}\label{subsec:Results on simulated data - Metrology}

\begin{figure*}[t]
\centering
\includegraphics[width = \textwidth]{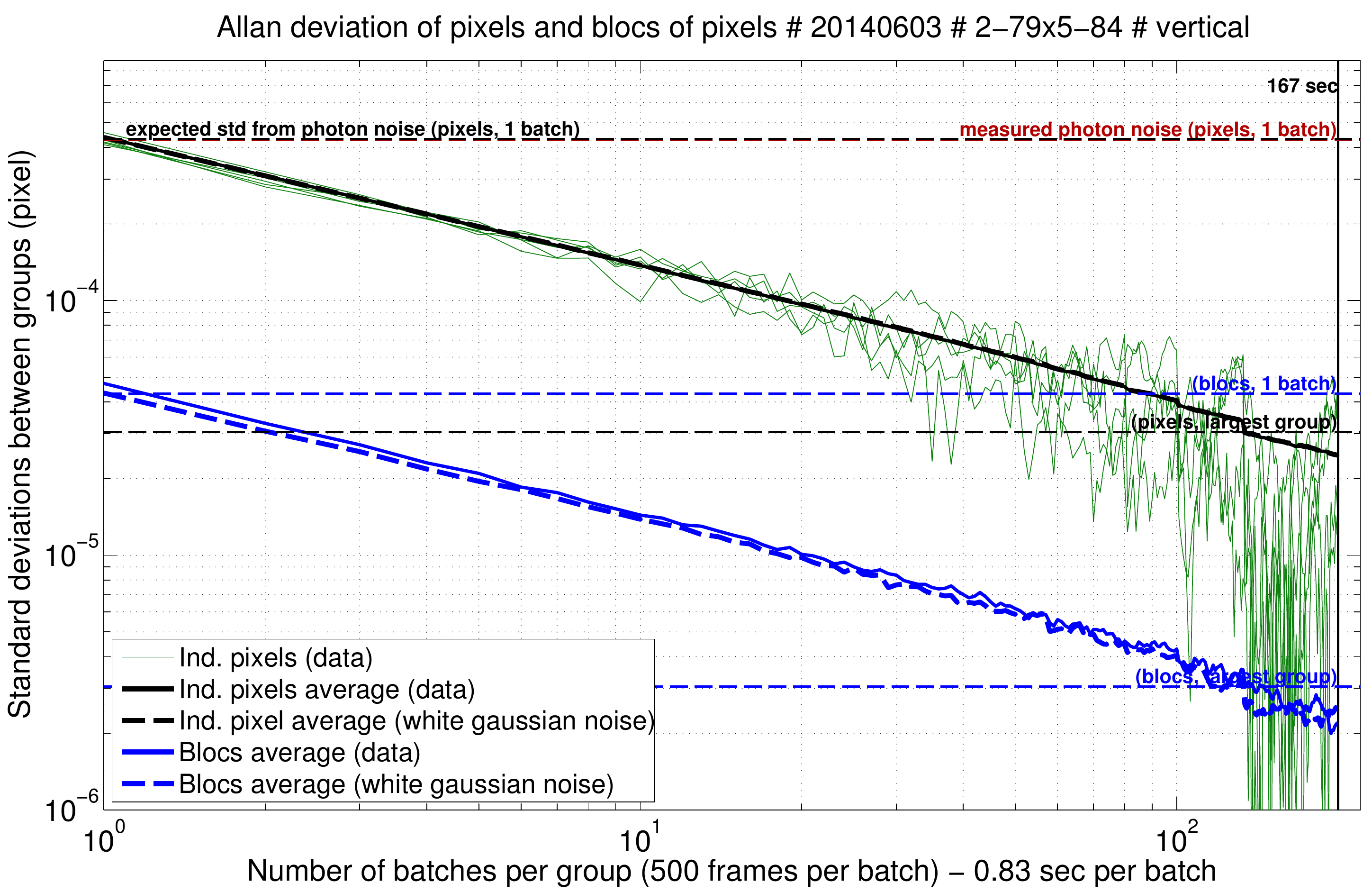}
\caption{Allan deviations of simulated data. Amplitude, visibility and photon noise of simulated fringes are adjusted to values typical of a real experiment (B = 10000 counts, A = 6000 counts, 1 count = 10 photo-electrons). Additional sources of noise are simulated, such as laser intensity ($1\e{-2}$ RSD), fringe phase ($1\e{-2}$ radian SD) and PRNU ($1\e{-5}$ RSD). The plot shows deviations for individuals pixels (plain green), their average deviation (plain black) and the average for blocks of 10 by 10 pixels in (plain blue). The dotted black and blue  curves are for a cube of white noise whose standard deviation is matched to the data for groups of 1 batch. Averaging Allan deviations over pixels or blocks is important because they tend to be noisy (plain green) when the final deviation is derived from very few groups. Horizontal dotted blue and black lines are estimations of the photon limit for individual pixels or blocks of 10 by 10 pixels.}
\label{allan_deviation_simData_example}
\end{figure*}

The metrology model is used in two ways. The first one is the comparison between the detector model and the results from the processing. The most important outputs here are the pixel offsets, the goal is to make sure that the processing does not introduce biases greater than ideally $5\e{-6}$ pixel for the pixel offsets. Since the exact solution is known, a simple subtraction between the model and the processing output yields the bias. 

The second output is the Allan deviation of pixel offsets. When working with actual data the exact solution is not accessible, other methods have to be used. The Allan deviation analysis gives information about the SD of the pixels offsets as a function of the number of frames used for the fit. For photon noise limited measures, the SD is expected to decrease proportionally with the square root of the number of frames. Allan deviations on simulated data also provide a mean to check that the data analysis method is well behaved: the SD should be in agreement with the theoretical photon noise limit and should decrease as the square root of the number of frames.

Figure \ref{allan_deviation_simData_example} shows the Allan deviation obtained after analysis of 200,000 simulated frames of metrology fringes. The goal is to validate the processing pipeline and the number of photons needed to reach the astrometric accuracy requirement.

For the largest groups (200 batches per group, i.e. 100,000 frames per group), the SD reaches $2\e{-6}$ pixel for blocks of 10 by 10 pixels and $2\e{-5}$ pixel for individual pixels. The expected SDs from photon noise are indicated by horizontal dashed lines on the plot and they coincide almost perfectly with the measured Allan deviations. The top black line shows what the deviation for groups of one batch should be: as expected, it crosses the left axis of the plot at the same place than the Allan deviation curve (black line). There are actually two lines near the top, nearly indistinguishable because on top on each other: the dark one is for theoretical photon noise, the red one is for measured photon noise using the first frame of the data cube. No red noise nor readout noise was included in the model. In the actual experiment the readout noise is negligible compared to photon noise.

\begin{figure}[t]
\centering
\subfigure[]{\label{offsetBiasSimu_qeSD5e-5}
\includegraphics[width=58mm]{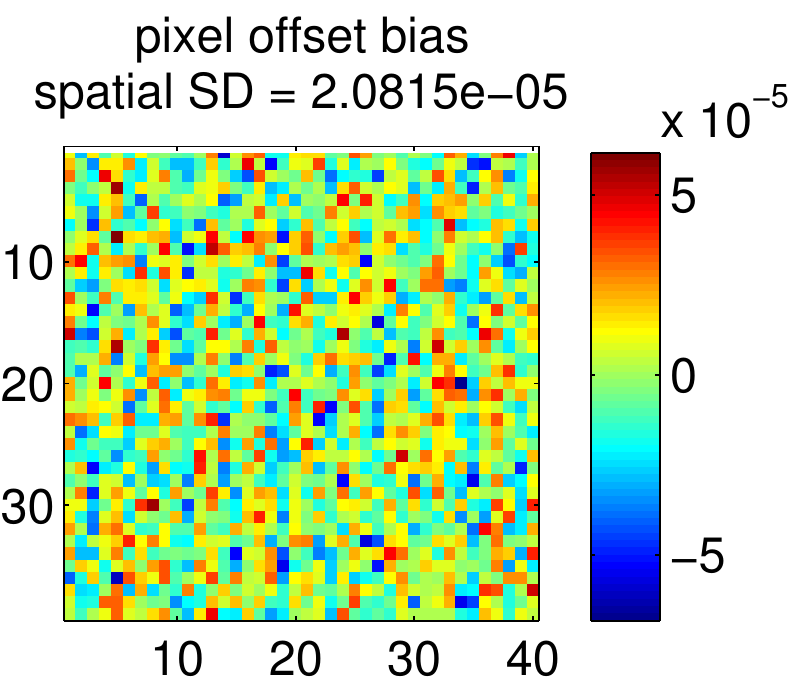}}
\hspace{5pt}
\subfigure[]{\label{offsetBiasSimu_qeSD2e-3}
\includegraphics[width=58mm]{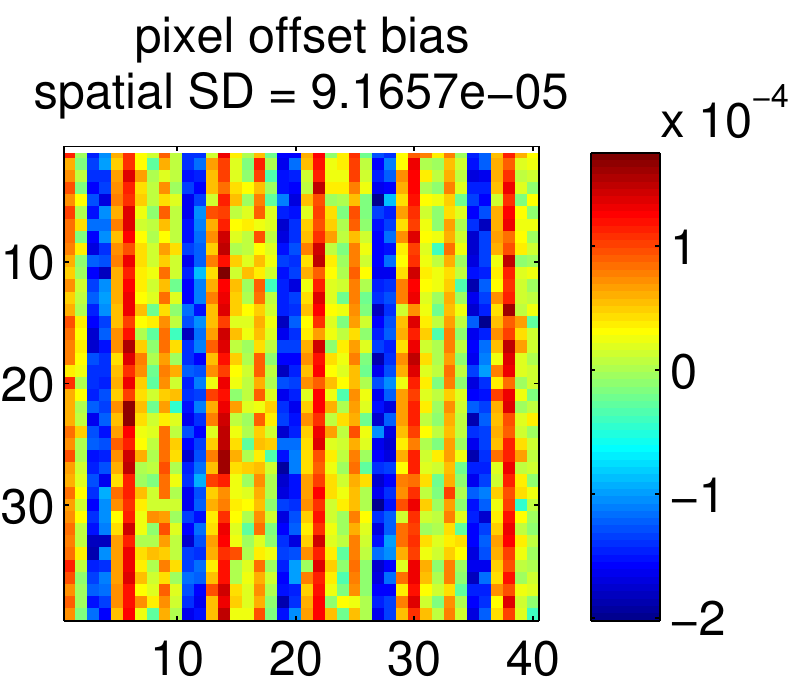}}
\caption{\label{pixelOffsetBias} Pixel offsets bias (in the horizontal direction) for different PRNU RSD. The maps show the difference between the pixel offsets found after processing and the solution of the simulation, for PRNU RSD of $5\e{-5}$(a), and $2\e{-3}$(b). The residuals for (a) are at the photon noise floor, while a stronger systematic bias of $1\e{-4}$ pixel is visible for (b). Both have the same photon noise.}
\end{figure}

Figure \ref{pixelOffsetBias} shows maps of the difference between the measured pixel offsets and the true solution (pixel offset simulation input) for different values of PRNU RSD (relative standard deviation). This is the ultimate metric to check the accuracy of the result, because biases constant in time are not visible in the Allan deviation. The latter only gives information about the stability of the pixel offset measurement. For low PRNU (RSD of $5\e{-5}$), the SD of $2\e{-5}$ pixel seen in Fig.~\ref{pixelOffsetBias} is in agreement to the value given by Allan deviation (i.e. the photon noise floor). However, tests with higher values of PRNU show residual offset bias well above photon noise floor, while the Allan deviation is unaffected. For a PRNU RSD of $2\e{-3}$, the residual SD of offsets rises to $10^{-4}$ pixel. No bias greater than $10^{-6}$ pixel has been observed for other types of noises, like fringe phase jitter or overall intensity variation.

\subsection{Pseudo stars simulation results}\label{subsec:Results on simulated data - Pseudo stars}

The first goal is to validate the reduction process itself, as capable of reaching $5\e{-6}$ pixel (centroid position error) in ideal conditions: the final accuracy is limited by simulated photon noise, all other noises are assumed negligible or well calibrated (e.g. PRNU). The accuracy has been validated in this way for both centroiding techniques (Gaussian and autocorrelation), with correction of pixel offsets from metrology data and without.

The second goal is to explore what are the impacts of different types of noise by injecting them one at a time into the simulated data, in Monte Carlo simulations. When using Gaussian PSFs, we can directly compare the input centroid locations with the fit results, thus relying on absolute positions (as opposed to relative ones) and the Procrustes analysis is not needed. The model was used to estimate the relations between the uncertainties on various parameters (PRNU, pixel offsets, photon noise, readout noise) and the pseudo star location error (final accuracy), the results are summarized in Table \ref{result_centroid_model}.

\begin{table}[t]
\caption{\label{result_centroid_model}Results from pseudo stars model. The error on centroids (i.e. pseudo stars measured locations) is always in pixel units.}
\tiny
\begin{center}
\begin{tabular}{  l  l  l }
\hline
Error type 		 							& Error normalization / definition				& \specialcell{Error on\\centroid}   \\ \hline\hline
PRNU: $\sigma_{\ml{PRNU}}$ & average pixel sensitivity = 1 & 0.40 $\sigma_{\ml{PRNU}}$ \\ \hline
Photon noise: $\sigma_{\ml{ph}}$	& \specialcell{relative photon noise\\calculated for the pixel\\with the highest value} & 0.55 $\sigma_{\ml{ph}}$\\ \hline
Pixel offset: $\sigma_{\ml{offset}}$ & \specialcell{offset expressed\\in pixel units} & 0.25 $\sigma_{\ml{offset}}$ \\ \hline
Pixel read noise: $\sigma_{\ml{read}}$ & \specialcell{relative read noise\\calculated for the pixel\\with the highest value} & 1.8 $\sigma_{\ml{read}}$ \\ \hline
\end{tabular}
\end{center}
\end{table}

Another useful aspect of these Monte Carlo simulations is the estimation of the proportion of centroid errors that are absorbed by the Procrustes superimposition. Among 10 observables ($(x,y)$ coordinates of 5 centroids), the superimposition technique allow for a fit with 5 degrees of freedom (2 translations, 2 scalings, 1 rotation), which will inevitably lead to underestimate the final noise. Monte carlo simulations of stars in the same geometrical configuration as in the real experiment with purely random and uncorrelated astrometric jitter have shown that Procrustes superimposition decrease uncorrelated location errors by a factor 1.4 The final accuracy expressed after the Procrustes superimposition is corrected (majored) to compensate for this factor.

\subsection{Conclusions from numerical simulations}

\subsubsection{Models convergence}

In addition to the numerical models, our set of tools also included an analytic error model \citep{2014SPIE.9150E..0IH}. The analytic error model resulted in a spreadsheet which can be used to understand error propagation quantitatively and determine some specifications on the stability or calibration accuracy needed on error sources like PRNU, offsets, laser intensity and wavelength stability. We have successfully checked for consistency between the two models, for the errors that we could characterize with both models, such has PRNU and offsets.

\subsubsection{Error propagation coefficients}

From the relations shown in Table \ref{result_centroid_model} we conclude that to reach an error below $5\e{-6}$ pixel on the centroid, the following calibrations must be fulfilled:
\begin{itemize}
\item PRNU to better than $1.2\e{-5}$ (relative QEs)
\item pixels offsets to better $2\e{-5}$ pixel
\end{itemize}
Here we have considered the errors independently, of course in the experiment the different kinds of errors will add up: it is the sum of their contributions than must yield a final accuracy of $5\e{-6}$. However the numbers presented above are meaningful as minimal requirements.

\subsubsection{Control of residual biases}

Sect. \ref{subsec:Results on simulated data - Metrology} mentioned the presence of a bias resulting from the data processing. There is a residual bias on the measured pixel offsets, which depends on the amplitude of the PRNU calibration error. We have presented bias maps for two different levels of PRNU RSD (low and high). These levels have not been chosen arbitrarily: $5\e{-5}$ (low) corresponds to the measured bias for the best sets of actual data, $2\e{-3}$ (high) is for uncalibrated pixels, which is characteristic of our detector. The bias discussed above is intrinsic to the data processing and is an possible point to improve, but a good pre-flat calibration of the actual metrology data can ensure that pixel offset bias remains lower than $5\e{-5}$ pixel. This is not an additional constraint as the bias RSD amplitude must be lower than $1.2\e{-5}$ pixel to be compatible with the $5\e{-6}$ pixel astrometric accuracy goal.

\subsubsection{Intrapixel responses}

With the metrology analysis methods presented here, using only two baselines of different directions, information of the PRF inside the pixel (intra-pixel calibration) can not be obtained. A single baseline produces a pure sinusoidal pixel response, so only the location and at best the pixel width are accessible. Characterizing the intra pixel response requires several (aligned) baselines \citep{2011RSPSA.467.3550Z}. Our model assumes that PRFs are adequately represented by parametric hyper-Gaussians, but we do not know the true shapes of PFRs. Higher frequency components of the PRFs are expected to generate very small errors.

\section{Experimental data analysis and discussion}\label{sec:Analysis of experimental data}


The experimental data presented here was obtained in September 2015. For some quantities there are significant deviations localized very close to the detector edges. When this is the case the images are cropped to avoid widening their dynamic range. Although the physical size of the detector is exactly 80 rows by 84 columns, the images always show smaller areas. A base "science area" (rows 2 to 79, columns 10 to 83) is used but for some quantities the images must be cropped slightly further. When numbers (e.g. a spatial STD) are associated with the images they are always calculated on the area shown in the pictures, that is after cropping. These issues have negligible impact on the results because the pseudo stars always stay far enough from the problematic edges.

\subsection{Experimental data}

\subsubsection{Dark and flat-fields}\label{xpdata:dark_and_flat}

\begin{figure*}[p]
\centering
\subfigure[Mean dark frame]{\label{darkXpData}
\includegraphics[width=0.42\textwidth]{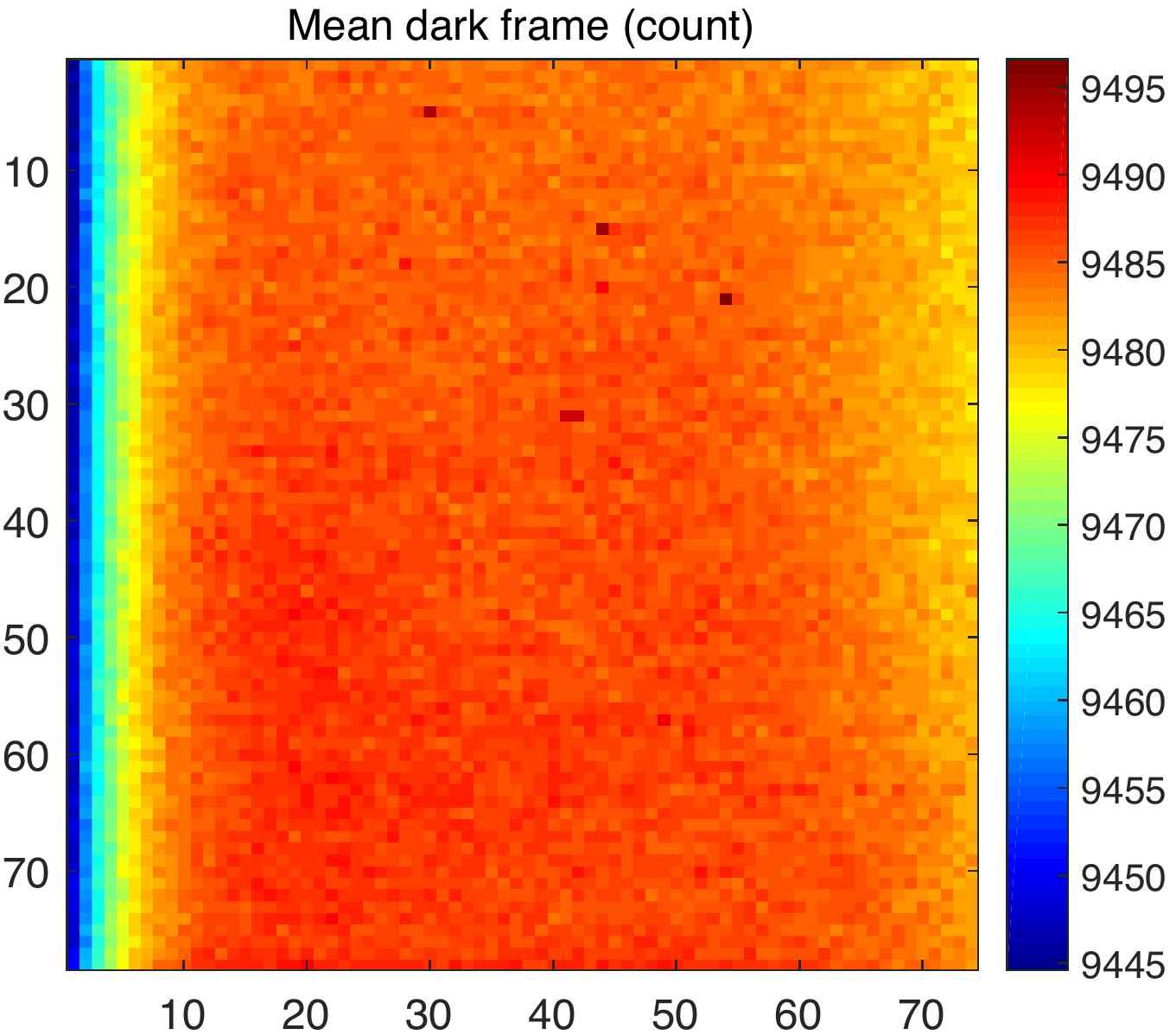}}
\subfigure[Temporal noise in dark data cube]{\label{darkNoise}
\includegraphics[width=0.405\textwidth]{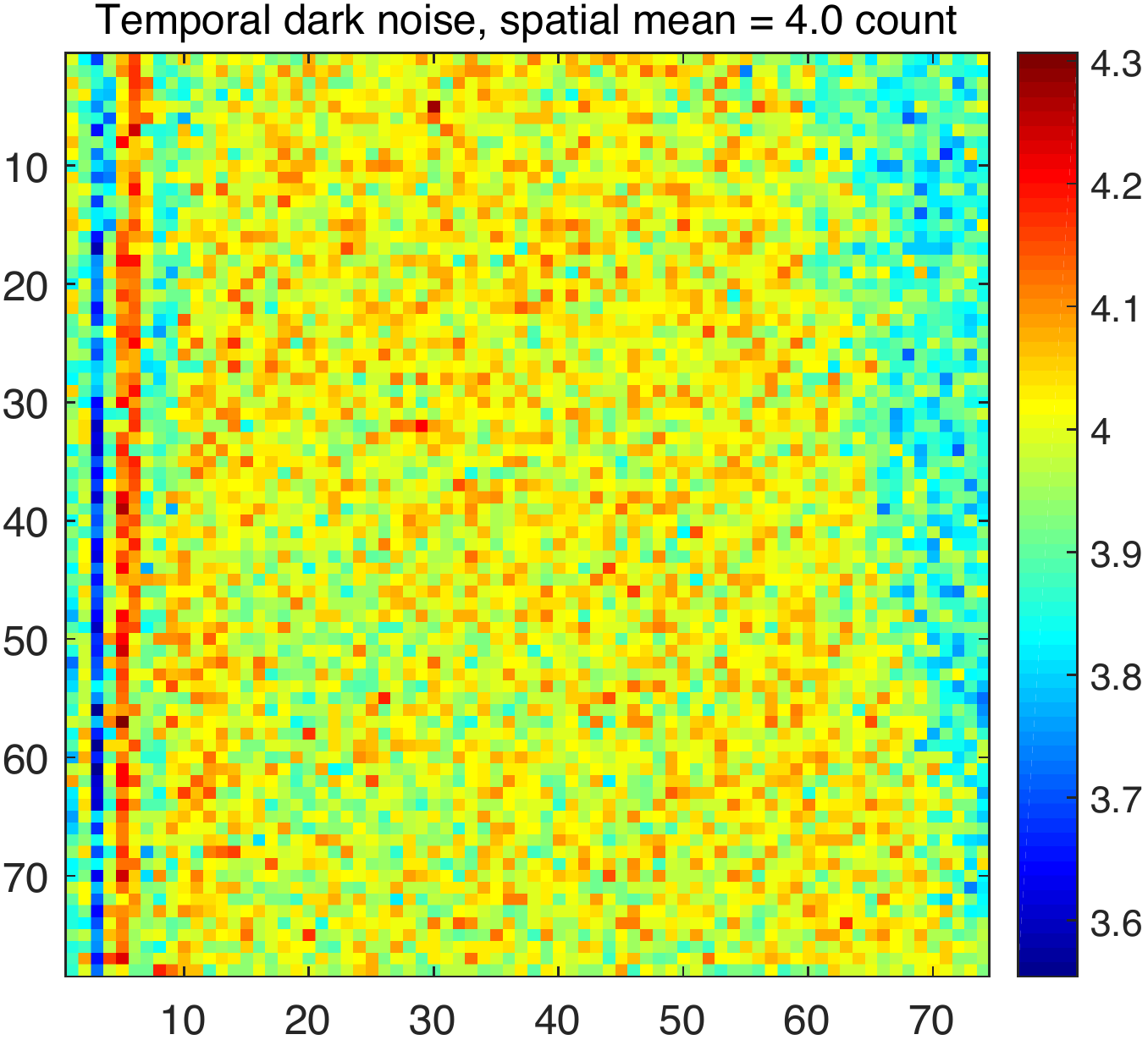}}
\vspace*{-4mm}
\caption{\label{darkCube}Temporal mean and noise of dark data cube}
\end{figure*}

Figure \ref{darkXpData} shows a mean dark frame. Figure \ref{darkNoise} shows the temporal standard deviation of the data cube which yielded the aforementioned mean dark frame. The dark plus readout noise (4.0 counts SD) is dominated by photon noise. The CCD is operated at fractions of full well that correspond to photon noises of 30 counts or more.

\begin{figure*}[p]
\centering
\subfigure[Measured PRNU with a white LED and multimode fiber]{\label{prnu}
\includegraphics[width=0.425\textwidth]{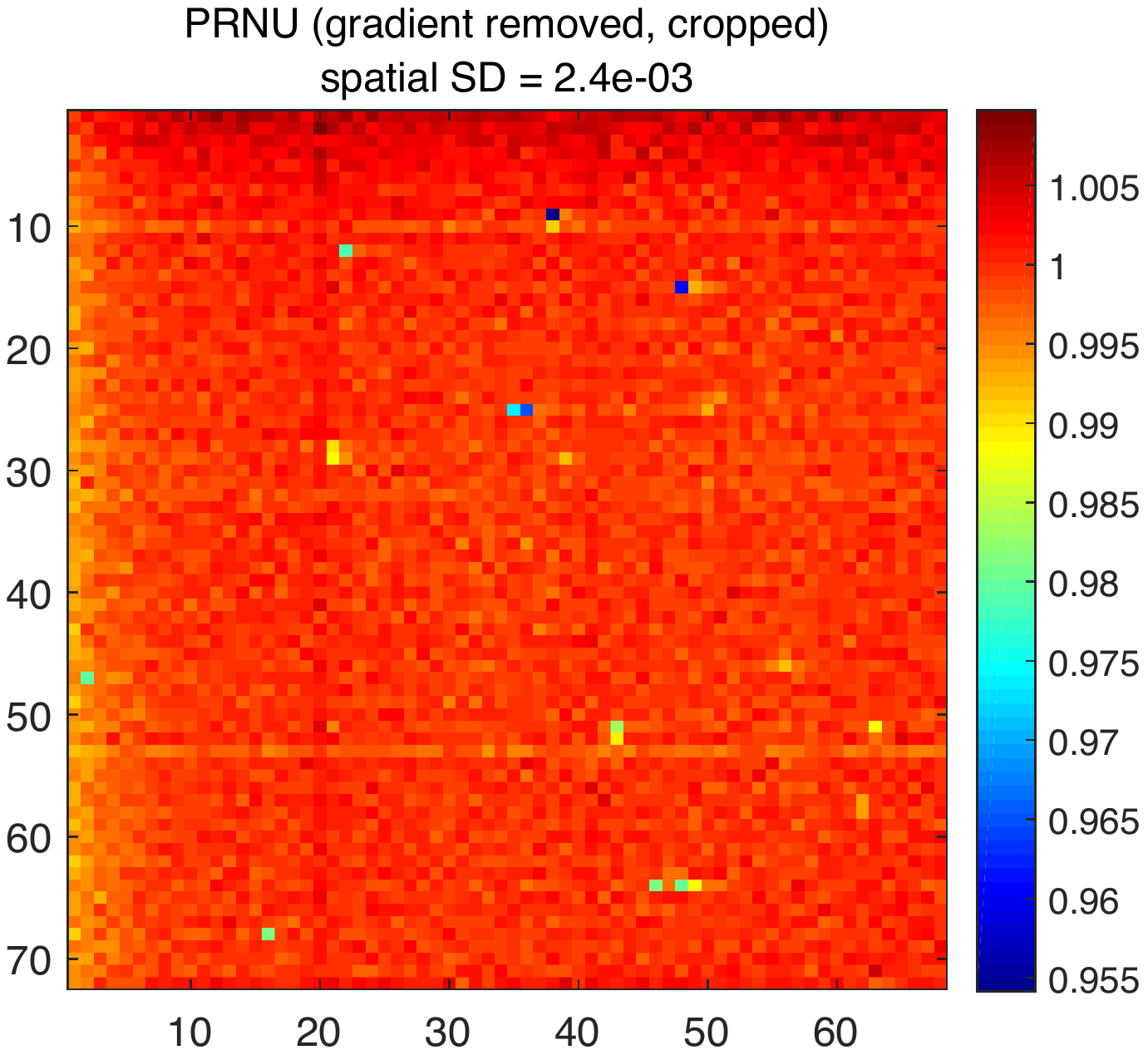}}
\subfigure[Difference between two measured PRNU, with a displacement of the fiber tip (photon noise limit: $2\e{-5}$)]{\label{prnuDiff}
\includegraphics[width=0.42\textwidth]{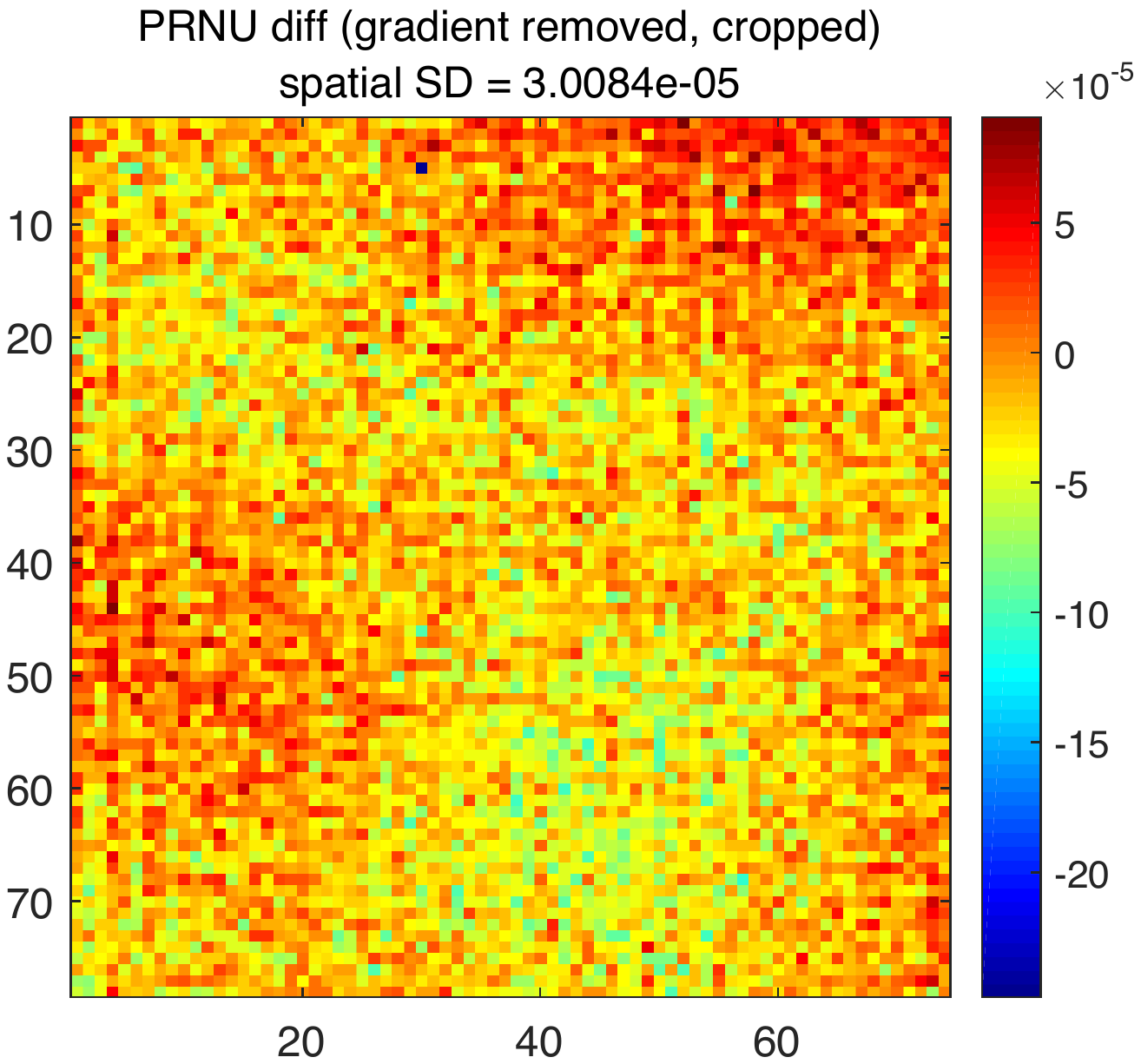}}
\vspace*{-4mm}
\caption{\label{mainLabel}PRNU and its difference with a displacement of the fiber tip.}
\end{figure*}

Figure \ref{prnu} shows the measured PRNU, after processing of the flat field obtained with the white LED and multimode fiber. The measured PRNU RSD is $2.4\e{-3}$. The PRNU distribution is not pure white noise, several features can be distinguished: 2 horizontal bright bands, dark pixels down to 95\% efficiency (caused by dust contamination), faint low frequency patterns and more visible biases close to the upper and left edges.

To estimate the systematics on this result, the PRNU difference between two different fibers positions spaced by 2 cm is computed (Fig. \ref{prnuDiff}). 2 cm correspond to 20\% of the baffle field of view (10 cm) and an angle of 1.9$\dg$ viewed from the detector. The magnitude of the difference is $3\e{-5}$, for a photon noise limit at $2\e{-5}$. Applying the Allan deviation method on the flat field data cube did not reveal any anomalies. Using flat fields from the metrology was attempted. But in this second case, PRNU differences between the two different fibers positions vary between $1\e{-3}$ and $5\e{-4}$ (depending on the residual level of stray light), but in all cases are significantly higher than for white flats.

There was a concern about the possibility of having systematic effects, and in particular speckles, caused by the large core (365 $\mu$m diameter) of the multimode fiber. To estimate these effects an experiment similar to the one described above was performed: a flat field difference after variation of the fiber bending (instead of a fiber displacement) was measured. It showed the same amplitude difference of about $3\e{-5}$, ruling out a significant impact from the bending.

\begin{figure*}[p]
\centering
\subfigure[Horizontal pixel offsets]{\label{xOffsetsXpData}
\includegraphics[width=0.42\textwidth]{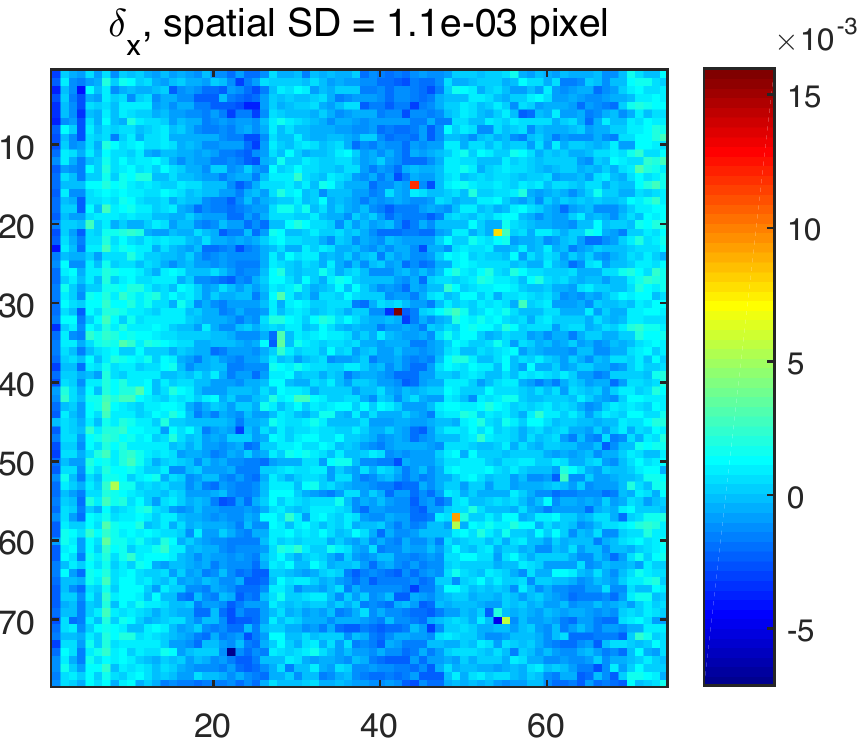}}
\subfigure[Vertical pixel offsets]{\label{yOffsetsXpData}
\includegraphics[width=0.42\textwidth]{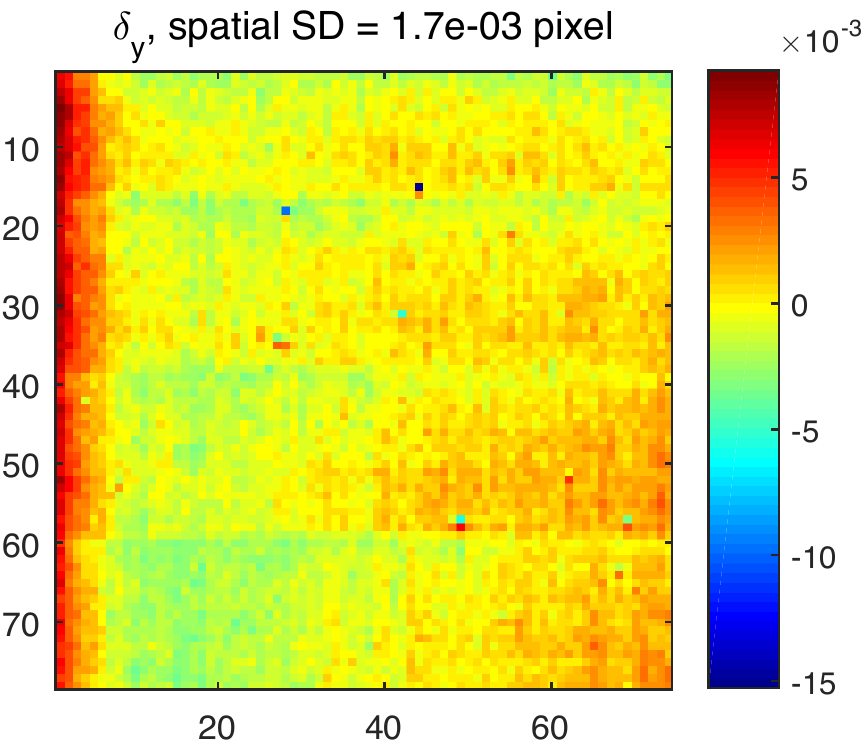}}
\vspace*{-4mm}
\caption{\label{mainLabel} Pixel offsets (in pixel units) as measured by the metrology (baseline H1H4-V1V4).}
\end{figure*} 

\subsubsection{Metrology}

One complete data set consists of a pair of baselines (vertical and horizontal). The results presented are with the laser coherence control system active (coherence $\approx$ 1 cm). Figures \ref{xOffsetsXpData} and \ref{yOffsetsXpData} show the measured pixel offsets for baseline H1H4/V1V4 (see Fig. \ref{metrology_baselines}), in respectively the horizontal and vertical directions.

\begin{figure*}[t]
\begin{center}
\includegraphics[width = \textwidth]{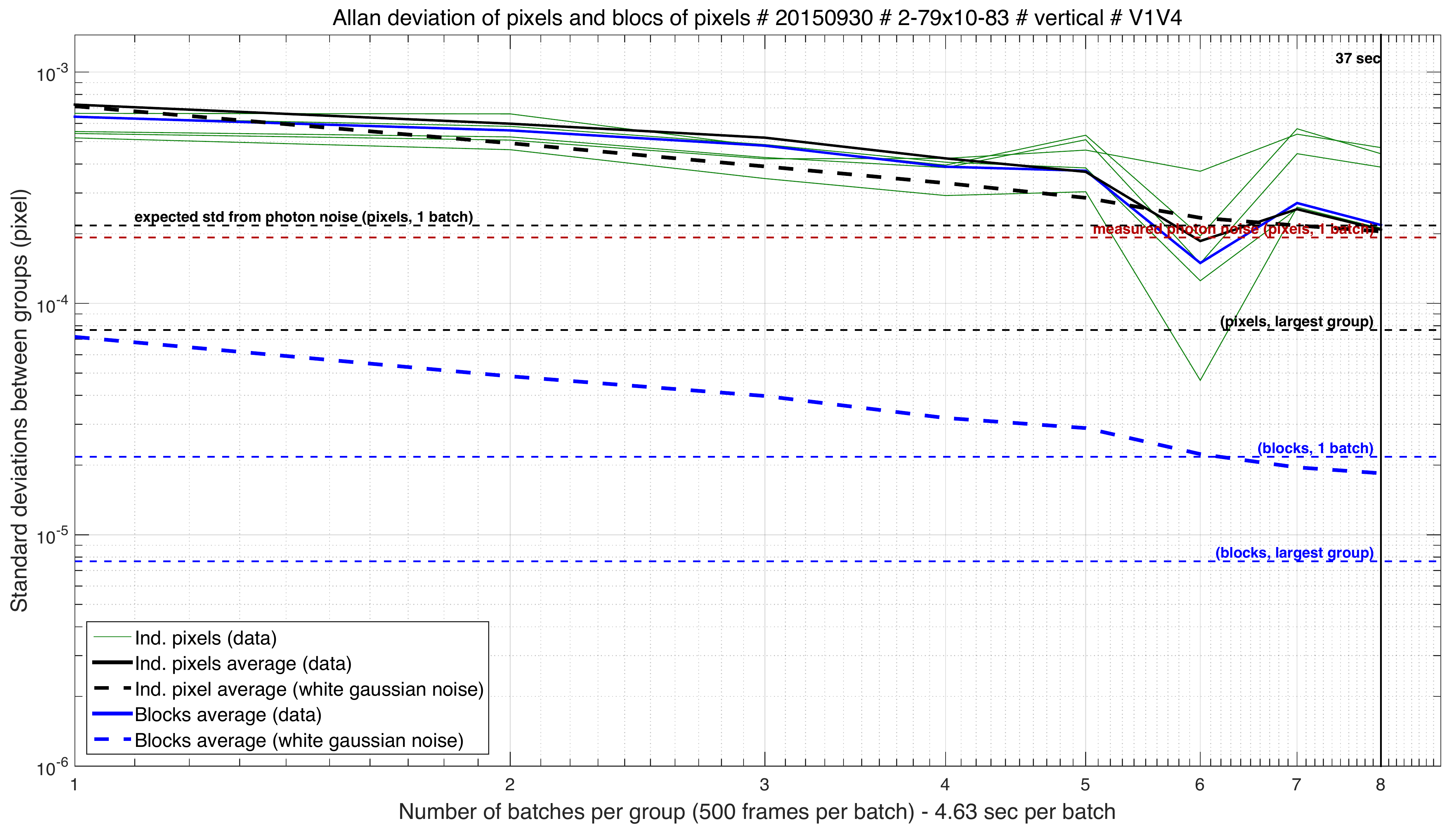}
\caption{\label{allanDevXpData} Allan deviations of projected pixel offsets (vertical baseline).}
\end{center}
\end{figure*}

Figure \ref{allanDevXpData} shows the Allan deviation of projected pixel offsets for the vertical baseline. At the maximum integration time, the deviation for individual pixels is $2\e{-4}$ pixel. In spite of all the measures used against stray light (coherence reduction and baffling), there is no improvement. The precision is far from the photon limit at $8\e{-5}$ pixel and correlated noise is still present. The result for the horizontal baseline are analogous (same final deviation, same issues with correlated noise). This suggests that stray light, although measurable, is not the largest source of correlated noise or systematics on the pixel offsets. Coherence reduction is expected to be ineffective if the stay light that is having an impact on the pixel offsets has a optical path difference smaller than 1 cm. In this case, further reduction of coherence below 1 cm would be beneficial.

\begin{figure*}[t]
\centering
\subfigure[Horizontal pixel offsets difference]{\label{xOffsetsDiff}
\includegraphics[width=0.48\textwidth]{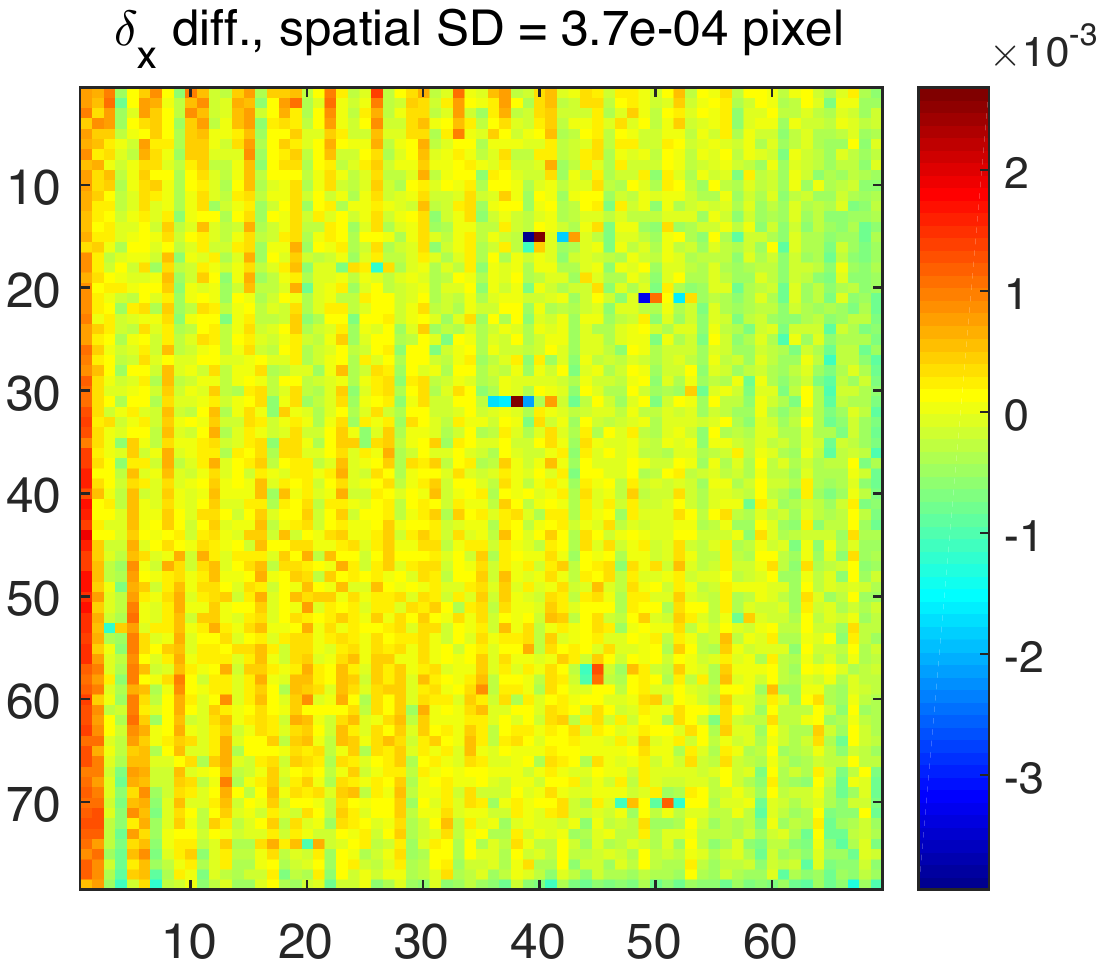}}
\subfigure[Vertical pixel offsets difference]{\label{yOffsetsDiff}
\includegraphics[width=0.48\textwidth]{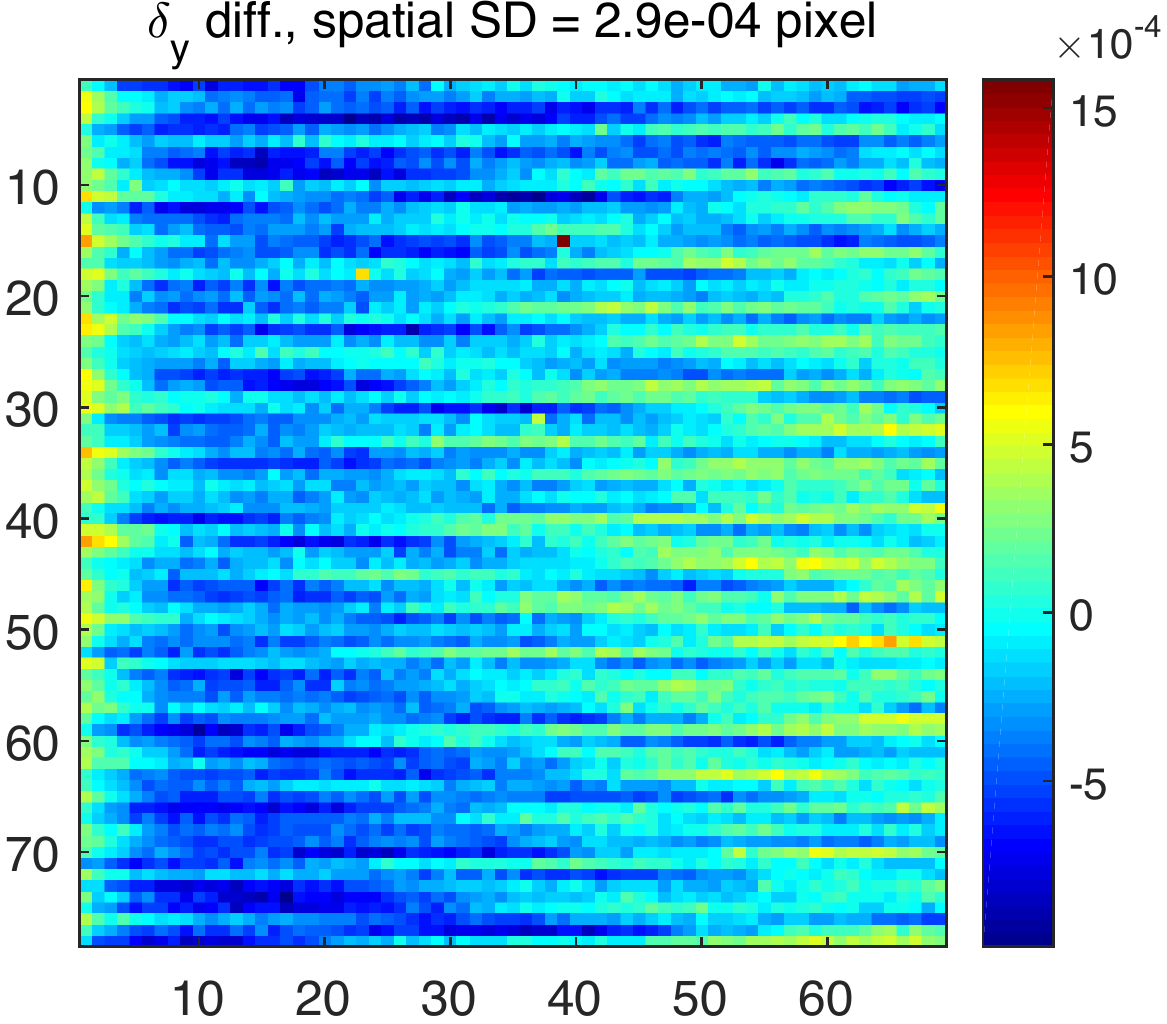}}
\vspace*{-2mm}
\caption{\label{offsetsDiff}Pixel offsets differences, between two different sets of metrology baselines.}
\end{figure*}

A more rigorous way to estimate the accuracy of pixel offsets measurements is to perform two independent analysis for two physically different pairs of baselines and to compare the results. Each pair have different separations, producing respectively fringes spacings of 2.4 and 4 pixels on the detector, and are located at different places in the vacuum chamber (the angles of incidence of metrology beams are changed). The stray light bias is different in each configuration. Figures \ref{xOffsetsDiff} and \ref{yOffsetsDiff} shows the difference between results for baselines H1H4-V1V4 and baselines H2H4-V2V4, for respectively horizontal and vertical pixel offsets.

Both the pixel offsets (Fig. \ref{xOffsetsXpData} and \ref{yOffsetsXpData}) and their differences (Fig. \ref{xOffsetsDiff} and \ref{yOffsetsDiff}) show that the dust contamination has biased the pixel offset measurements to at least several parts per thousand. However the effect is localized (one or a few pixels), its effect on astrometric measurements is identifiable and avoidable: it affects centroids independently and at precise and known locations.

The amplitudes of the differences have standard deviations up to $4\e{-4}$ pixel, versus an Allan deviation value of $2\e{-4}$ pixel. This indicates that static systematics (constant in time) are present in the data. In this case the visible systematics depend of the physical position of the metrology fibers. Moreover, this map of the difference shows structures and speckles. This indicates that stray light is still an issue, but other unknown systematics are possible as well. This data only gives a lower bound for the amplitude of systematics in the measured pixel offsets. The confirmation that the metrology measurements are correct is when the addition of pixel offsets into the data pipeline improves the astrometric accuracy (see next Section).

\subsubsection{Pseudo stars}\label{sec:xp data pseudo stars}

The pseudo star data which yields the best results is the one presented here (taken in ambient air). The translation stage supporting the detector was moved into 90 different positions by small steps of 0.17$\pm$0.01 pixel onto a vertical line. To each detector position corresponds a data cube of pseudo star data (the detector is not moved during acquisition). The \emph{single-position} analysis has not revealed any problem with the data. The precision obtained is $6\e{-5}$ pixel for all detector positions, when splitting each data cube into 4 batches. This corresponds to the photon noise limit. Extrapolating the photon noise to the whole cubes yields an expected photon noise accuracy limit at $3\e{-5}$ pixel.

\begin{figure}[t]
\begin{center}
\includegraphics[width = 0.5\textwidth]{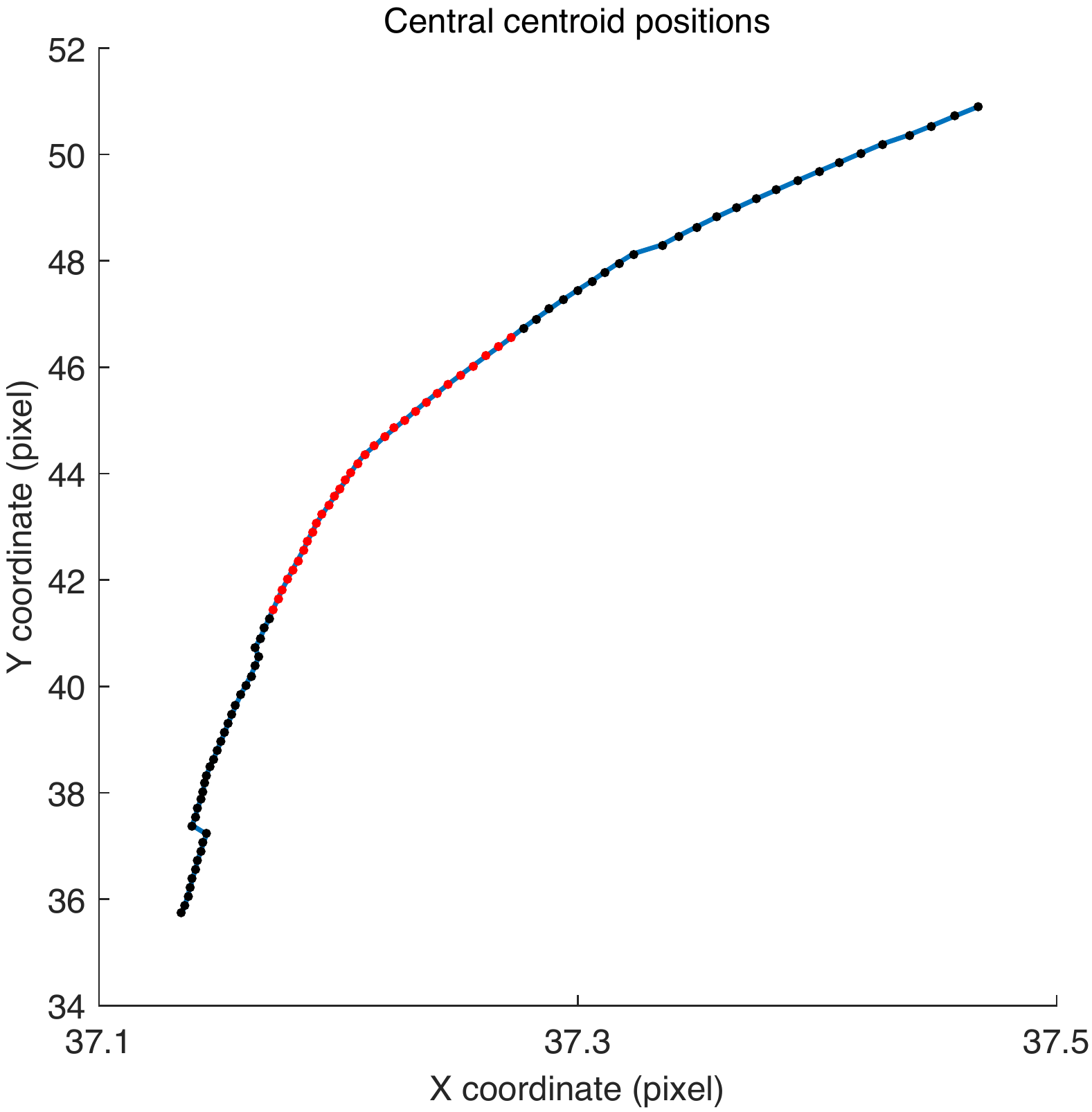}
\caption{\label{ccdPositionsAutocorr} Central centroid coordinates for the 26 Sep 2015 pseudo star dataset. The scale is much smaller for the horizontal axis. Indexes between 25 to 57 (final dataset for accuracy) are in red.}
\end{center}
\end{figure}

The cause of the non uniform detector motion ($\pm$0.01 pixel, and the line is actually not perfectly vertical and straight) is the erratic behavior of the translation stage which uses piezoelectric motors in an open loop. Figure \ref{ccdPositionsAutocorr} shows the pixel coordinates of the central centroid from positions 1 to 90.  The commands sent to the stage for each step were identical and for a vertical axis motion only.

The non uniformity of the translation has no major consequence for the analysis. However there is another related issue with the translation stage, which is intrinsic to its design. The piezoelectric motors are pushing against springs. To produce a translation, two parallel sliders are pushed by one motor each. This design allows for rotation motion by using opposite directions with each slider. But for translation, as each motor pushes in an irregular and unequal manner, this also creates significant unwanted tip tilt. When discussing the astrometric accuracy, this crucial point will be developed. 

\begin{figure}[t]
\begin{center}
\includegraphics[width = 0.5\textwidth]{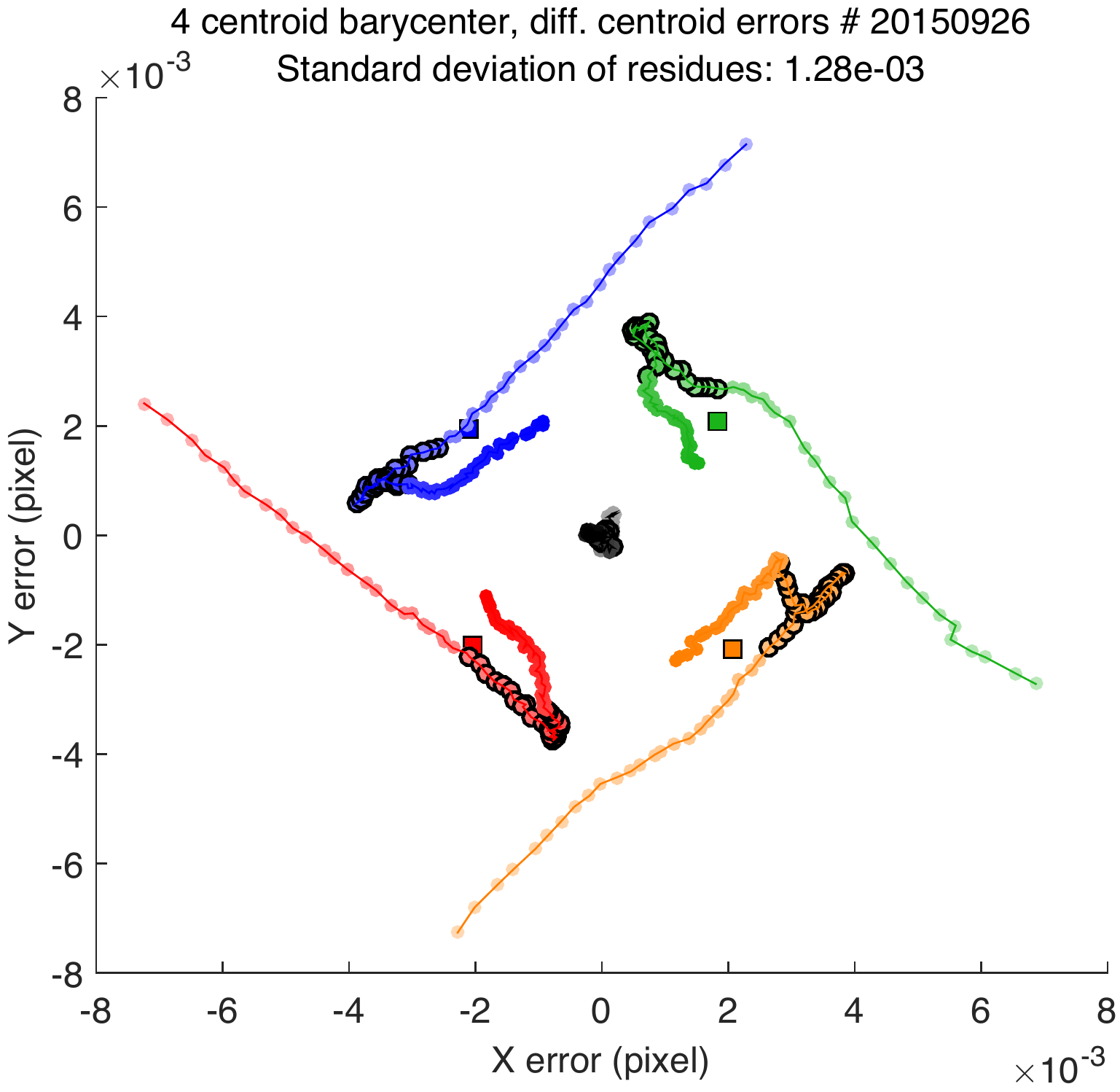}
\caption{\label{20150926_barycenter_residualsXvY_pos1to90} Barycenter residuals for each position (from index 1 to 90). Each centroid has a different color (black centroid is the central one). The relative positions between pseudo stars (illustrated by the squares) have been downscaled to correspond to the magnitude of the residuals. Indexes between 25 to 57 (final dataset for accuracy) are circled in black.}
\end{center}
\end{figure}

\begin{figure*}[t]
\begin{center}
\includegraphics[width = 0.85\textwidth]{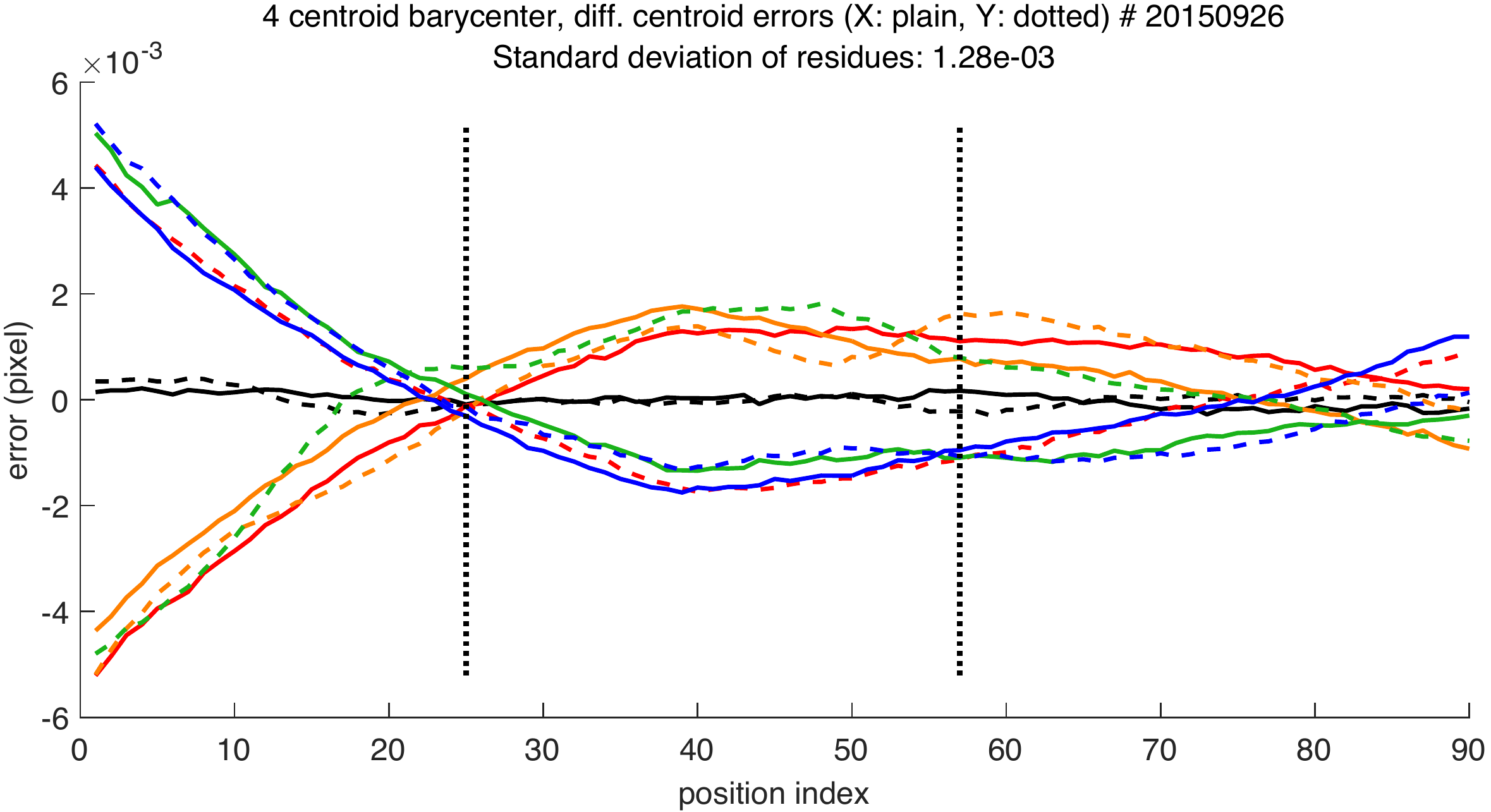}
\caption{\label{20150926_barycenter_residuals_pos1to90} Barycenter residuals for each position (from index 1 to 90). Each centroid has a different color (black for the central one). Plain lines are X axis residuals, dotted lines are Y axis residuals. Indexes between 25 to 57 (final dataset for accuracy) are indicated by the vertical dotted lines.}
\end{center}
\end{figure*}

Figures \ref{20150926_barycenter_residualsXvY_pos1to90} and \ref{20150926_barycenter_residuals_pos1to90} show the residuals obtained with a simple barycenter method, for all positions (1 to 90). More precisely, the residuals are the variations of the projected distances (on the horizontal and vertical axis) between each star and the barycenter of all the other ones. The large amplitude drifts of residuals are produced by a corresponding drift of the translation stage tip/tilt axis as a function of the position index. Correlations and anti-correlations between the stars are based on the geometric layout, as show by Fig. \ref{20150926_barycenter_residualsXvY_pos1to90}, where the same residuals have been plotted in a parametric manner (X versus Y). The unwanted motion is mainly a roll, but some tip-tilt is present as well.

\begin{figure*}[t]
\begin{center}
\includegraphics[width = \textwidth]{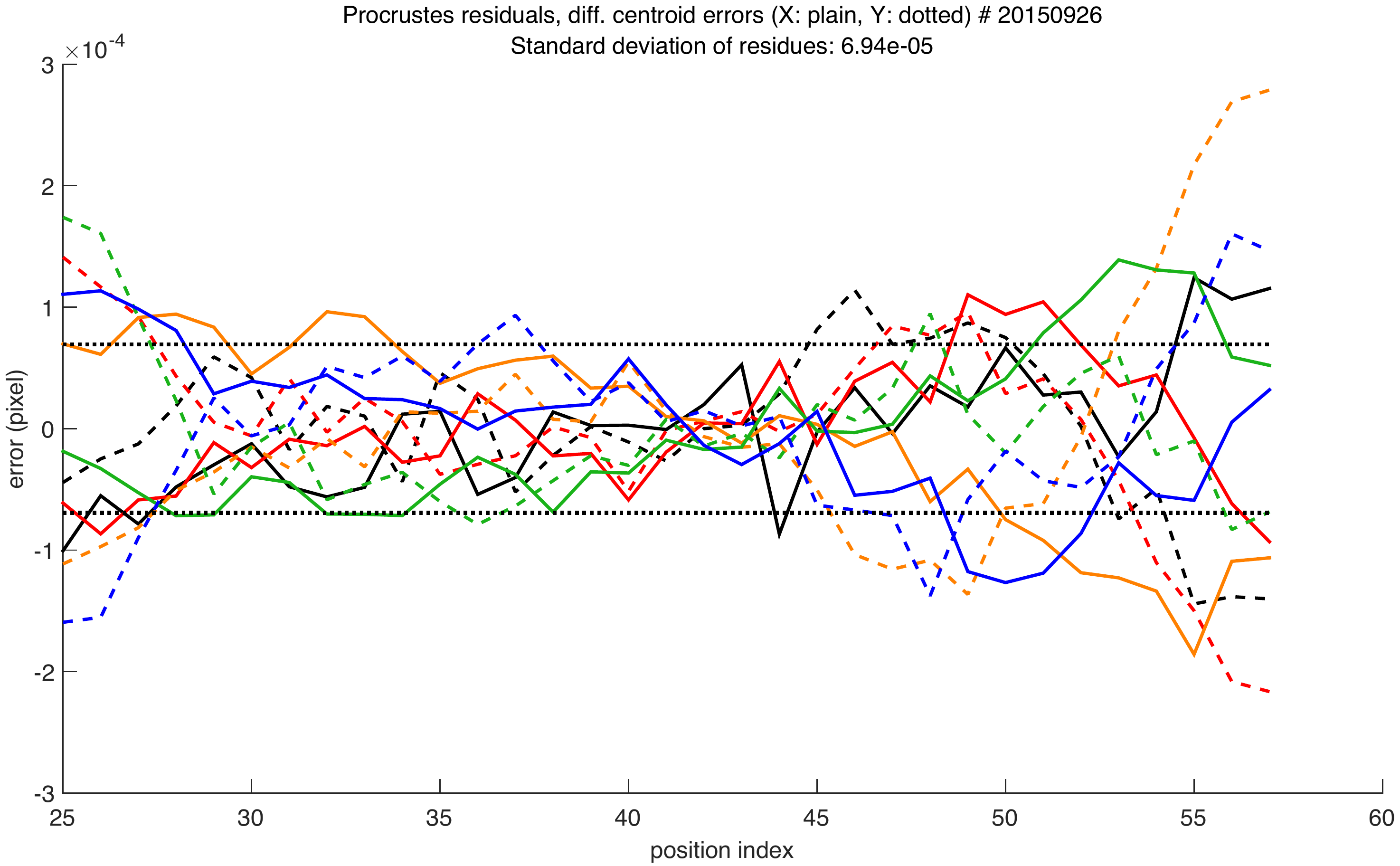}
\caption{\label{20150926_procrustes_residuals_pos25to57} Procrustes residuals for each position (from index 25 to 57). In this case the PRNU and pixel offset calibration are used. Each centroid has a different color (black centroid is the central one). The horizontal dotted lines indicate the dispersion of the residues ($\pm \sigma$).}
\end{center}
\end{figure*}

To minimize the impact of this issue, an area with relatively small tip tilt errors was selected, corresponding to positions 25 to 57, spanning 5.4 pixels. Figure \ref{20150926_procrustes_residuals_pos25to57} shows the Procrustes residuals for this range of indexes. The purpose of the Procrustes method is to compensate for the geometrical effects, after PRNU and pixel offsets corrections. After Procrustes, the residuals are smaller than with the simple barycenter technique: $6.9\e{-5}$ pixel versus $1.3\e{-3}$ pixel.

Table \ref{tab:Accuracies with different types of calibrations} shows the final accuracies (after Procrustes residuals) for the 25-57 position indexes, in 4 different calibration setups. The PRNU calibration (from flat field) and the pixel offset calibration (from metrology fringes) can each be activated independently. The final accuracy corresponding to each case is given in the table. It is important to note that these accuracies are the result of a \textit{multi-position} analysis and not a \textit{single-position} one (these terms are defined in Sect. \ref{Differential astrometric dispersion metrics}).

A test of spatial averaging over several detector positions was done. Instead of considering the residuals from each position separately, the centroids are averaged over groups of positions. The accuracy is then given by the Procrustes residuals over the centroids corresponding to each group. The 33 positions are binned into groups in the following way: positions 1 to 11, positions 12 to 22, positions 23 to 33, effectively forming 3 juxtaposed segments, without interlacing. In table \ref{tab:Accuracies with different types of calibrations}, accuracies corresponding to individual positions and groups of positions are respectively flagged by (p) and (g). For the groups, the photon limit ($3\e{-5}$ for individual positions) is below $1\e{-5}$.

\begin{table}[t]
\caption{Accuracies with different types of calibrations (in pixel units).}\label{tab:Accuracies with different types of calibrations}
\vspace{-0.5cm}
\begin{center}
\begin{tabular}{c c c}
\hline
- & No flat (pixel) & flat  (pixel)\\
\hline
\hline
No metrology & \specialcell{(p) $2.1\pm0.12\e{-4}$\\(g) $1.4\pm0.27\e{-4}$} & \specialcell{(p) $1.7\pm0.10\e{-4}$\\(g) $1.6\pm0.30\e{-4}$} \\
\hline
Metrology & \specialcell{(p) $2.9\pm0.17\e{-4}$\\(g) $2.1\pm0.40\e{-4}$} & \specialcell{(p) $9.7\pm0.56\e{-5}$\\(g) $5.9\pm1.1\e{-5}$}  \\
\hline

\multicolumn{3}{l}{\specialcell{The (p) and (g) flags respectively indicate accuracies over\\positions~/~groups of positions.}}
\end{tabular}
\end{center}
\end{table}

The best result was obtained when the flat field and metrology calibration were combined, yielding a final astrometric accuracy of $9.7\pm0.56\e{-5}$ pixel for individual positions and $5.9\pm1.1\e{-5}$ pixel when binning into 3 groups. The error bars displayed on the table are from Monte Carlo simulations with random astrometric jitter, independent for each star, as is expected from photon plus pixelation noise. This result is a good upper limit of the residual pixel calibration noise of a space instrument with the same CCD chip and metrology system. If there are other uncalibrated systematics caused by some shortcomings specific to the testbed (e.g. translation stage tip-tilt), these systematics will be absent on the real instrument.

\subsection{Discussion}

\subsubsection{Bias on pixel offsets}

There are indications that stray light is a significant source of errors for the pixel offset measurement:firstly flat field differences with fiber tip motion in coherent light show high bias: speckles with a relative contrast up to $1\e{-3}$ are visible in this case. Secondly when comparing the pixel offsets obtained with two different pairs of baselines, there is a difference significantly above photon noise. While reducing the coherence length results in a significant attenuation of the speckle contrast (about a factor 3), no gain was observed on the pixel offsets with reduced coherence. A possible explanation to this surprising result is that diffraction and/or diffusion on the baffle vane edges is the main source of noise on pixel offsets. In this case the OPD between direct and stray light is around 1 cm plus or minus a factor 2 (depending on which vane is considered), so the contrast attenuation obtained is small to non existent. Another attempt to improve the metrology was to average the results from 9 different fiber tip positions to provide angle diversity, but no gain was obtained. More theoretical work and simulations are needed to explain all these observations. 

The pixel offset difference is 3 times smaller than the pixel offset SD, so useful calibration information was obtained. The comparison of the astrometric accuracy with and without pixel offset calibration (after PRNU correction) has confirmed this: the accuracy was improved by about a factor two (when used with the flat field).

\subsubsection{Issue with tip-tilt error correction}

A critical issue was discovered with the Procrustes technique: it is in fact not powerful enough to correctly address the geometrical distortions down to the required accuracy. As each axis can change separately, plate scale changes can occur separately on the X and Y axis. The standard Procrustes technique only has one global scaling parameter, the technique used in this experiment is thus already a modified version with 2 independent scaling axis. But this is still insufficient. As the light beams corresponding to each star are coming from different angles, the effect of tip-tilts are not strictly homogeneous plate scale changes. A geometrical analysis of the problem reveals that the plate scale inhomogeneity between two opposite stars is proportional to $\alpha\theta$, with $\alpha$ the angular separation between the stars and $\theta$ the tip-tilt amplitude (see appendix \ref{append:Scaling inhomogeneity}). Fig. \ref{20150926_barycenter_residuals_pos1to90} and \ref{20150926_barycenter_residualsXvY_pos1to90} show these motions. Fig.~\ref{20150926_barycenter_residualsXvY_pos1to90} shows a roll, which impacts the astrometric accuracy and which is in principle not permitted by the translation stage. However this roll motion is not problematic since it is easily subtracted. The tip-tilt is the real issue.

A way more sophisticated than Procrustes superimposition to inverse the projection was implemented. The new method uses 6 parameters to characterize the detector position: X, Y, Z positions, and tip, tilt and roll angles. From these parameters and centroid angular coordinates (2 per centroid) their positions on the detector can be obtained through exact projection in 3D. A gradient descent method is used to simultaneously minimize the centroid angular coordinates and the detector parameters for each data cube (one data cube per detector position). The number of parameters to fit is $10+6 N_{\ml{cubes}}$, for $10 N_{\ml{cubes}}$ data points. At this point in the development process the proper working of the deprojection algorithm is uncertain. Monte Carlo simulations seem to indicate that only 5 centroids are not enough to properly retrieve the deprojection without overfitting. The problem may be too poorly constrained with too many free parameters. The problem is nonlinear and the correct parameters of the geometric projection are hard to constrain because there are other errors of the same magnitude affecting the centroids. An accurate retrieval of the projection parameters is possible only if the tip tilt favorably samples the tip-tilt parameter space. If two data cubes have close projection parameters, such that the resulting difference on centroids is smaller than the photon noise, they are really equivalent to a single cube. Because the tip-tilt is caused by dry friction, it can be the case. However it is hard to be certain, as in this case (tip tilt parameter space not properly sampled), the retrieved values of the projection parameters are not reliable. There was no independent mean of measuring the tip tilt to the required accuracy available on the experiment.

In our setup, $\alpha=1.6\e{-3}$ rad (separation defined as the side of the square formed by the pseudo stars, which is 40 pixels). The translation stage tip-tilt amplitude is about $5\e{-3}$ rad (peak to valley), as measured with both Procrustes and the projection inversion technique. So astrometry systematic noise induced can go up to $3\e{-4}$ pixel in the worst case and between opposite stars.


\subsubsection{Effectiveness of calibrations}

Table \ref{tab:Accuracies with different types of calibrations} displays several interesting features:
\begin{enumerate}
\item Spatial averaging of the centroid always improves the accuracy, although the gains from one configuration to the next are unequal. However it is hard to conclude anything given the error bars associated with these measurements.
\item PRNU calibration improves the accuracy for individual positions, and further improvement is obtained by adding the metrology calibration.
\item Using the metrology alone deteriorates the accuracy.
\end{enumerate}

Having on the same dataset observations two and three is surprising, so far no explanation has been found. On most previous data sets both the PRNU and metrology have had impacts on the astrometric accuracy of the same sign whether used alone or combined. Non linearities seem to arise from the biases present in the metrology calibration result and are not well understood. This does not question the validity of the final astrometric result, which has been checked through numerical simulations and on other datasets. However the simplifying assumptions used in the numerical simulations are limiting the ability to interpret the observations. Another startling result is the accuracy before PRNU, which is already good. From a PRNU RSD measured at $2.4\e{-3}$, this was not expected. This result has been consistently obtained over all datasets. It seems that the good pixel to pixel homogeneity of the back-illuminated CCD chosen as a detector allow for a good baseline astrometric accuracy (without calibration). With the flat field only, an accuracy of $1.7\e{-4}$ is reached. A possible explanation for this good pre-flat accuracy is that the PRNU measured has a non Gaussian distribution (dust contamination, detector edge effects), it could thus have a different impact on astrometry than expected. The astrometric impact is minimized if the PSF stays clear of the dust and edges.

The gain in astrometric accuracy obtained from the flat field plus pixel offset calibrations (without spatial averaging) is moderate: a factor 2.2. One possible reason for this moderate gain is the spectral dependency of the pixel responses: the pixel offsets are measured at 633 nm whereas the pseudo stars cover all of the visible spectrum. CCDs, and in particular back-illuminated CCDs are known to show measurable spectral dependency \citep{1995PASP..107.1094R}. This spectral effect was not investigated in the experiment, the metrology system was designed for operation at 633 nm and the only source available was the HeNe laser.

\subsubsection{Spatial averaging}

The experimental results show that astrometric accuracy can be increased by spatial averaging, thus spreading the pixelation errors over more pixels than would be allowed by a single PSF. The gain is however limited, for example from $9.7\e{-5}$ to $5.9\e{-5}$ pixel with all calibrations active. A much more useful data set for this technique would be a grid with a large number of detector positions. In fact an analysis was done over such a dataset (a grid of 340 positions, spacing step of one pixel), with an otherwise identical experimental setup, and using the same method as described in Sect. \ref{sec:xp data pseudo stars}. A gain in accuracy of about a factor 30 (from $2.4\e{-3}$ to $7\e{-5}$ pixel) with groups of interlaced positions was obtained. With a spacing as small as 1 pixel, a good fraction of systematics cancels out in the differential astrometry, including in particular pixelation and tip tilt errors. Because of poor starting accuracy in this case (large tip tilt errors), the final accuracy after spatial averaging is not better than for the other dataset. It does however suggest a way to mitigate pixelation systematics for a space mission. The differential astrometric measurement of each epoch can be an average of several different astrometric measurements done with centroid at different places on the CCD.

%

\section{Conclusion}

The calibration system yielded the pixel positions to an accuracy estimated at $4\e{-4}$ pixel. After including the pixel position information, an astrometric accuracy of $6\e{-5}$ pixel was obtained, for a PSF motion over more than 5 pixels. Without the (flat and metrology) calibrations the astrometric accuracy is $1.4\e{-4}$ pixel  (all other things equal). With the \textit{single-position} mode (small jitter motion of less than $1\e{-3}$ pixel), a photon noise limited precision of $3\e{-5}$ pixel was reached.

The \textit{single-position} result shows that the detector and electronics dark and readout noises are well behaved and do not prevent reaching higher accuracies. The number that is relevant for an astrometric mission is the \textit{multi-position} analysis result: $6\e{-5}$ pixel. It characterizes the residual noise from pixelation errors after calibrations. As this accuracy was obtained for a motion over 5.4 pixels, a distance larger than the PSF diameter, it can be extrapolated to the whole CCD, considering only pixelation noise and assuming no spatial correlation of pixels properties. In the DICE experiment the translation stage tip tilt is responsible for the larger errors associated with wider motions. 

The \textit{single-position} results confirm that turbulence (in the closed vacuum chamber) is not an issue. A photon noise of $3\e{-5}$ pixel was reached for individual data cubes (results in Sect. \ref{sec:xp data pseudo stars}), which correspond to the expected photon noise. This first test on the main dataset presented here validates the absence of a lot systematics that could have been an issue, before even considering the \textit{multi-position} analysis, but still has a photon limit higher than the final requirement. Other tests on special datasets, also in air, and with much longer acquisition time showed that this \textit{single-position} precision can be improved further, at least to $10^{-5}$ pixel, while remaining at the photon limit. Some of the earlier datasets were taken in air and in vacuum, all other conditions unchanged, but no gain on the final astrometric accuracy was measured when going from air to vacuum.

One of the main sources of noise for the metrology seems to be stray light, even after numerous baffle upgrades attempted to solve the issue. More work is needed to identify other possible sources of systematics and to understand what are the best ways to mitigate stray light. The limited effectiveness of the metrology calibration on astrometric accuracy ($\sim$2 fold improvement) could be caused by a spectral dependency of the PRFs. To first quantitatively assess and then mitigate the issue the same interferometric calibration should be performed at several wavelengths distributed across the visible spectrum.

The final objective was set at $5\e{-6}$ pixel. In reality, the exact requirement depends on the spacecraft parameters and scientific objectives. For the new mission concept Theia, it is lightly easier: $10^{-5}$ pixel \citep{2015pathwaysMalbet,Malbet2016}. Theia can still detect nearby habitable Earths, even if it is restricted to slightly fewer and closer stars. Further progress is needed to reach the required Theia accuracy, but several leads have been identified to improve the metrology calibration. Additionally, as the experimental data showed, it could be possible to significantly enhance the final accuracy by averaging the relative star coordinates over several detector positions (previously refereed to as spatial averaging). In a real mission, one can reasonably conceive that up to 100 different positions (per epoch) could be used, resulting in the best case into a relaxing of the specification up to a factor 10 (the maximum gain is given by the square root of the number of positions). However spatial averaging is not desirable as a first approach to increase accuracy as it could impose significant additional constraints on the instrument capabilities, such as fast re-pointing, higher bandwidth or on-board processing (e.g. to "shift and add" images) and could decrease the overall instrument efficiency, by impacting the number of observable targets, or percentage of time spent collecting photons versus doing maneuvers). The experiment was put into storage in October 2015, but it is still functional and can be restarted if interest arises, for example in the context of a mission phase A study.



\bibliographystyle{aa}
\bibliography{article_DICE_AandA_v1}


\clearpage
\begin{appendix}

\section{Deprojection of pixel offsets}\label{append:Deprojection of pixel offsets}

\begin{figure}[t]
\includegraphics[width = 0.45\textwidth]{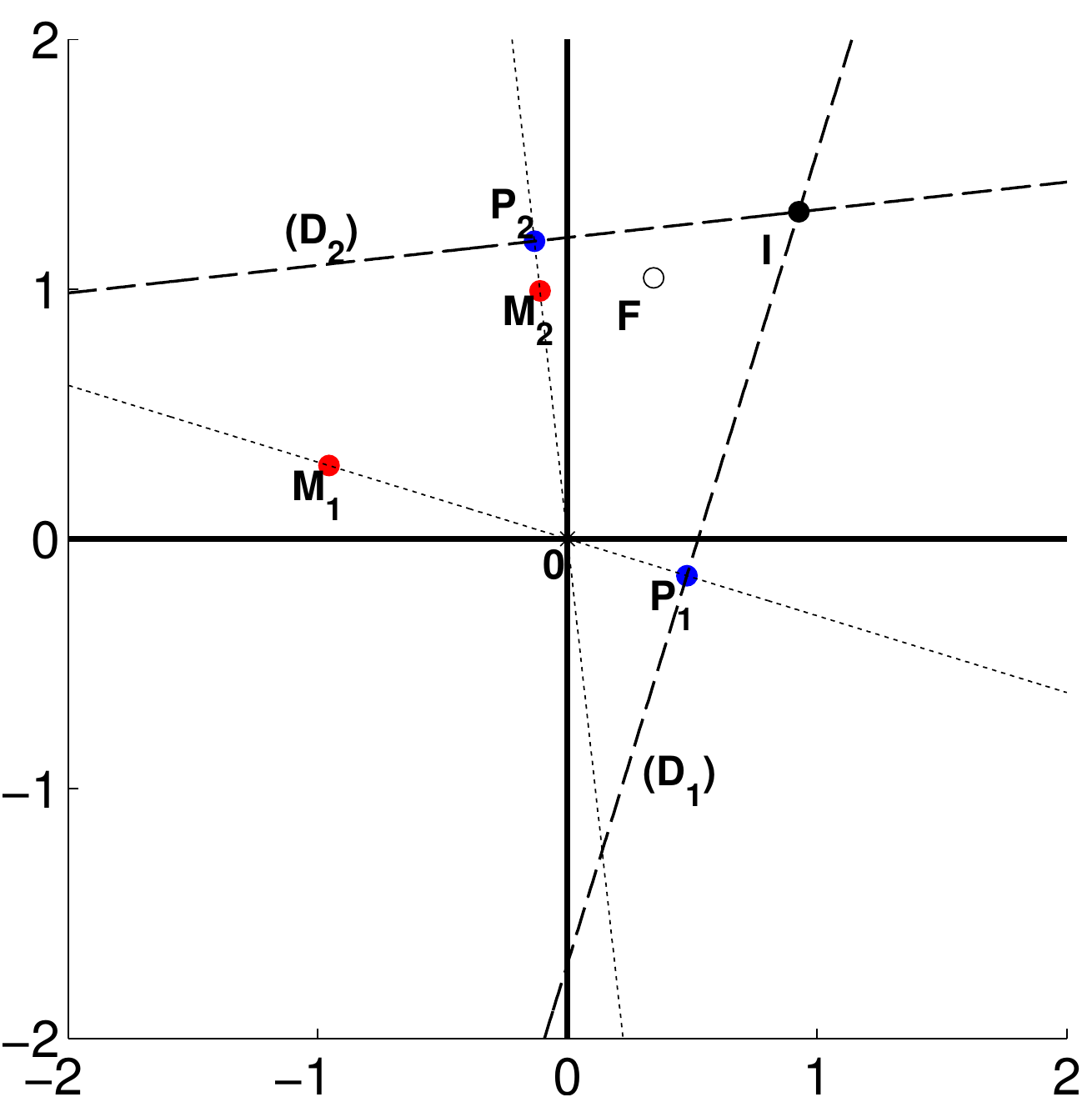}
\caption{Illustration of the deprojection problem. $\protect\overrightarrow{OM_1}$ and $\protect\overrightarrow{OM_2}$ are the wave vectors normalized to a length of one pixel. The points $P_1$ and $P_2$ represent the measured offsets ($\pmb{\delta_{p,1}}, \pmb{\delta_{p,2}}$), i.e. the projections of the true pixel offset (point $I$), unto the lines generated by the wave vectors. The axis units are in pixels. Simply summing the projected offsets gives a wrong answer (point $F$).}
\label{deprojection_schematic}
\end{figure}

\begin{figure}[t]
\centering
\includegraphics[width = 0.35\textwidth]{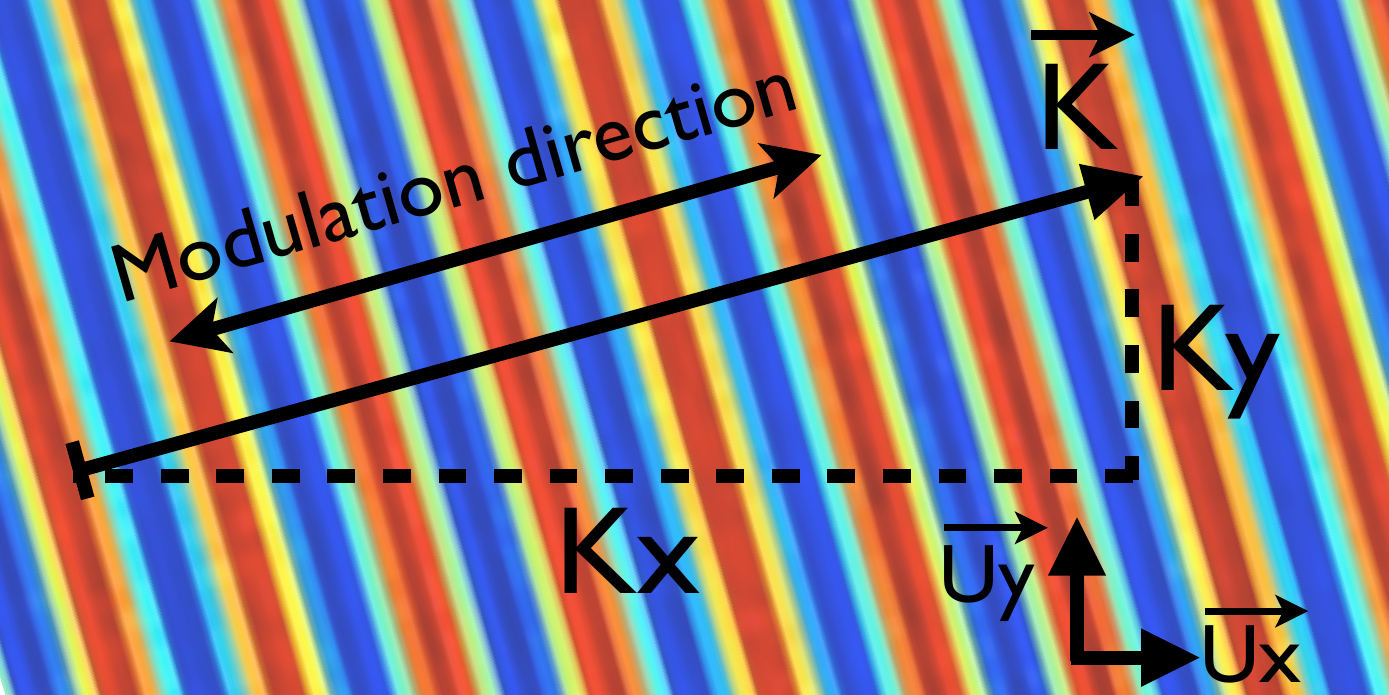}
\caption{Definition of the wave vector. For a given metrology orientation, the wave vector is defined as $\vec{K} = (K_x(t),K_y(t))$, for a plane wave of the form: $I(x,y,t)\propto\sin(x K_x(t)+y K_y(t) + \phi(t))$. The wave vector is by definition perpendicular to the fringes and aligned with the modulation direction, i.e. the apparent motion of the fringes when the phase changes. The default distance unit is the pixel, so the magnitude of the vector is $|\vec{K}|=2\pi/\lambda$, with lambda in pixel units.} 
\label{wavevector}
\end{figure}

Figure~\ref{deprojection_schematic} illustrates the ‘‘deprojection" problem, the solution is found by straightforward application of Euclidean geometry in the detector plane. The definition of the wavevector in shown by Fig.~\ref{wavevector}).

The goal is to transform a set of two non-aligned projected offsets into X,Y offset coordinates. Let ($\alpha_1$,$\alpha_2$) be the coefficients such that $\overrightarrow{OP_1} = \alpha_1 \overrightarrow{OM_1}$ and $\overrightarrow{OP_2} = \alpha_2 \overrightarrow{OM_2}$: they are the offsets measured by the metrology. The wave vectors $\pmb{K_1}$, $\pmb{K_2}$ and the projections $\alpha_1$, $\alpha_2$ are known, we search the Cartesian coordinates of the point $I=(x,y)$ which by construction lies at the intersection of lines $(D_1)$ and $(D_2)$ in fig\ref{deprojection_schematic}. One can not use a simple base change: it would yield the coordinates of the point $F = \overrightarrow{OP_1} + \overrightarrow{OP_2}$, which has a different location than $I$ if the metrology baselines are not orthogonal.
\\

Let the coordinates of the normalized wave vectors be $k_{x,1}, k_{y,1}, k_{x,2}, k_{y,2}$, for respectively metrology cube 1 (first baseline and corresponding orientation) and cube 2 (second baseline and corresponding orientation). The direction vectors of lines $(D_1)$ and $(D_2)$ are noted $c_{x/y,1/2}$ and have trivial relations with the wave vectors coordinates:

\begin{equation}
\left\{ \begin{array}{ll} 
& c_{x,1} = k_{y,1}\\
& c_{y,1} = - k_{x,1}\\
& c_{x,2}= k_{y,2}\\
& c_{y,2}= - k_{x,2}\\
\end{array}\right.
.
\end{equation}

The lines $(D_1)$ and $(D_2)$ are passing thought respectively $P_1 = (d_{x,1},d_{y,1})$ and $P_2 = (d_{x,2},d_{y,2})$, of coordinates:

\begin{equation}
\left\{ \begin{array}{ll} 
d_{x,1} = k_{x,1}\alpha_1\\
d_{y,1} = k_{y,1}\alpha_1\\
d_{x,2} = k_{x,2}\alpha_2\\
d_{y,2} = k_{y,2}\alpha_2\\
\end{array}\right.
.
\end{equation} 

The parametric equation of line $(D1)$, passing thought $P_1$, of direction vector $(c_{x,1},c_{y,1})$, is:
\begin{equation}
\left\{ \begin{array}{ll}  
x = c_{x,1} t + d_{x,1}\\
y = c_{y,1} t + d_{y,1}
\end{array}\right.
.
\end{equation}

$t$ is removed from the system above:
\begin{eqnarray}\nonumber
\left\{ \begin{array}{ll}  
c_{y,1} x = c_{y,1} c_{x,1} t + c_{y,1} d_{x,1}\\
c_{x,1} y = c_{x,1} c_{y,1} t + c_{x,1} d_{y,1}
\end{array}\right.
\Rightarrow c_{y,1} x - c_{x,1} y &=& c_{y,1} d_{x,1} - c_{x,1} d_{y,1}
.
\end{eqnarray}

The second line has the same equation, but with index 2. The coordinates of the point P is at the lines intersection and thus verifies the system:
\begin{eqnarray}
\label{eq:intersection_eq_system}
\left\{ \begin{array}{ll}  
c_{y,1} x - c_{x,1} y = c_{y,1} d_{x,1} - c_{x,1} d_{y,1}\\
c_{y,2} x - c_{x,2} y = c_{y,2} d_{x,2} - c_{x,2} d_{y,2}
\end{array}\right.
.
\end{eqnarray}

In Eq. (\ref{eq:intersection_eq_system}), by multiplying line 1 by $c_{x,2}$ and line 2 by $c_{x,1}$, we find $x$:
\begin{eqnarray}
&\Rightarrow x(c_{x,2} c_{y,1} - c_{x,1} c_{y,2}) = c_{y,1} c_{x,2} d_{x,1} - c_{y,2} c_{x,1} d_{x,2}\\
&\Rightarrow x = \frac{c_{x,2} c_{x,1} d_{y,2} + c_{y,1} c_{x,2} d_{x,1} - c_{y,2} c_{x,1} d_{x,2} - c_{x,2} c_{x,1} d_{y,1}}{c_{x,2} c_{y,1} - c_{x,1} c_{y,2}}
.
\end{eqnarray}

The solution for $y$ is analog (indexes 1 and 2 are switched). The final solution is:
\begin{eqnarray}
x &=& \frac{c_{x,2} c_{x,1} d_{y,2} + c_{y,1} c_{x,2} d_{x,1} - c_{y,2} c_{x,1} d_{x,2} - c_{x,2} c_{x,1} d_{y,1}}{c_{x,2} c_{y,1} - c_{x,1} c_{y,2}}\\
y &=& \frac{c_{x,1} c_{x,2} d_{y,1} + c_{y,2} c_{x,1} d_{x,2} - c_{y,1} c_{x,2} d_{x,1} - c_{x,1} c_{x,2} d_{y,2}}{c_{x,1} c_{y,2} - c_{x,2} c_{y,1}}
.
 \end{eqnarray}

\section{Linear sine wave fit}\label{append:Linear sine wave fit}

In Eq. \ref{temporal_sinwave}, the sine wave is rewritten as a sine + cosine functions. To simplify formulas, we assume $A=1$ and $B=1$ (this does not impact the generality of the result).
\begin{equation}\label{linear_sinwave_coefficients}
\begin{split}
I(i,j,t) &=& \iota(i,j) + \alpha(i,j) \sin\left[\phi(t) + \phi(i,j)\right]  \\ 
		 &=& a_{i,j} \sin(\phi(t)) + b_{i,j} \cos(\phi(t)) + c_{i,j}
\end{split}
.
\end{equation}

A least square minimization of the sum:
\begin{equation}
S_{i,j} = \sum_{t=0}^{N-1} \left[I(i,j,t)- a_{i,j}\sin(\phi(t)) - b_{i,j}\cos(\phi(t)) - c_{i,j}\right]^2.
\end{equation}
yields the values for $a$, $b$ and $c$ from which $\alpha(i,j)$, $\phi(i,j)$ and $\iota(i,j)$ are derived for each pixel. \\

For each pixel, we have introduced three coefficients $(a,b,c) \in\mathbb{R}^3$. The equality is obtained when $a,b,c$ verify:

\begin{equation}
\left\{ \begin{array}{ll} 
& a = \alpha \cos\phi_{i,j} \\
& b =  \alpha \sin\phi_{i,j} \\
& c = \iota
\end{array}\right.
.
\end{equation} 

For simplicity we have dropped most of the $i,j$ indexes and have rewritten the phases $\phi(i,j) = \phi_{i,j}$ and $\phi(t) = \phi_t$. The system above is equivalent to:
\begin{equation}
\left\{\begin{array}{ll} 
& \alpha = \sqrt{a^2 + b^2} \\
& \phi_{i,j}  =  \ml{arctan2}(b,a)   \\
& \iota = c
\end{array}\right.
.
\end{equation} 

We fit the constants (a,b,c) by minimizing S, the sum of the quadratic errors between the model and
the data points: $S=\sum_{t=0}^{N-1}\left((I_t)-(a\sin(\phi_t)+b\cos(\phi_t)+c)\right)^2$. The minimum verifies: 

\begin{equation*}
\left\{ \begin{array}{ll}  
\frac{\partial S}{\partial c} = 0\\
\frac{\partial S}{\partial a} = 0\\
\frac{\partial S}{\partial b} = 0      
\end{array}\right.
\Leftrightarrow
\left\{ \begin{array}{ll}  
\sum_{t=0}^{N-1} I_n - a\sin(\phi_t) - b\cos(\phi_t) - c\!\!\!\! &=0\\
\sum_{t=0}^{N-1} \sin(\phi_t) (I_n - a\sin(\phi_t) - b\cos(\phi_t) - c)\!\!\!\! &=0\\
\sum_{t=0}^{N-1} \cos(\phi_t) (I_n - a\sin(\phi_t) - b\cos(\phi_t) - c)\!\!\!\! &=0
\end{array}\right.
.
\end{equation*}

\begin{equation}
\nonumber
\Leftrightarrow
\begin{bmatrix}
1 & \sum_{t=0}^{t-1} \sin(\phi_t) & \sum_{i=0}^{t-1} \cos(\phi_t)\\
\sum_{t=0}^{t-1} \sin(\phi_t) & \sum_{i=0}^{t-1} \sin^2(\phi_t) & \sum_{i=0}^{n-1} \cos(\phi_t) \sin(\phi_t) \\
\sum_{t=0}^{t-1} \cos(\phi_t) & \sum_{i=0}^{t-1} \cos(\phi_t) \sin(\phi_t) & \sum_{i=0}^{n-1} \cos^2(\phi_t)
\end{bmatrix}
\times
\begin{bmatrix}
c \\
a \\
b
\end{bmatrix}
.
\end{equation}

\begin{equation}
\centering
\:\:\:\:\:\:\:\:\:\:\:\:\:\:\:\:\:\:\:\:\:\:\:\:\:\:\:\:\:\:\:\:\:\:\:\:=
\begin{bmatrix}
\sum_{t=0}^{n-1} I_n\\
\sum_{t=0}^{n-1} I_n \sin(\phi_t) \\
\sum_{t=0}^{n-1} I_n \cos(\phi_t) 
\end{bmatrix}
.
\end{equation}

\section{Coherent and incoherent stray light}\label{append:Coherent and incoherent stray light}

In the case of incoherent light, we have: $\ml{SNR} \propto \frac{I_0}{I_1}$, where $I_0$ is the light intensity from the source (direct path) and $I_1$ is the stray light intensity, that is after reflection(s) inside the vacuum chamber or the baffle. 

The stray light is a much more sensitive issue when the light is coherent: consider the fringe pattern created by two beams of respective intensities $I_0$ and $I_1$, the resulting intensity can be written as:
$$ I_0 + I_1 + 2\sqrt{I_0 I_1} \cos (2\pi \delta_{12}(M)).$$
Here $\delta_{12}(M)$ is the path-length difference between the beams 1 and 2 and the point M. 

In the case $I_0 >> I_1$, where $I_0$ is the main beam and $I_1$ is the stray light (the intensity, after (multiple) reflection(s), on the CCD), we have: $\ml{SNR} \approx \frac{I_0}{2\sqrt{I_0I_1}} = \frac{1}{2}\sqrt{\frac{I_0}{I_1}}.$ $I_0$ is the intensity created by the main beam and $2\sqrt{I_0I_1}$ is the secondary fringe pattern created by the stray light. This means that this kind of stray light must be taken care off very minutely.

\section{Scaling inhomogeneity caused by 3D projection}\label{append:Scaling inhomogeneity}

\begin{figure}[t]
\begin{center}
\includegraphics[width = 0.5\textwidth]{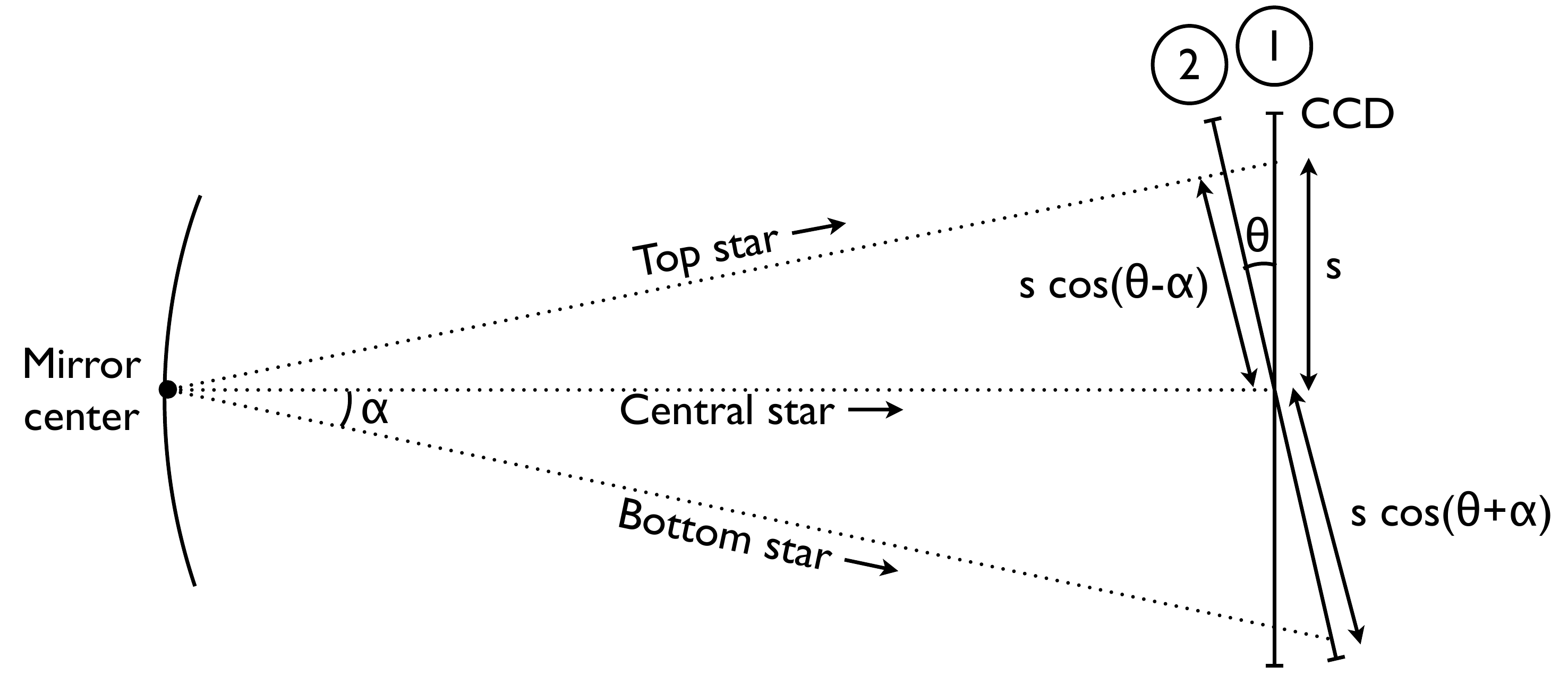}
\caption{\label{non_homogeneous_scaling_projection} Illustration of the non-homogeneous projection effect.}
\end{center}
\end{figure}

The $\alpha\theta$ inhomogeneity of scaling is illustrated by Fig.~\ref{non_homogeneous_scaling_projection}, where $\alpha$ and $s$ are the separation between the stars (in respectively angular and pixel units) and $\theta$ is the tilt with respect to the normal configuration. $s_0$ is a reference length (in pixels). When the CCD is tilted from position 1 (normal) to position 2 (tilt = $\theta$), the effective separations measured on the CCD are modified in the following way:

\bit
\item Between the top and central star, the effective separation $s_{\ml{top}}$ is scaled by:\\
$\gamma_{\ml{top}}=1/\cos(\theta-\alpha) \approx 1+\frac{\theta^2}{2}-\alpha\theta+\frac{\alpha^2}{2}$.
\item Between the bottom and central star, the effective separation $s_{\ml{bot}}$ is scaled by:\\ $\gamma_{\ml{bot}}=1/\cos(\theta+\alpha) \approx 1+\frac{\theta^2}{2}+\alpha\theta+\frac{\alpha^2}{2}$.
\ei 

The non homogeneity of the scaling is thus given at the first order by $\abs{s_{\ml{bot}}/s_{\ml{top}}} \approx 2\alpha\theta$. There is a $\frac{\theta^2}{2}$ term affecting the whole field uniformly, but the additional $2\alpha\theta$ term originating by the star separation produces another non uniform scaling effect. In the example above, the upper part of the CCD FoV undergoes a slight differential scale up, and vice versa for the lower part.

\end{appendix}

\end{document}